\documentclass{aa}  

\usepackage{graphicx}
\usepackage{txfonts}
\usepackage[colorlinks=true, allcolors = blue]{hyperref}
\usepackage{subfigure}

\usepackage{xcolor}
\usepackage{pdflscape}
\usepackage{todonotes}
\usepackage{multirow}
\usepackage{threeparttable}
\usepackage{capt-of}
\usepackage{longtable}
\usepackage[colorlinks=true, allcolors = blue]{hyperref}
\usepackage[normalem]{ulem}

\newcommand{\blue}[1]{\textcolor{black}{#1}}

\newcommand{\grey}[1]{\textcolor{gray}{#1}}

\begin{document}

	\title{X-Shooting ULLYSES: Massive stars at low metallicity }\subtitle{XV. 
		On the metallicity dependence of B-supergiant mass-loss rates}
	
	
	\author{O.\ Verhamme
		\inst{\inst{\ref{inst:KUL}, \ref{inst:KASI}}},
		J.O.\ Sundqvist\inst{\ref{inst:KUL}},
		A.\ de Koter \inst{\ref{inst:KUL}, \ref{inst:UAmst}},
		H.\ Sana\inst{\ref{inst:KUL}},
		F.\ Backs \inst{\ref{inst:KUL}},
		S. A. Brands \inst{\ref{inst:UAmst}},
        D. Debnath \inst{\ref{inst:KUL}},
        N. Moens \inst{\ref{inst:cmpa}},
        P. Schillemans \inst{\ref{inst:KUL}},
        C. Van der Sijpt \inst{\ref{inst:KUL}},
        S. R. Berlanas \inst{\ref{inst:CSIC-INTA}},
        M.\ Bernini-Peron \inst{\ref{inst:ari}},
        P.A.\ Crowther \inst{\ref{inst:sheffield}},
        A. C.\ Gormaz-Matamala \inst{\ref{inst:ceske}},
        R.\ Kuiper \inst{\ref{inst:UDE}}, 
        C.\ Hawcroft \inst{\ref{inst:STSCI}},
        F. Najarro \inst{\ref{inst:CSIC-INTA}},
        D. Pauli \inst{\ref{inst:KUL}},
        A.A.C.\ Sander \inst{\ref{inst:ari}},
        J.Th.\ van Loon \inst{\ref{inst:LJL}},
        J.S.\ Vink \inst{\ref{inst:armagh}},
        H.\ Todt \inst{\ref{inst:Pots}},
        F.\ Tramper \inst{\ref{inst:CSIC-INTA}},
        and Xshootu collaboration
	}
	\institute{
    {Institute of Astronomy, KU Leuven, Celestijnenlaan 200D, 3001, Leuven, Belgium}\label{inst:KUL}
    \and
    {Korea Astronomy and Space Science Institute (KASI) 776, Yuseong-gu, Daejeon 34055, Republic of Korea}\label{inst:KASI}\\
		\email{olivierverhamme@kasi.re.kr}
		\and
		{Anton Pannekoek Institute for Astronomy, University of Amsterdam, Science Park 904, 1098 XH Amsterdam, The Netherlands \label{inst:UAmst}}
        \and
        {Centre for mathematical Plasma Astrophysics, Department of
        Mathematics, KU Leuven, Celestijnenlaan 200B, Leuven, 3001, Belgium \label{inst:cmpa}}
        \and 
        {Departamento de Astrofísica, Centro de Astrobiología, (CSIC-
        INTA), Ctra. Torrejón a Ajalvir, km 4, 28850 Torrejón de Ardoz,
        Madrid, Spain \label{inst:CSIC-INTA}}
         \and 
        {Zentrum f{\"u}r Astronomie der Universit{\"a}t Heidelberg, 
        Astronomisches Rechen-Institut, M{\"o}nchhofstr. 12-14, 69120 
        Heidelberg, Germany \label{inst:ari}}
        \and 
        {Department of Physics \& Astronomy, Hounsfield Road, University of Sheffield, Sheffield, S3 7RH United Kingdom \label{inst:sheffield}}
         \and
         {Astronomický ústav, Akademie věd Ceské republiky, Fričova 298, 251 65 Ondřejov, Czech Republic\label{inst:ceske}}
          \and 
         {Faculty of Physics, University of Duisburg-Essen, Lotharstra{\ss}e 1, D-47057 Duisburg, Germany \label{inst:UDE}}
          \and 
         {Space Telescope Science Institute, 3700 San Martin Drive, Baltimore, MD, 21218, USA \label{inst:STSCI}}
        \and 
        {Lennard-Jones Laboratories, Keele University, ST5 5BG, UK \label{inst:LJL}}
        \and
        {Armagh Observatory and Planetarium, College Hill, BT61 9DG Armagh, UK \label{inst:armagh}}
         \and 
         {Institut f\"ur Physik \& Astronomie, Universit\"at Potsdam,Karl-Liebknecht-Str. 24/25, 14476 Potsdam, Germany \label{inst:Pots}}
	}
	\date{Received December 1, 2025; accepted February 16, 2026}
	
	
	\abstract
    {Our limited understanding of the winds of massive stars hampers our ability to predict their evolution.
    We currently often rely on mass-loss rate prescriptions that show strong features that lack a firm empirical confirmation, such as the bi-stability jump. This bi-stability jump is a pronounced increase in mass loss in the decreasing temperature regime $T_{\rm eff} \sim$28-21\,kK, i.e., spectral types B1-B2.
    Different prescriptions also give different metallicity dependences; these sometimes include a  dependence on effective temperature, luminosity or both, that may also differ among prescriptions.
    Although recent papers have compared empirical results to prescriptions, a large observational sample of B stars for which the wind has been scrutinised over different metallicity ranges is still lacking.
    }	
	{Through the modelling of both ultraviolet (from the ULLYSES programme) and optical (from the XShootU programme) high-resolution spectra, we determined the stellar and wind parameters, including detailed clumping parameters, of 24 SMC B stars ranging in $T_{\rm eff}$ from 13 to 29\,kK. 
    By combining this sample with LMC studies, we compared the wind behaviour of B stars in two different metallicity regimes.
    We compared our results to common mass-loss rate prescriptions to test features present in these models and their metallicity dependence.}
   	{We have used the model atmosphere code {\sc fastwind} and the genetic algorithm code Kiwi-GA to fit the UV and optical spectra simultaneously. This allows us to determine wind properties including clumping behaviour.} 
	{The moderate to strong metallicity trends present in the mass-loss prescriptions ($\dot{M} \propto Z^{0.41-1.4}$) explored here overestimate the empirical metallicity dependence in the B-star regime, which appears very weak.
    Similarly to a previous study of B supergiants in the LMC, we do not find a sudden increase in mass-loss rate at approximately spectral type B1, but only a weak temperature trend. 
    As in this LMC study, we show that on average around 40\% of the wind mass is located in the wind medium between the clumps.
    Using other studies from this paper series, we compiled a sample of 80+ O and B stars in the SMC and LMC. 
    From a comparison we find a clear difference in O- and B-type metallicity dependence.}    
    {The apparent lack of a bi-stability jump in the B-star regime and a very weak metallicity dependence for the same stars offers new empirical constraints to theoretical models of line-driven winds. As differences between these models are large (reaching orders of magnitude in parts of parameter space) such constraints are much needed. More studies exploring the mass-loss rates of B stars and cooler objects will be helpful to our understanding of hot star winds.
    }

	\titlerunning{On the metallicity dependence of B-supergiant mass-loss rates}
	\authorrunning{O.\ Verhamme, J.\ Sundqvist, A.\ de Koter, H.\ Sana }
	
	\maketitle
	%
	\section{Introduction}
	Mass loss strongly impacts the evolution of massive stars \citep{josiek_impact_2024, gormaz-matamala_evolution_2024}, and hence an accurate implementation of this process in evolution codes is essential.
    The mass loss in hot massive stars is driven by the absorption and scattering of photons in spectral lines, supported by Thomson scattering on free electrons 
	\citep{castor_radiation-driven_1975}. As momentum transfer in line transitions dominates the process,
	relatively abundant atoms with many energy transitions, such as iron, are the main driving contributors. 
	Currently, the mass-loss rates used in evolutionary modelling are prescriptions based on 1D models of the stellar atmosphere \citep{vink_nature_1999, vink_mass-loss_2001, krticka_new_2021, krticka_new_2024, bjorklund_new_2023}.
	The most widely used prescription for stellar mass loss in the hot upper Hertzsprung--Russell diagram, and a benchmark, is the pioneering work by \citet{vink_mass-loss_2001}, which features a bi-stability jump. 
    This is a pronounced upward trend in mass-loss behaviour from $\sim$\blue{28\,kK} down to 21\,kK superimposed on a generally downward trend of mass loss with decreasing temperature, i.e., along the evolutionary sequence. 
	The cause of this upward jump (with decreasing temperature) is thought to be the recombination to a lower ionisation stage of iron, Fe\,{\sc iv} to Fe\,{\sc iii}, which, in principle, can increase the efficiency of line-driven winds as more line transitions become available in the region of the stellar flux maximum. 

    Tentative empirical evidence for a bi-stability jump may come from the temporal behaviour in mass-loss properties in luminous blue variable stars during their S\,Doradus-type excursions in the Hertzsprung--Russell diagram \citep{2002A&A...393..543V,2011A&A...531L..10G}. However, it has not been confirmed in regular B-type stars, even though multiple studies have investigated the temperature range covering the bi-stability jump over a large range of luminosities. 
    In the Small Magellanic Cloud (SMC) these are \citep{trundle_understanding_2004,trundle_understanding_2005,bernini-peron_x-shooting_2024}; in the Large Magellanic Cloud (LMC) \citep{verhamme_x-shooting_2024}, and in the Milky way \citep{crowther_physical_2006,markova_bright_2008, rubio-diez_upper_2022, de_burgos_iacob_2024_mass_loss}. 
	In recent years, a focussed spectroscopic study, that uses both ultraviolet (UV) and optical lines and takes into account the clumping behaviour of the winds \citep{verhamme_x-shooting_2024,alkousa_x-shooting_2025} as well as a large-scale approach using optical spectroscopy that uses a more simplified model of the wind \citep{de_burgos_iacob_2024_mass_loss} did not find compelling evidence for a bi-stability jump.
	All research groups come to the conclusion that the \citet{vink_mass-loss_2001} prescription should be used cautiously, as it may overestimate the mass loss by up to a factor of 10--100 for stars cooler than 25\,kK, agreeing more closely with the trends predicted by \citet{bjorklund_new_2021, krticka_new_2024}.	
    
    As the wind is mostly driven by metals, such as carbon, nitrogen, oxygen, and iron-group elements, the mass-loss rate $\dot{M}$ is predicted to be metallicity dependent, specifically $\dot{M} \propto Z^{0.6-1.4}$ \citep{vink_mass-loss_2001, bjorklund_new_2023, krticka_new_2024}. 
	Spectroscopic studies confirm this behaviour for O stars \citep{2007A&A...473..603M,backs_x-shooting_2024}. 
	For B stars, however, recent in-depth  studies at different metallicities that take a homogeneous approach and use data with the same spectral coverage and resolution while accounting for clumping are lacking. Hence, to date, theoretical results in this temperature range have not been comprehensively tested. 

    What complicates analyses of hot-star wind properties is that the mass-loss rates are found to depend on the degree of structure in the wind outflow. These density (and velocity) inhomogeneities are referred to as wind clumping \citep[e.g.,][]{2008A&ARv..16..209P,2022ARA&A..60..203V}. 
    Only in the case that the scattering-dominated lines (notably resonance lines in the UV) and recombination-dominated lines (notably H$\alpha$ and He\,{\sc ii}\,$\lambda$4686) can be modelled simultaneously is it possible to distinguish between clumping properties and mass-loss rate. 
    To date, most studies have relied on optical/IR data alone \citep{trundle_understanding_2004, trundle_understanding_2005, markova_bright_2008, rubio-diez_upper_2022, de_burgos_iacob_2024_mass_loss}, and hence are affected by this degeneracy. 
	\citet{bernini-peron_x-shooting_2024} have the UV data needed to derive clumping parameters, but use a simplified clumping description that assumes that all wind mass is located in the dense regions of the wind and none in the interclump medium. 
    At least for O stars, several studies in both the LMC and SMC accounting for a detailed description of clumping show that a significant fraction of the wind mass is located in the interclump region \citep{hawcroft_empirical_2024, brands_r136_2022, brands_x-shooting_2025}. 
    A study of LMC B stars adopting the same clumping model shows that 40\% of the wind mass is located in the medium between the clumps \citep{verhamme_x-shooting_2024}. 
    What is currently lacking is a similar detailed study for SMC B stars to better understand the clumping behaviour over different metallicities.  
	
	The UV Hubble database, {\it UV Legacy Library of Young Stars as Essential Standards} (ULLYSES) \citep{roman-duval_ultraviolet_2020, roman-duval_uv_2025}, and the follow-up optical X-Shooter/VLT observations of {\it Xshooting ULLYSES} (XShootU) \citep{vink_x-shooting_2023}, secured high spectral resolution data of a sample of 250 O- and B-type stars in both the LMC and SMC to study mass loss in the hot upper Hertzsprung--Russell diagram in detail at different metallicities.
	The availability of high-quality optical and UV spectroscopy allows us to consistently derive essential stellar and wind parameters, such as terminal wind speed and clumping properties. 
	This paper is part of a series in which the ULLYSES sample has been studied using a fitting method that relies on a genetic algorithm (GA) and the spectral synthesis code {\sc fastwind} \citep{puls_atmospheric_2005, verhamme_x-shooting_2024, backs_x-shooting_2024, hawcroft_empirical_2024, brands_x-shooting_2025}. 
	We focus on the results of the SMC B stars, and address the presence or absence of a bi-stability jump and the $\dot{M}(Z)$ dependence in the B-star regime, while also discussing trends found in the entire homogenously studied sample of 84 stars over a $T_{\rm eff}$ regime from 12\,kK to 60\,kK.  

    We discuss the method of spectral analysis and the SMC sample in Sect.~\ref{sec:method}. Our results are presented in Sect.~\ref{sec:results} and discussed in Sect.~\ref{sec:discussion}. We summarize our findings and draw conclusions in Sect.~\ref{sec:summary}.

	\section{Method and sample}\label{sec:method}
    We studied the spectra of 24 hot stars in the SMC. In this section we give details on how we extracted this information from the spectra and how these spectra were obtained. 
	\subsection{Sample}

    For this paper we analysed 24 SMC B-type stars ranging from spectral type B0 to B8. 
    The stars are mostly supergiants, with a few bright giants and one giant star. 
    The UV spectra are from the Hubble Space Telescope programme ULLYSES \citep{roman-duval_ultraviolet_2020, roman-duval_uv_2025} in which 250+ O- and B-type stars in the SMC and LMC were observed using the Space Telescope Imaging Spectrograph (STIS) and Cosmic Origins Spectrograph (COS). 
	This sample was complemented with optical spectra obtained using the X-shooter spectrograph installed on the Very Large Telescope in the XshootU programme \citep{vink_x-shooting_2023, sana_x-shooting_2024}.
    Although our aim was to select single B stars, the first nine epochs of the BLOeM survey, which has overlapping targets, showed AzV\,488 to be a binary \citep{shenar_binarity_2024,britavskiy_binarity_2025}. It is possible that a radial velocity analysis of the full projected 25 epochs of this programme will reveal more of our targets to be binaries.

	\begin{table}[]
        \tiny
		\centering
        \setlength{\tabcolsep}{4pt} 
        \caption{Sample used in this paper.}
		\begin{tabular}{lllll}
			Name&BLOeM ID$^{\text{a}}$& SpT$^{\text{b}}$ & $K_s$ [mag]$^{\text{c}}$& Instrument   \\
			\hline
			AzV 215&  4-020 &B0 Ia &13.04 &FUSE, STIS\\
		AzV 488&  6-080 &B0 Ia & 12.24 & FUSE, STIS \\
		AzV 235&  7-064 &B0 Ia & 12.28& FUSE, STIS \\ 
		AzV 104& &B0.5 Ia & 13.73 & FUSE, STIS\\
		AzV 506& &B0.5 II &14.12 & FUSE, STIS \\ 
		AzV 43 & &B0.5 III:n & 14.48 & COS \\
		AzV 266& &B0.7 Ia & 12.91 & FUSE, STIS \\
		AzV 410& &B0.7 Iab& 13.83&COS, STIS \\
		AzV 264&  1-009&B1 Ia & 12.75 & FUSE, STIS \\
		AzV 96 &  8-008 &B1 Iab & 12.95 & FUSE, STIS\\
		AzV 210& &B1.5 Ia & 12.79 & FUSE, STIS\\
		SK 191& &B1.5 Ia  & 11.9  & COS, FUSE, STIS \\
		AzV 175& &B1.5 Ib  & 13.89 & COS, FUSE \\
		AzV 18 & &B2 Ia & 12.32 & FUSE, STIS\\
		AzV 374& &B2 II & 13.53 & COS,FUSE,STIS\\
		NGC330 ELS 4& & B2.5 Ib & 13.46& COS,STIS \\
		AzV 234 & 4-066& B2.5 Ib &13.22 &COS,STIS\\
		AzV 22 & &B3 Ia & 12.15 & STIS \\
		AzV 445& &B3 Iab & 12.94 & COS,STIS\\
		AzV 314& &B3 Ib & 13.23 & COS,STIS \\
		NGC330 ELS 2& & B3 Ib & 13.08 & STIS \\
		SK 179& &B3 II & 13.37 & COS,STIS\\
		AzV 343& &B8 Iab & 13.07 & COS,STIS \\
		AzV 324&  1-062 &B8 Ib &13.1& COS,STIS \\
			\hline
		\end{tabular}
		\begin{tablenotes}
			\item \textbf{Notes}:(a)Corresponding BLOeM ID; (b) Spectral types from \citet{bestenlehner_x-shooting_2025}; (c) K-band magnitudes compiled by \citet{vink_x-shooting_2023}.
		\end{tablenotes}
		\label{tab:sample}
	\end{table}

	The UV spectral range of COS spans from $940-1783\,\AA$ with a spectral resolving power $R = 11000-19000$ depending on the grating used. 
    The STIS data cover $1141-2366\,\AA$ with $R = 30000-48500$ depending on the grating. 
    The UV data has a signal-to-noise ratio around 20( for more details see \citet{roman-duval_uv_2025}). 
    As our method only uses small line windows, we carried out local normalisation using the same method as \citet{hawcroft_x-shooting_2024}.
    For the optical spectra the full spectral range is $3100-8000\,\AA$. This range is covered by two different arms; the UVB arm, which runs from 3300\,$\AA$ to 5500\,$\AA$ with $R = 6700$, and the VIS arm, which covers $5500-8000\AA$ with  $R = 11400$. 
    The optical spectra have a signal-to-noise ratio that is typically higher than 100; more details are available in \citet{vink_x-shooting_2023} (for details on the reduction and normalisation of the optical data, see \citet{sana_x-shooting_2024}).
    To facilitate the reproducibility of these results, we did not renormalise any spectral windows in the optical regime.

	\subsection{Model and fitting}
    \label{sec:model_fitting}
	We created synthetic spectra using the 1D stellar atmosphere and spectral synthesis code {\sc fastwind} \citep{santolaya-rey_atmospheric_1997,puls_atmospheric_2005,sundqvist_atmospheric_2018}. 
    {\sc fastwind} solves for the state of the gas assuming non-local thermodynamic equilibrium (NLTE) in a spherically symmetric and stationary extended stellar envelope comprising both the photosphere and out-flowing wind. 
    Here we use version 10.5 (see \citealt{puls_atmospheric_2005, rivero-gonzalez2012, sundqvist_atmospheric_2018, carneiro_carbon_2018}), which accounts for the accumulative feedback effects from the multitudes of metallic spectral lines (e.g. C, N, O, Ne, Mg, Si, S, Ar, Fe, Ni) upon the radiation field and atmospheric structure by means of a computationally efficient statistical method.
    Only a part of these chemical elements are used for detailed spectroscopy (here H, He, Si, C, N, O). 
    The lines of the remaining elements may blend with the diagnostic lines. 
    The clearest cases of this are the iron lines that overlap with the carbon and silicon lines in the UV region. To ensure that we indeed fit the intended lines, we removed the non-modelled lines from the observed spectra.
    This method has been used in other recent analyses of early-type spectra \citep{brands_r136_2022, verhamme_x-shooting_2024, backs_x-shooting_2024, brands_x-shooting_2025}.

    Clump optical depths for the spectral lines are calculated from the input parameters using the \citet{sobolev_1960} approximation. 
    This means that we do not assume that clumps are either optically thin or thick; instead, we compute the opaqueness for all lines according to the structure parameters. 
    For this ensemble of clumps, we evaluated the impact upon the ionisation balance and spectrum formation.
    For details on the version of {\sc fastwind} used and details on the implementation of X-rays, we refer to section 2.2 in \cite{verhamme_x-shooting_2024}, where we use the exact same set-up, with the exception of the metallicity which is set to the SMC value of 0.2 times the solar metallicity, defined by \citet{asplund_2009}. 
    This commonly used scaling of the abundance of the SMC is not perfect as all elements scale differently compared to the Galactic abundance. For instance, the nitrogen abundance in the SMC, according to \citet{vink_x-shooting_2023}, is 0.05 the galactic value. However, in this work, CNO are fitted, not scaled. 

	The {\sc fastwind} models used here have 18 free wind and stellar parameters, which are constrained by comparing them to the UV plus optical spectra.
	The 18 parameters are effective temperature ($T_{\rm eff}$), effective surface gravity ($g_{\rm eff}$); helium abundance ($Y_{\rm He}$); CNO abundances; upper limit on projected rotational velocity ($\varv_{\rm max, rot} \sin i$); mass-loss rate ($\dot{M}$); terminal wind speed ($\varv_{\infty}$); wind acceleration parameter ($\beta$); clumping factor ($f_{\rm cl}$); interclump density in units of mean density ($f_{\rm ic} = \rho_{\rm ic} / \langle \rho \rangle$); velocity filling factor ($f_{\rm vel}$); onset velocity of clumping as a fraction of terminal wind speed ($\varv_{\rm cl,start}$); velocity at which full clumping is achieved ($\varv_{\rm cl,max}$); maximum wind turbulence (max $\varv_{\rm turb}$), which starts at the set micro-turbulence ($\varv_{\xi} =  10$\,km/s) at the transition point (0.1 times sound speed at the surface) and increases similarly to the clumping factor until it reaches max $\varv_{\rm cl,max}$; a transformation of the X-ray filling factor ($f_X$); and the maximum jump velocity ($u_{\infty}$). 
    We note that the radius is not one of the fit parameters because it is constrained using the temperature and the observed K-band magnitude.
    
	The effective surface gravity is the measured surface gravity affected by the rotation of the star itself, which can be converted to the value we use to compute the stellar mass by $g_\star = g_{\rm eff} +(\varv_{\rm rot} \sin i)^2 /R_\star $ \citep{herrero_mass_1992}.
	We do not separate the mechanisms broadening spectral lines into macro-turbulence and rotational broadening as these are nearly impossible to separate for all but the fastest rotators \citep{sundqvist_rotation_2013}.
	Therefore, here we report total broadening, which includes the stellar macro-turbulence motion and the rotational speed. 
    The reported value is therefore the upper limit for possible rotation. 
	\blue{Both the wind acceleration parameter and the terminal wind speed determine the radial velocity field: $\varv(r) = \varv_\infty (1 - b R_\ast/r)^\beta$ with $R_\ast$ being the radius at the photospheric boundary and $b = 1-(\varv_0/\varv_{\infty})^{1/\beta}$ ensures a smooth velocity transition from the (quasi-)hydrostatic atmosphere to the wind \citep{santolaya-rey_atmospheric_1997}. $\varv_0$ is set to the standard value 0.1 times the sonic speed at $T_{\rm eff}$.} 
	The clumping factor denotes the overdensity of the denser parts of the wind compared to the average, defined as $f_{\rm cl} = \langle \rho^2\rangle / \langle \rho \rangle^2$.
	The velocity filling factor shows the fraction of the velocity field filled with clumps. 
	In these {\sc fastwind} models clumping does not start in the quasi-hydrostatic atmosphere, but instead the onset velocity ($\varv_{\rm cl, start}$). 
	From this onset velocity the clumping increases \blue{linearly with velocity} until it reaches $\varv_{\rm cl,max}$ (\citealt[see figure 16]{verhamme_x-shooting_2024}). 
	
    We included X-rays in most of our stars following a prescription explained in \cite{carneiro_atmospheric_2016}, where the X-rays are created due to differential motion of material in the wind leading to shocks. 
	The parameter $f_X$ is defined as $f_X = 16e_{\rm s}^2$, where $e_{\rm s}$ is the X-ray volume filling factor. 
	Finally, $u_{\infty}$ is the maximum jump velocity, which sets the actual jump velocity to $u(r) = u_{\infty} \left[ \varv (r)/ \varv_{\infty} \right]^{1.45}$.
	
	There are two parameters not included in the list above, which we did not fit, but instead set to a chosen value. 
    The first is the micro-turbulence, which may potentially have an impact on derived parameters. 
    For this study, just as in the LMC B-star study by \citep{verhamme_x-shooting_2024}, we set the micro-turbulence to $\varv_{\xi} =  10$\,km/s. Tests have shown that the parameters we value most here, notably $T_{\rm eff}$ and $\dot{M}$, are not noticeably sensitive to $\varv_{\xi}$ and vary only due to the randomness of the GA method when changing the micro-turbulence. 
    The CNO abundances may be influenced by micro-turbulence, however only within error margins (see the appendices in \citealt{brands_r136_2022, verhamme_x-shooting_2024}). 
    The second fixed parameter is the Si abundance, which is set to the abundance suggested by \cite{vink_x-shooting_2023}. Unlike CNO, Si is not involved in the CN/CNO cycle and is not expected to change as the star evolves. 
    \citet{verhamme_x-shooting_2024} find only minor changes in $T_{\rm eff}$ when fitting the Si abundance rather than keeping it to one value.
    Si \blue{is set to a number fraction of $\epsilon_{\rm Si} = 6.72$, defined as $\epsilon_{\rm X} = 12+\log_{10}(n_X/n_H)$}, i.e. 0.14\,$Z_{\odot}$ \citep{vink_x-shooting_2023}. 
	
	To fit these 18 parameters simultaneously we used a genetic algorithm (GA) as previously explored by \citet{mokiem_spectral_2005, tramper_properties_2014, abdul-masih_clues_2019, hawcroft_empirical_2021}. 
    More specifically. here we use Kiwi-GA \citep{brands_r136_2022, brands_x-shooting_2025}. 
    In our fitting, we employed a two-step approach where we used the optical-only fits to inform the initial guess of the GA when fitting both UV and optical data \cite[see section 2.4 of][]{verhamme_x-shooting_2024}. 
    The aims was to improve the stability of the method, while not influencing the results. The list of diagnostic spectral lines is given in Table \ref{tab:linelist}.
	
	Though the methodology used follows \cite{verhamme_x-shooting_2024}, we needed to adjust it in one respect as for a number of stars we were unable to determine the terminal wind speed. 
    The spectral features that typically allow us to measure $\varv_{\infty}$ are the C\,{\sc iv}\,1550 and Si\,{\sc iv}\,1400 UV-resonance lines. 
    For some of the cooler, low mass-loss rate stars the wind is insufficiently ionised to have these species abundantly present, thus removing their sensitivity to wind properties far into the outflow.
    An example of this is AzV\,324 (B8\,Ib) shown in Fig.~\ref{fig:line_compare}. Its effective temperature is too low to produce a 
    C\,{\sc iv} profile originating in the wind. The narrow observed profile is mostly of interstellar origin. 
    The Si\,{\sc ii}-{\sc iii} lines show that the derived temperature is acceptable. 
    Due to the version of {\sc fastwind} that we used, we did not have access to the Al\,{\sc iii} features in UV, which might have given more constraints in these cases. 
    Therefore, to obtain $\varv_{\infty}$ we resorted to using the SMC effective temperature-terminal wind speed relation of \cite{hawcroft_x-shooting_2024} derived using stars hotter than 22500\,K,
	\begin{align}\label{eq:hawcroft}
		\varv_{\infty} = 0.089 \cdot T_{\rm eff} - 1560,       
	\end{align}
    where $\varv_{\infty}$ is in km/s.
    This empirical formula yields negative wind speeds for stars cooler than 17528\,K. To avoid unrealistically low values, we required that $\varv_{\infty}$ is at least the effective escape speed at the stellar surface, i.e. $\varv_{\rm esc}^2 = 2GM_* (1-\Gamma_{\rm e})/R_*$. 
    Here $M_*$ and $R_*$ are stellar mass and radius and $\Gamma_{\rm e} = \kappa_{\rm e} \,L_*/(4 \pi c \,G M_*)$ is the Eddington factor for electron scattering opacity ($\kappa_{\rm e}$) only, with $L_*$ the stellar luminosity.

    The density in the wind scales as $\langle \rho \rangle \sim \dot{M}/\varv_{\infty}$.
    Therefore, if a star has a terminal velocity lower than its surface escape velocity, we overestimate the terminal wind speed. To recover the strengths of wind sensitive lines, set by $\langle \rho \rangle$, we therefore overestimate the mass-loss rate. This overestimation is modest.
    Assuming the actual terminal wind speed were $0.75\cdot \varv_{\rm esc}$, we would be overestimating the mass-loss rate by 0.1 dex.
    Typical error margins on the mass-loss rates are much higher, ranging from 0.3 to 0.6 dex.
	
	\section{Results}\label{sec:results}
	Here, we present the results of the spectral fitting of the sample of SMC stars described in Table \ref{tab:sample}. Examples of the fits of important diagnostic lines of four stars covering the full temperature range are shown in Fig. \ref{fig:line_compare}.

\begin{figure}
		\centering
		\includegraphics[width=0.5\textwidth]{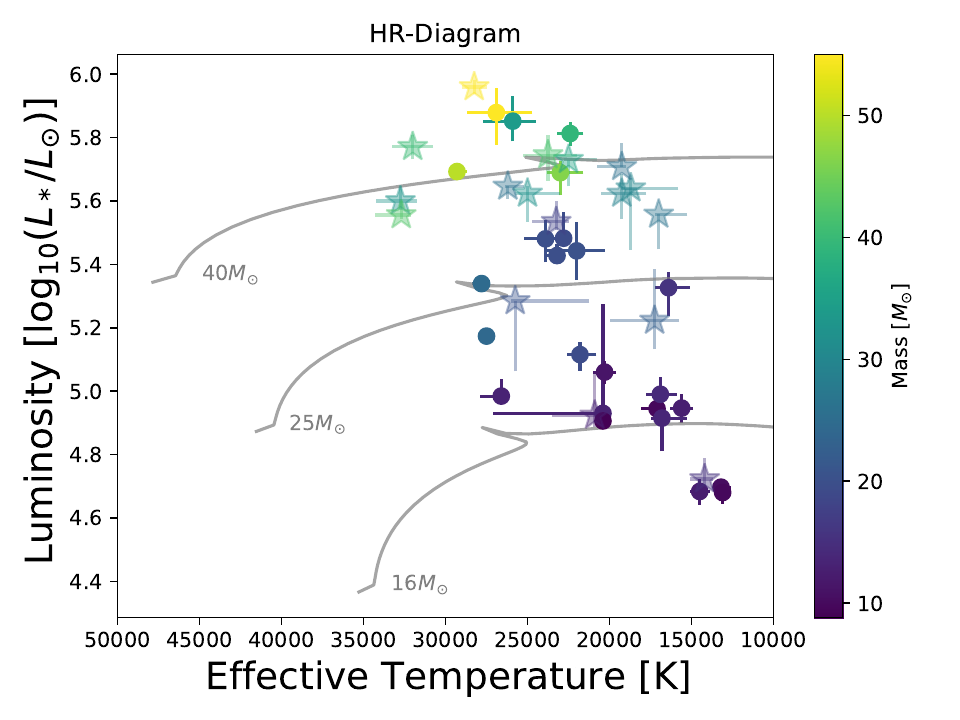}
		\caption{Hertzsprung--Russell diagram showing the SMC sample (circles) and the LMC sample discussed in \citet{verhamme_x-shooting_2024} (stars).  The symbol colour is a measure of the derived spectral mass.  The non-rotating stellar evolution tracks are from \citet{schootemeijer_constraining_2019} and have $\alpha_{\rm co} = 0.33$ and $\alpha_{\rm sc} = 1$. \blue{The metallicity of the evolutionary tracks is set to 0.2 $Z_{\odot}$.}}
		\label{fig:HR-diagram_LMC+SMC}
	\end{figure}

	The SMC sample is shown in the Hertzsprung--Russell diagram (figure \ref{fig:HR-diagram_LMC+SMC}), marked as circles. 
    Also in this figure is the LMC sample discussed in \citet{verhamme_x-shooting_2024} shown as partially transparent star markers. 
    Theoretical evolutionary pathways by \citet{schootemeijer_constraining_2019} are also plotted for initially 16, 25, and 40\,$M_{\odot}$ stars with a metallicity of 0.2 $Z_{\odot}$. 
    These tracks are for non-rotating stars and adopt a core-overshooting $\alpha_{\rm co} = 0.33$ and semi-convection $\alpha_{\rm sc} = 1$, following \citet{2011A&A...530A.115B}. 
    The luminosity range of our sample implies initial masses between about 15 and 50\,$M_{\odot}$. The star's temperatures range from 12.5 to 30\,kK. 
    We find that the majority of the sample stars, among which all relatively low-mass sources, are located beyond the terminal age main sequence (TAMS) in the Hertzsprung gap. 
    This conclusion on evolutionary status is, however, dependent on the adopted value for $\alpha_{\rm co}$. 
    For instance, for $\alpha_{\rm co} = 0$ the entire SMC sample is positioned beyond the TAMS. 
    We note that this assessment of evolutionary status also ignores a potentially unnoticed binary interaction history (for more information on the effect of binary mergers on the evolution of massive stars, see \citet{bellinger_potential_2024}).
    
	\subsection{Stellar properties}
	The stellar and wind parameters are determined by fitting the spectral lines listed in table \ref{tab:linelist}. 
    For the seven atmospheric parameters of $T_{\rm eff},\,g_{\rm eff},\,Y_{\rm He},\,\varv_{\rm max, rot} \sin i$, and CNO abundance the results are listed in Table \ref{tab:stellar_param}. 
    As magnesium lines are not considered, the effective temperature of the coolest stars is derived solely from Si\,{\sc ii} and He\,{\sc i}. 
    Additionally, the O\,II 4088 line blends with the Si\,IV 4088 line potentially causing a shift in the derived $T_{\rm eff}$ \citep{de_burgos_iacob_2024_survey}. 
    This causes the coolest stars to have higher $T_{\rm eff}$ uncertainties compared to the hottest stars. 
    The surface gravity, from which the spectroscopic mass is derived, ranges from $\log_{10}g = 2.0-3.5$, with typical error margins of 0.1 dex. 
    It is known that temperature and gravity are to some extent degenerate, which impacts their uncertainties. 
    Due to the statistical fitting process the errors quoted in this paper take these degeneracies into account automatically.
	
	The maximum rotational velocity ranges from 40 to 120 km/s with one outlier at 230 km/s; the typical error margin is around 10 km/s. We note again that these values are a combination of macro-turbulence and rotational velocity. Determining the exact value of the macro-turbulence using the modest resolution of X-shooter is difficult and determining the macro-turbulence separately from the rotational velocity is almost impossible for all but the fastest of rotators \citep{sundqvist_rotation_2013}.

    The helium abundance given is $Y_{\rm He} = n_{\rm He}/n_{\rm H}$, i.e., the ratio of number density of He nuclei to H nuclei. Values range from lower than 0.08 to 0.23 with error margins typically not larger than 0.02. In three cases (AzV445, AzV314, and AzV343) the best fit helium abundance is below the Big Bang nucleosynthesis value of 0.08 \citep{Schramm_1977_abundance}. Even though they agree with this value within their $1\sigma$ uncertainty, we explore whether the helium abundance may impact other stellar properties.  
    To this end, as a test we refitted these stars while preventing $Y_{\rm He}$ from going below a 0.08 threshold. This did not change the other parameters outside of the random variation due to the use of the GA method. 
    The low $Y_{\rm He}$ is mostly found in the lower temperature stars.
    Undetected companions are known to potentially artificially lower the derived helium abundance, although for the one confirmed binary (AzV\,488) this does not appear to be the case.   
	The quoted CNO abundance is $\epsilon_x = 12+\log_{10}(n_x/n_{\rm H})$, where $n_x$ is the number abundance of element $x$. 
    We note that this convention implies that our quoted CNO abundances depend on the derived hydrogen, and hence helium abundance.
    Due to the small number of data points for CNO lines compared to other lines, the loss function \citep[i.e. the root mean square error of approximation or RMSEA; see, e.g.,][]{steiger_notes_2016} does not change significantly when varying CNO. 
    This causes the error margins on the CNO abundances to be relatively high ranging from 0.5 to 1.5 dex. 
    The carbon abundance ranges from 6 to 8 dex, the nitrogen abundance from 6 to 7.5 dex, and the oxygen abundance from 7 to 8.5 dex. 
    These should be assessed relative to the SMC baseline abundance of 7.42, 6.66, and 8.05, for C, N, and O, respectively \citep{vink_x-shooting_2023}.
    The errors could be reduced by fixing all parameters except for the abundances of interest and only fitting lines of those species, but this would ignore all degeneracies with other parameters. 
    Because of the high uncertainty of the CNO abundances and the small influence of their lines on the fit function, CNO abundances are not the focal point of this paper \citep[for a closer look into abundances of B-stars, see e.g.][]{przybilla_quantitative_2006, przybilla_cosmic_2008, urbaneja_lmc_2017, urbaneja_metallicity_2023}.
    Here, we focus on the wind parameters. 
    In estimating the error margins of a parameter, degeneracies with other parameters are automatically accounted for. 
    Within the mass-loss rate error margin, for example, are models covering the full error margins of the CNO abundances.
	\begin{table*}[]
		\centering
		\caption{Photospheric parameters of the sample.}
        \tiny
		\begin{tabular}{llllllll}
			Object Name & $T_{\rm eff}$[K] & $\log_{10}{g}_{\rm eff}$ [$\log_{10}$(cm/s$^2$)]& $Y_{\rm He}[n_{\rm He}/n_{\rm H}]$ & $\varv_{\text{max, rot}} \sin i$[km/s] &$\epsilon_{\rm C}$ [dex] &$\epsilon_{\rm N}$ [dex] &$\epsilon_{\rm O}$ [dex] \\
			\hline
			AzV215&$29300_{-600}^{+200}$&$3.25_{-0.35}^{+0.05}$&$0.16_{-0.01}^{+0.02}$&$92_{-13}^{+1}$&$6.42_{-0.4}^{+0.5}$&$7.56_{-1.75}^{+0.05}$&$8.45_{-0.85}^{+0.25}$\\
			AzV488&$25900_{-1400}^{+1800}$&$2.7_{-0.15}^{+0.2}$&$0.2_{-0.12}^{+0.04}$&$78_{-13}^{+5}$&$7.02_{-0.2}^{+1.2}$&$6.56_{-0.95}^{+1.05}$&$7.3_{-0.2}^{+1.4}$\\
			AzV235&$26900_{-2200}^{+1800}$&$3.05_{-0.05}^{+0.4}$&$0.12_{-0.02}^{+0.04}$&$81_{-7}^{+16}$&$5.82_{-0.25}^{+0.95}$&$7.46_{-0.65}^{+0.05}$&$8.6_{-0.7}^{+0.2}$\\
			AzV104&$27800_{-400}^{+600}$&$3.2_{-0.05}^{+0.1}$&$0.12_{-0.01}^{+0.02}$&$90_{-14}^{+6}$&$7.87_{-0.05}^{+0.45}$&$7.56_{-1.65}^{+0.05}$&$7.95_{-0.7}^{+0.35}$\\
			AzV506&$27500_{-100}^{+500}$&$3.35_{-0.05}^{+0.1}$&$0.12_{-0.04}^{+0.0}$&$59_{-1}^{+4}$&$7.57_{-0.05}^{+0.2}$&$7.61_{-1.65}^{+0.05}$&$8.2_{-0.95}^{+0.05}$\\
			AzV43&$26600_{-400}^{+1300}$&$3.05_{-0.05}^{+0.15}$&$0.12_{-0.01}^{+0.04}$&$234_{-1}^{+34}$&$7.82_{-0.1}^{+0.2}$&$7.06_{-1.0}^{+0.4}$&$8.2_{-0.15}^{+0.35}$\\
			AzV266&$23900_{-1500}^{+1300}$&$2.7_{-0.2}^{+0.25}$&$0.22_{-0.08}^{+0.04}$&$78_{-17}^{+6}$&$6.27_{-0.4}^{+0.4}$&$6.06_{-0.45}^{+0.6}$&$8.05_{-0.4}^{+0.4}$\\
            AzV410&$20400_{-100}^{+6700}$&$2.75_{-0.05}^{+0.6}$&$0.09_{-0.01}^{+0.03}$&$120_{-15}^{+28}$&$7.92_{-0.3}^{+0.2}$&$6.61_{-1.0}^{+1.1}$&$8.1_{-0.55}^{+0.55}$\\
			AzV264&$22000_{-1700}^{+1800}$&$2.6_{-0.25}^{+0.1}$&$0.16_{-0.01}^{+0.05}$&$71_{-3}^{+20}$&$5.72_{-0.35}^{+0.55}$&$7.66_{-0.8}^{+0.05}$&$7.65_{-0.1}^{+0.7}$\\
			AzV96&$23200_{-500}^{+800}$&$2.7_{-0.15}^{+0.1}$&$0.2_{-0.0}^{+0.06}$&$73_{-10}^{+12}$&$7.12_{-0.3}^{+0.4}$&$7.41_{-1.3}^{+0.2}$&$8.0_{-0.55}^{+0.5}$\\
			AzV210&$22800_{-200}^{+1700}$&$2.6_{-0.1}^{+0.2}$&$0.19_{-0.0}^{+0.04}$&$75_{-13}^{+13}$&$7.02_{-0.3}^{+0.65}$&$7.11_{-1.25}^{+0.35}$&$8.2_{-0.85}^{+0.1}$\\
			SK191&$22400_{-800}^{+800}$&$2.55_{-0.05}^{+0.2}$&$0.23_{-0.04}^{+0.0}$&$72_{-1}^{+16}$&$7.57_{-0.05}^{+0.55}$&$6.61_{-0.65}^{+0.05}$&$7.3_{-0.05}^{+0.35}$\\
			AzV175&$20400_{-200}^{+200}$&$2.65_{-0.15}^{+0.4}$&$0.09_{-0.02}^{+0.0}$&$46_{-3}^{+11}$&$8.07_{-0.25}^{+0.15}$&$7.31_{-1.1}^{+0.25}$&$7.7_{-0.05}^{+0.85}$\\
			AzV18&$23000_{-1400}^{+800}$&$2.8_{-0.25}^{+0.2}$&$0.18_{-0.02}^{+0.1}$&$67_{-7}^{+10}$&$5.77_{-0.4}^{+0.5}$&$6.86_{-1.1}^{+0.65}$&$7.5_{-0.2}^{+0.4}$\\
			AzV374&$21800_{-1000}^{+800}$&$2.9_{-0.2}^{+0.2}$&$0.08_{-0.0}^{+0.02}$&$69_{-27}^{+1}$&$7.87_{-0.2}^{+0.5}$&$6.81_{-0.25}^{+0.8}$&$7.3_{-0.45}^{+1.05}$\\
			NGC330-ELS-04&$20300_{-700}^{+700}$&$2.6_{-0.15}^{+0.2}$&$0.14_{-0.04}^{+0.05}$&$49_{-11}^{+13}$&$7.82_{-0.2}^{+0.45}$&$6.91_{-1.25}^{+0.5}$&$6.6_{-0.1}^{+1.35}$\\
			AzV234&$17100_{-400}^{+1000}$&$2.35_{-0.1}^{+0.2}$&$0.08_{-0.01}^{+0.02}$&$45_{-5}^{+11}$&$8.12_{-0.05}^{+0.35}$&$6.31_{-0.7}^{+0.55}$&$7.1_{-0.2}^{+1.25}$\\
			AzV22&$16400_{-1300}^{+800}$&$2.1_{-0.2}^{+0.2}$&$0.09_{-0.01}^{+0.02}$&$46_{-8}^{+6}$&$6.37_{-1.0}^{+0.75}$&$6.91_{-1.2}^{+0.4}$&$6.75_{-0.25}^{+0.95}$\\
			AzV445&$15600_{-700}^{+700}$&$2.3_{-0.15}^{+0.15}$&$0.065_{-0.02}^{+0.04}$&$41_{-15}^{+6}$&$7.72_{-0.25}^{+0.35}$&$6.46_{-0.65}^{+0.8}$&$7.3_{-0.65}^{+1.1}$\\
			AzV314&$16800_{-1500}^{+700}$&$2.5_{-0.25}^{+0.15}$&$0.075_{-0.045}^{+0.005}$&$39_{-7}^{+21}$&$8.12_{-0.65}^{+0.05}$&$6.36_{-0.75}^{+1.3}$&$7.6_{-0.7}^{+1.05}$\\
			NGC330-ELS-02&$16900_{-1000}^{+900}$&$2.45_{-0.1}^{+0.15}$&$0.09_{-0.06}^{+0.0}$&$44_{-8}^{+4}$&$6.42_{-0.05}^{+0.75}$&$6.51_{-0.3}^{+0.6}$&$7.85_{-0.05}^{+1.0}$\\
			SK179&$14500_{-600}^{+600}$&$2.35_{-0.15}^{+0.15}$&$0.09_{-0.01}^{+0.04}$&$116_{-6}^{+15}$&$7.57_{-0.2}^{+0.35}$&$6.96_{-0.5}^{+0.75}$&$7.45_{-0.45}^{+1.6}$\\
			AzV343&$13200_{-600}^{+300}$&$2.15_{-0.15}^{+0.05}$&$0.04_{-0.0}^{+0.05}$&$61_{-3}^{+5}$&$7.07_{-0.25}^{+1.0}$&$6.86_{-0.4}^{+0.6}$&$7.55_{-0.35}^{+1.1}$\\
			AzV324&$13100_{-500}^{+500}$&$2.15_{-0.15}^{+0.15}$&$0.09_{-0.01}^{+0.04}$&$42_{-28}^{+18}$&$7.67_{-0.25}^{+0.7}$&$5.81_{-0.15}^{+1.2}$&$7.3_{-0.3}^{+1.75}$\\
			\hline
		\end{tabular}
		\label{tab:stellar_param}
		\begin{tablenotes}
			\item \textbf{Notes}: For $Y_{\rm He}$ and $\varv_{\text{max, rot}} \sin i$ the errors are constrained 
            from the optical-only fits as the allowed ranges of these parameters in optical plus UV fitting is limited in order to prevent unrealistic error bounds.
           The SMC base abundance of carbon, nitrogen, and oxygen is 7.42, 6.66, and 8.05, respectively \citep{vink_x-shooting_2023}.
		\end{tablenotes}
	\end{table*}
	
	\subsection{Wind properties}
\begin{figure}
		\centering
		\subfigure{\includegraphics[width=0.45\textwidth]{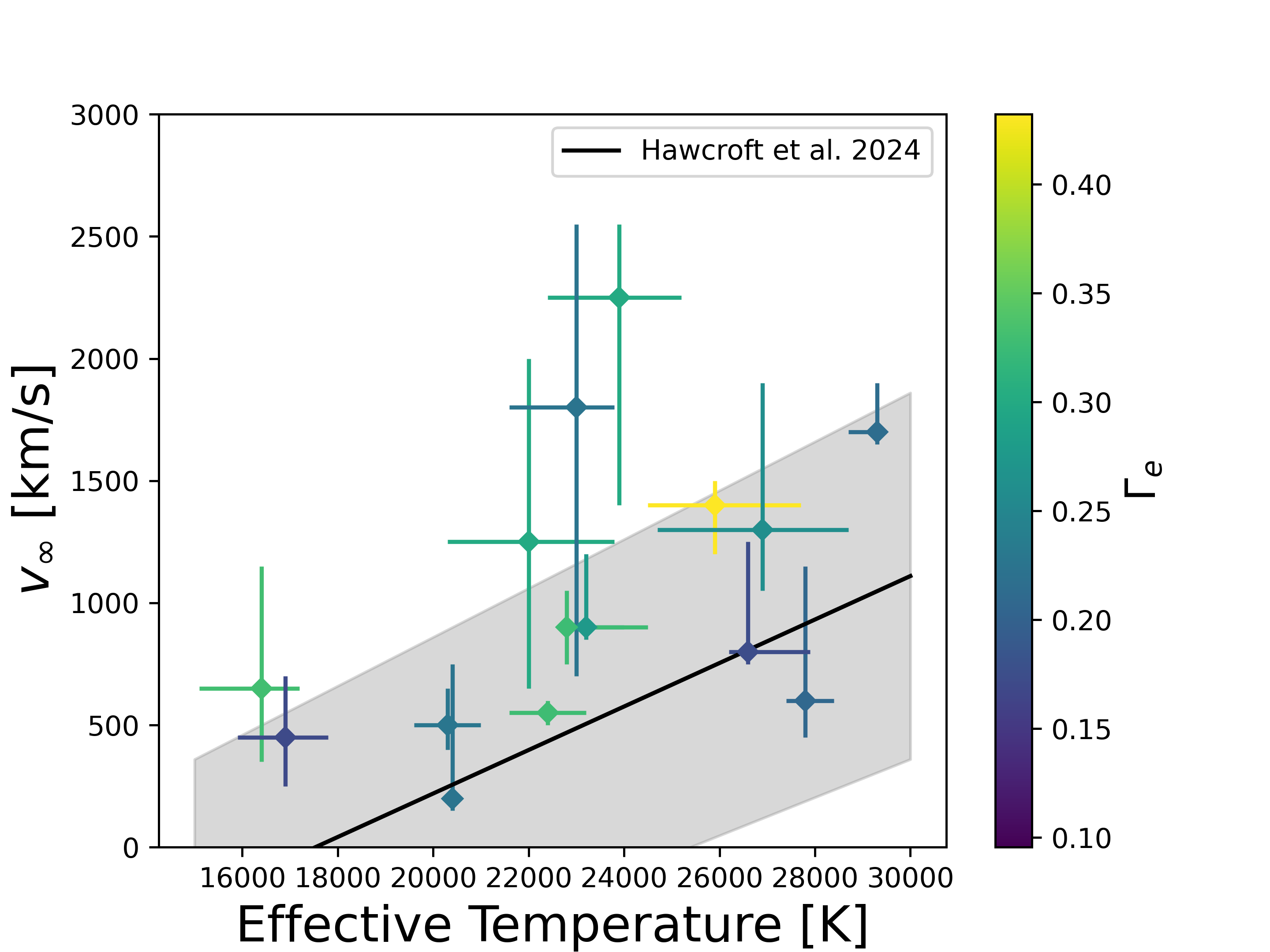}}
		\subfigure{\includegraphics[width=0.45\textwidth]{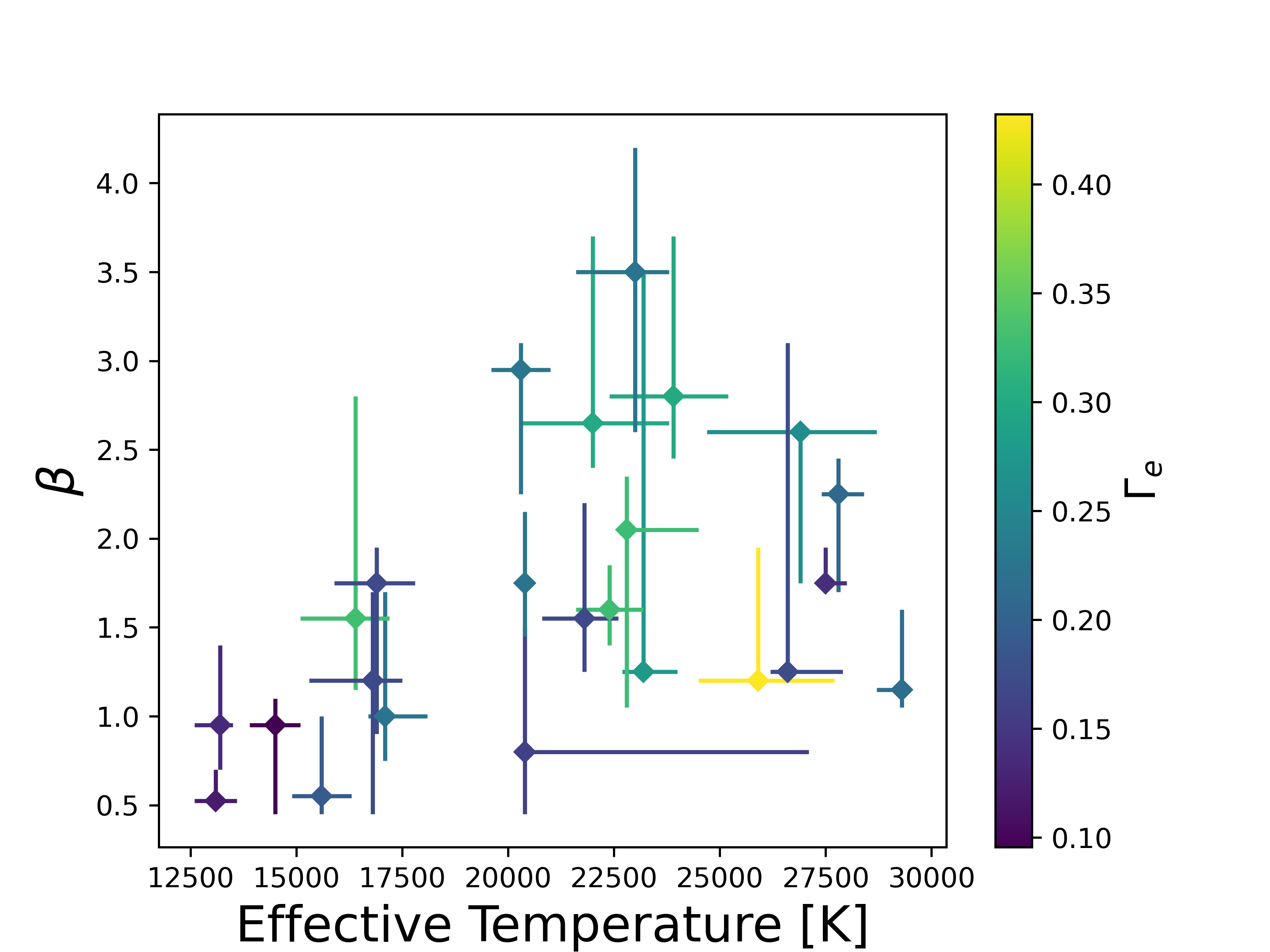}}
		\subfigure{\includegraphics[width=0.45\textwidth]{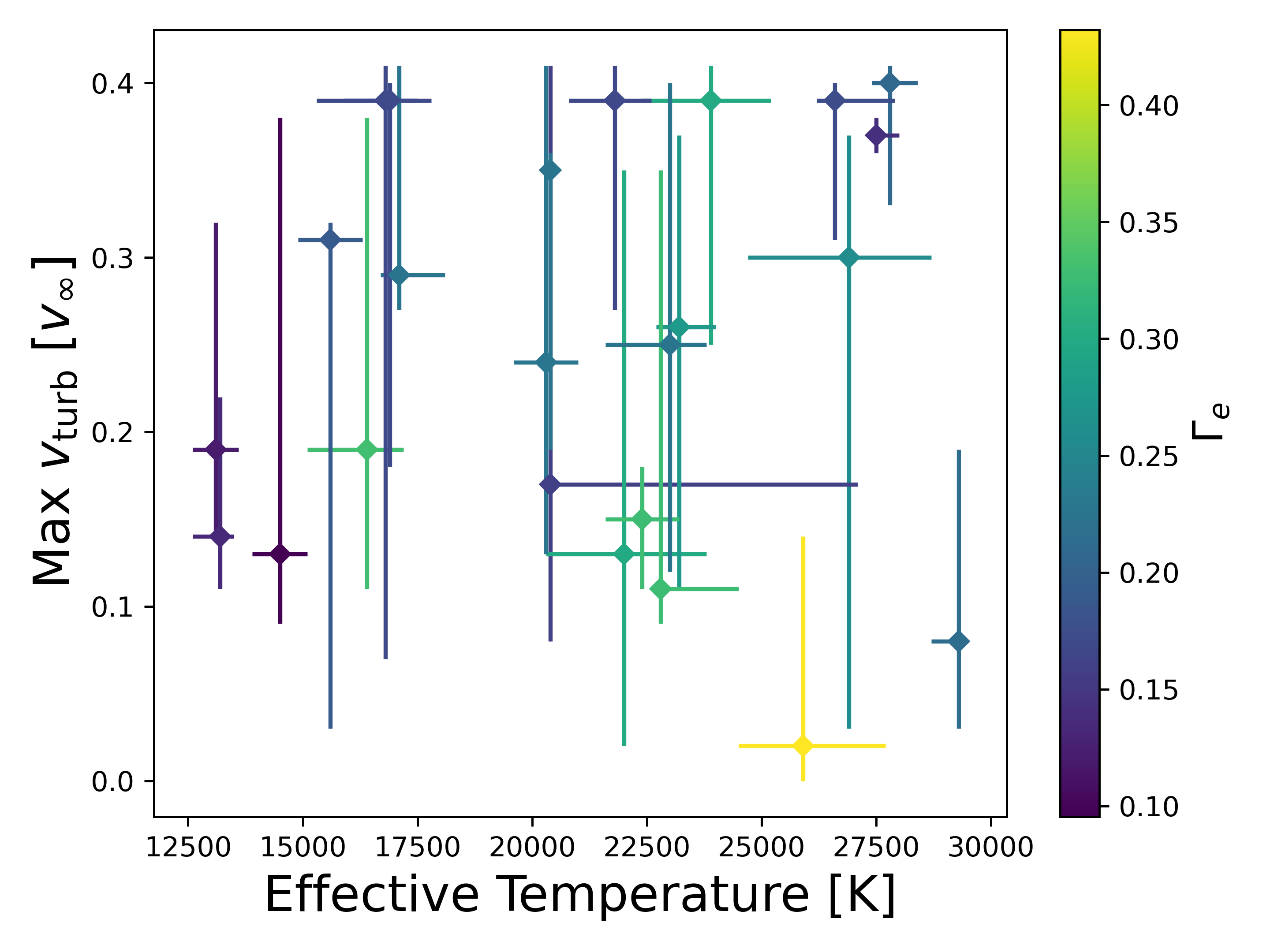}}
		\caption{Derived terminal velocities, wind acceleration parameter, and maximum wind micro turbulence, as function of effective temperature. The top panel excludes stars for which we adopted $\varv_{\infty}$ from \citet{hawcroft_empirical_2024} (shown in black with its 1$\sigma$ uncertainty). }
		\label{fig:velocity+acceleration}
	\end{figure}

    For the nine wind parameters of $\dot{M}$, $f_{\rm cl}$, $f_{\rm ic}$, $f_{\rm vel}$, $\varv_{\rm cl, start}$, $\varv_{\rm cl, max}$, $\varv_{\infty}$, $\beta$, and max $\varv_{\rm turb}$ of the results are listed in Table \ref{tab:wind_param}. 
    The top panel of figure \ref{fig:velocity+acceleration} shows the terminal wind speed over $T_{\rm eff}$; the colour of the symbol is a measure of the Eddington factor for electron scattering. 
    Only stars for which we extracted $\varv_{\infty}$ from our analysis are shown. 
    We made a comparison to the empirical relation for SMC stars by \citet{hawcroft_x-shooting_2024} (black line, with grey area showing the 1$\sigma$ uncertainty). 
    Except for four stars with relatively large uncertainties due to weak wind features, our values are within the uncertainty of the \citeauthor{hawcroft_x-shooting_2024} relation, though on average somewhat larger.
    Appendix \ref{sec:vinf_teff_discussion} gives more details on $\varv_{\infty}$ and a comparison of this sample with \citet{hawcroft_x-shooting_2024} including a very weak effect of metallicity on $\varv_{\infty}$.
	In the second panel of figure \ref{fig:velocity+acceleration}, we show the wind acceleration parameter. Both the scatter (from 0.5 to 3.5) and the uncertainties (typically 0.25, but sometimes larger) in $\beta$ are fairly large. 
    Especially in the region from 20\,kK to 25\,kK the spread is large and uncertainties are sizeable. 
    This might indicate that particularly here the $\beta$-law performs poorly.
    This could indicate that for this regime a different velocity-law prescription is needed, for example the $\delta$-slow solution \citep{cure_influence_2004,cure_slow_2011}.
    The high $\beta$ values only show for stars with low rotational velocity compared to their critical rotational velocity ($\varv_{\rm max, rot} \sin i \leq 0.4 \,\varv_{\rm crit}$), meaning they could potentially be classified as low $\Omega$, high $\delta$ slow solutions \citep{cure_slow_2011}.
    However, the $\varv_{\infty}$ of the slow solution typically shows terminal velocities of 100--200\,km/s, which is lower than almost all of the sample.

    For the case of AzV22 (B3\,Ia), we show in Fig.~\ref{fig:corr_mass_fcl} the relation between clumping factor and mass-loss rate for the many models calculated in the Kiwi-GA fitting run. The yellow region shows ($f_{\rm cl},\dot{M}$) combinations for which the fits are of good quality. It reveals that up to $f_{\rm cl} \sim 20$ models of similar quality can be created by reducing the mass loss and increasing the clumping factor. This is expected for optically thin clumps embedded in a void interclump medium \citep[e.g.][]{puls_atmospheric_2020}. We note, however, that for $f_{\rm cl} \ga 20$ only a narrow range of mass loss is allowed for a wider range of, and more poorly constrained, clumping factors. 
    This is the case for many of the objects.
    We find rather high error margins on the clumping parameters (see Fig. \ref{fig:clumping}) and reasonably low margins on the mass-loss rate (see figure \ref{fig:Mass-loss rate emp}).
    For completeness, we present the clumping properties $f_{\rm cl}$, $f_{\rm ic}$, and $f_{\rm vel}$ in Fig. \ref{fig:clumping}. All three show a large spread accompanied by large uncertainties, with no apparent trends.

    In the top panel of figure \ref{fig:Mass-loss rate emp}, mass-loss rate as function of temperature is plotted. 
    As our target stars span a sizeable range in luminosity and mass \citep[strongly impacting $\dot{M}$; see e.g][]{vink_mass-loss_2001}, we assessed our findings by comparing them directly to the predictions. 
    This is shown in the bottom panel of the figure. 
    On a star-by-star basis, we compare our results to four different prescriptions:
    in red to \citet{vink_mass-loss_2001}, in black to \citet{bjorklund_new_2023}, in blue to \citet{krticka_new_2024}, and in green to LIME \citep{sundqvist_--fly_2025}.
	For the comparisons we used the empirically derived spectroscopic mass, $\varv_{\infty}$, luminosity, and escape speed as input for the theoretical mass-loss estimate. 
	
	\begin{table*}[]
		\centering
		\caption{Wind parameters of the sample.}
        \tiny
		\begin{tabular}{llllllllll}
			Object Name& $\log_{10} \dot{M} [\log_{10}({\rm M}_{\odot}/$yr)] & $\varv_{\infty}$ [km/s] & $\beta$ & $f_{\rm cl}$ & $f_{\rm ic}$& $f_{\rm vel}$ & $\varv_\text{cl, start} [\varv_\infty]$& $\varv_\text{cl, max} [\varv_\infty]$ & $\varv_\text{turb}$ [$\varv_\infty$]  \\
			\hline
			AzV215&$-5.9_{-0.4}^{+0.05}$&$1700_{-50}^{+200}$&$1.15_{-0.1}^{+0.45}$&$10.0_{-4.0}^{+19.0}$&$0.15_{-0.1}^{+0.45}$&$0.45_{-0.15}^{+0.1}$&$0.01_{-0.0}^{+0.06}$&$0.15_{-0.01}^{+0.18}$&$0.08_{-0.05}^{+0.11}$\\
			AzV488&$-5.85_{-0.6}^{+0.05}$&$1400_{-200}^{+100}$&$1.2_{-0.05}^{+0.75}$&$6.0_{-1.0}^{+36.0}$&$0.92_{-0.38}^{+0.03}$&$0.25_{-0.22}^{+0.7}$&$0.01_{-0.01}^{+0.08}$&$0.16_{-0.07}^{+0.08}$&$0.02_{-0.02}^{+0.12}$\\
			AzV235&$-5.95_{-0.05}^{+0.25}$&$1300_{-250}^{+600}$&$2.6_{-0.85}^{+0.05}$&$26.0_{-17.0}^{+1.0}$&$0.42_{-0.20}^{+0.58}$&$0.38_{-0.22}^{+0.25}$&$0.08_{-0.08}^{+0.01}$&$0.24_{-0.02}^{+0.11}$&$0.3_{-0.27}^{+0.07}$\\
			AzV104&$-7.35_{-0.1}^{+0.35}$&$600_{-150}^{+550}$&$2.25_{-0.55}^{+0.2}$&$17.0_{-4.0}^{+33.0}$&$0.02_{-0.02}^{+0.02}$&$0.18_{-0.15}^{+0.05}$&$0.03_{-0.01}^{+0.05}$&$0.3_{-0.06}^{+0.08}$&$0.4_{-0.07}^{+0.00}$\\
			AzV506&$-7.4_{-0.1}^{+0.05}$&$\grey{888_{-10}^{+44}}$&$1.75_{-0.05}^{+0.2}$&$19.0_{-1.0}^{+9.0}$&$0.02_{-0.02}^{+0.02}$&$0.02_{-0.02}^{+0.08}$&$0.07_{-0.01}^{+0.04}$&$0.18_{-0.01}^{+0.12}$&$0.37_{-0.01}^{+0.01}$\\
			AzV43&$-7.45_{-0.2}^{+0.05}$&$800_{-50}^{+450}$&$1.25_{-0.05}^{+1.85}$&$26.0_{-9.0}^{+3.0}$&$0.08_{-0.08}^{+0.08}$&$0.1_{-0.02}^{+0.5}$&$0.07_{-0.06}^{+0.02}$&$0.2_{-0.05}^{+0.01}$&$0.39_{-0.08}^{+0.01}$\\
			AzV266&$-6.7_{-0.35}^{+0.2}$&$2250_{-850}^{+300}$&$2.8_{-0.35}^{+0.9}$&$18.0_{-13.0}^{+8.0}$&$0.38_{-0.22}^{+0.52}$&$\grey{0.08_{-0.08}^{+0.92}}$&$0.02_{-0.02}^{+0.06}$&$0.14_{-0.04}^{+0.07}$&$0.39_{-0.14}^{+0.02}$\\
            AzV410&$-7.55_{-1.8}^{+0.05}$&$\grey{423_{-0}^{+428}}$&$0.8_{-0.35}^{+0.7}$&$17.0_{-15.0}^{+33.0}$&$0.8_{-0.75}^{+0.2}$&$0.62_{-0.6}^{+0.4}$&$0.07_{-0.07}^{+0.04}$&$0.26_{-0.05}^{+0.14}$&$0.17_{-0.09}^{+0.24}$\\
			AzV264&$-6.9_{-0.3}^{+0.25}$&$1250_{-600}^{+750}$&$2.65_{-0.25}^{+1.05}$&$19.0_{-4.0}^{+17.0}$&$0.88_{-0.35}^{+0.12}$&$0.2_{-0.2}^{+0.77}$&$0.01_{-0.01}^{+0.06}$&$0.23_{-0.05}^{+0.15}$&$0.13_{-0.11}^{+0.22}$\\
			AzV96&$-6.85_{-0.3}^{+0.05}$&$900_{-50}^{+300}$&$1.25_{-0.05}^{+2.25}$&$34.0_{-1.0}^{+13.0}$&$0.38_{-0.03}^{+0.6}$&$0.6_{-0.18}^{+0.4}$&$0.01_{-0.01}^{+0.07}$&$0.1_{-0.01}^{+0.23}$&$0.26_{-0.15}^{+0.11}$\\
			AzV210&$-7.0_{-0.05}^{+0.25}$&$900_{-150}^{+150}$&$2.05_{-1.0}^{+0.3}$&$24.0_{-10.0}^{+26.0}$&$0.82_{-0.75}^{+0.18}$&$0.65_{-0.4}^{+0.35}$&$0.07_{-0.07}^{+0.02}$&$0.19_{-0.1}^{+0.13}$&$0.11_{-0.02}^{+0.24}$\\
			SK191&$-6.3_{-0.2}^{+0.1}$&$550_{-50}^{+50}$&$1.6_{-0.2}^{+0.25}$&$7.0_{-5.0}^{+3.0}$&$0.55_{-0.25}^{+0.17}$&$0.42_{-0.15}^{+0.35}$&$0.04_{-0.02}^{+0.01}$&$0.3_{-0.01}^{+0.07}$&$0.15_{-0.04}^{+0.03}$\\
			AzV175&$-8.3_{-0.05}^{+0.55}$&$200_{-50}^{+550}$&$1.75_{-0.3}^{+0.4}$&$12.0_{-1.0}^{+8.0}$&$0.22_{-0.22}^{+0.28}$&$0.1_{-0.08}^{+0.12}$&$0.04_{-0.01}^{+0.04}$&$0.22_{-0.09}^{+0.04}$&$0.35_{-0.16}^{+0.01}$\\
			AzV18&$-6.15_{-0.15}^{+0.15}$&$1800_{-1100}^{+750}$&$3.5_{-0.9}^{+0.7}$&$3.0_{-2.0}^{+8.0}$&$0.05_{-0.05}^{+0.52}$&$0.02_{-0.02}^{+0.6}$&$0.07_{-0.07}^{+0.03}$&$0.27_{-0.09}^{+0.07}$&$0.25_{-0.13}^{+0.15}$\\
			AzV374&$-7.75_{-0.5}^{+0.05}$&$\grey{492_{-93}^{97}}$&$1.55_{-0.3}^{+0.65}$&$35.0_{-12.0}^{+12.0}$&$0.02_{-0.02}^{+0.15}$&$0.18_{-0.12}^{+0.2}$&$0.09_{-0.08}^{+0.01}$&$0.38_{-0.06}^{+0.03}$&$0.39_{-0.12}^{+0.02}$\\
			NGC330-ELS-04&$-7.55_{-0.3}^{+0.3}$&$500_{-100}^{+150}$&$2.95_{-0.7}^{+0.15}$&$20.0_{-12.0}^{+18.0}$&$0.08_{-0.08}^{+0.25}$&$0.42_{-0.4}^{+0.23}$&$0.05_{-0.03}^{+0.05}$&$0.28_{-0.1}^{+0.12}$&$0.24_{-0.11}^{+0.17}$\\
			AzV234&$-7.5_{-0.35}^{+0.15}$&$\grey{292_{-18}^{+70}}$&$1.0_{-0.25}^{+0.7}$&$10.0_{-4.0}^{+12.0}$&$0.65_{-0.18}^{+0.35}$&$0.57_{-0.45}^{+0.43}$&$0.1_{-0.08}^{+0.01}$&$0.34_{-0.06}^{+0.02}$&$0.29_{-0.02}^{+0.12}$\\
			AzV22&$-6.8_{-0.3}^{+0.15}$&$650_{-300}^{+500}$&$1.55_{-0.4}^{+1.25}$&$12.0_{-8.0}^{+20.0}$&$0.25_{-0.2}^{+0.2}$&$0.9_{-0.52}^{+0.10}$&$\grey{0.02_{-0.02}^{+0.08}}$&$0.24_{-0.14}^{+0.12}$&$0.19_{-0.08}^{+0.19}$\\
			AzV445&$-7.35_{-0.25}^{+0.25}$&$\grey{308_{-33}^{+36}}$&$0.55_{-0.1}^{+0.45}$&$7.0_{-6.0}^{+6.0}$&$0.7_{-0.15}^{+0.32}$&$0.62_{-0.6}^{+0.08}$&$0.08_{-0.07}^{+0.03}$&$0.32_{-0.01}^{+0.08}$&$0.31_{-0.28}^{+0.01}$\\
			AzV314&$-7.8_{-0.55}^{+0.4}$&$\grey{357_{-74}^{+47}}$&$1.2_{-0.75}^{+0.5}$&$17.0_{-12.0}^{+33.0}$&$0.7_{-0.65}^{+0.3}$&$\grey{0.85_{-0.85}^{+0.15}}$&$0.04_{-0.03}^{+0.07}$&$0.28_{-0.18}^{+0.1}$&$0.39_{-0.32}^{+0.02}$\\
			NGC330-ELS-02&$-7.9_{-0.3}^{+0.45}$&$450_{-200}^{+250}$&$1.75_{-0.85}^{+0.2}$&$17.0_{-15.0}^{+23.0}$&$0.1_{-0.05}^{+0.52}$&$0.5_{-0.35}^{+0.08}$&$\grey{0.06_{-0.06}^{+0.04}}$&$0.38_{-0.18}^{+0.01}$&$0.39_{-0.21}^{+0.01}$\\
			SK179&$-7.25_{-0.3}^{+0.15}$&$\grey{351_{-27}^{+34}}$&$0.95_{-0.5}^{+0.15}$&$2.0_{-1.0}^{+9.0}$&$0.82_{-0.7}^{+0.18}$&$0.55_{-0.45}^{+0.45}$&$0.07_{-0.06}^{+0.03}$&$0.28_{-0.07}^{+0.07}$&$0.13_{-0.04}^{+0.25}$\\
			AzV343&$-7.62_{-0.53}^{+0.03}$&$\grey{282_{-29}^{+2}}$&$0.95_{-0.25}^{+0.45}$&$6.0_{-5.0}^{+5.0}$&$0.28_{-0.15}^{+0.22}$&$0.5_{-0.45}^{+0.15}$&$0.04_{-0.03}^{+0.05}$&$0.38_{-0.15}^{+0.02}$&$0.14_{-0.03}^{+0.08}$\\
			AzV324&$-7.4_{-0.05}^{+0.3}$&$\grey{278_{-38}^{+33}}$&$0.52_{-0.05}^{+0.18}$&$1.0_{-0.0}^{+3.0}$&$0.42_{-0.22}^{+0.48}$&$0.35_{-0.3}^{+0.3}$&$0.01_{-0.0}^{+0.09}$&$0.25_{-0.04}^{+0.15}$&$0.19_{-0.05}^{+0.13}$\\
			\hline
		\end{tabular}
		\begin{tablenotes}
			\item \textbf{Notes}: For stars lacking strong P\,Cygni lines, terminal wind speeds are obtained from the scaling law by \cite{hawcroft_empirical_2024}. These are given in grey. 
            Additionally, any parameters for which the error margins cover the full allowed parameter range are shown grey as well.  
            These values should be considered unconstrained.
		\end{tablenotes}
		\label{tab:wind_param}
	\end{table*}
	
	\begin{table*}[]
		\centering
		\caption{Derived and X-ray parameters of all objects in our sample.}
        \tiny
		\begin{tabular}{llllllll}
			Object Name&  $M_\text{spec} [M_\odot]$ &  $\log_{10}(L/L_\odot)$ & Radius$ [R_\odot]$ & $\Gamma_{\rm e}$ & $u_\infty [km/s]$ & $\log_{10}$($f_x$) & $\log_{10}(L_x/L_\text{bol})$\\
			\hline
			AzV215&$50.24_{-24.19}^{+0.04}$&$5.69_{-0.02}^{+0.0}$&$27.49_{-0.05}^{+0.27}$&$0.22_{-0.01}^{+0.18}$&$667.5_{-487.5}^{+87.5}$&$-0.5_{-0.05}^{+0.4}$&$-5.83_{-2.62}^{+0.29}$\\
			AzV488&$33.99_{-5.19}^{+11.57}$&$5.85_{-0.06}^{+0.08}$&$42.25_{-1.51}^{+1.27}$&$0.43_{-0.1}^{+0.14}$&$205.0_{-75.0}^{+150.0}$&$-1.05_{-0.25}^{+0.75}$&$-8.8_{-0.63}^{+0.74}$\\
			AzV235&$68.2_{-0.0}^{+82.56}$&$5.88_{-0.1}^{+0.08}$&$40.4_{-1.39}^{+1.93}$&$0.26_{-0.15}^{+0.0}$&$117.5_{-37.5}^{+87.5}$&$0.9_{-1.05}^{+0.05}$&$-8.6_{-1.23}^{+0.54}$\\
			AzV104&$24.76_{-0.08}^{+2.58}$&$5.34_{-0.01}^{+0.02}$&$20.33_{-0.2}^{+0.13}$&$0.21_{-0.02}^{+0.0}$&$180.0_{-12.5}^{+112.5}$&$0.5_{-0.75}^{+0.3}$&$-8.07_{-0.25}^{+0.76}$\\
			AzV506&$24.39_{-0.0}^{+2.53}$&$5.17_{-0.0}^{+0.02}$&$17.17_{-0.14}^{+0.0}$&$0.140_{-0.000}^{+0.003}$&$255.0_{-12.5}^{+75.0}$&$0.0_{-1.3}^{+0.05}$&$-7.95_{-1.04}^{+0.0}$\\
			AzV43&$13.15_{-0.0}^{+3.53}$&$4.98_{-0.01}^{+0.05}$&$14.76_{-0.37}^{+0.1}$&$0.17_{-0.04}^{+0.0}$&$205.0_{-12.5}^{+500.0}$&$-0.35_{-0.3}^{+0.35}$&$-8.52_{-0.0}^{+1.84}$\\
			AzV266&$20.18_{-4.54}^{+9.34}$&$5.48_{-0.07}^{+0.06}$&$32.36_{-0.91}^{+1.16}$&$0.3_{-0.06}^{+0.08}$&$367.5_{-37.5}^{+437.5}$&$-0.4_{-0.45}^{+0.3}$&$-8.06_{-0.45}^{+1.13}$\\
            AzV410&$13.18_{-0.0}^{+17.51}$&$4.93_{-0.0}^{+0.35}$&$23.57_{-3.55}^{+0.0}$&$0.16_{-0.05}^{+0.11}$&$380.0_{-325.0}^{+462.5}$&$-1.55_{-0.45}^{+2.55}$&$-8.14_{-9.14}^{+1.31}$\\
			AzV264&$20.38_{-5.92}^{+2.37}$&$5.44_{-0.09}^{+0.09}$&$36.56_{-1.56}^{+1.67}$&$0.3_{-0.04}^{+0.12}$&$692.5_{-412.5}^{+100.0}$&$-0.35_{-0.5}^{+0.4}$&$-7.15_{-0.85}^{+0.41}$\\
			AzV96&$20.0_{-4.26}^{+2.35}$&$5.43_{-0.02}^{+0.04}$&$32.32_{-0.55}^{+0.33}$&$0.28_{-0.04}^{+0.08}$&$355.0_{-212.5}^{+112.5}$&$0.3_{-0.7}^{+0.05}$&$-6.81_{-1.96}^{+0.0}$\\
			AzV210&$19.5_{-1.9}^{+6.61}$&$5.48_{-0.01}^{+0.08}$&$35.64_{-1.38}^{+0.09}$&$0.33_{-0.06}^{+0.03}$&$255.0_{-150.0}^{+362.5}$&$-0.55_{-1.1}^{+0.85}$&$-8.52_{-1.1}^{+1.45}$\\
			SK191&$39.21_{-0.0}^{+16.2}$&$5.81_{-0.04}^{+0.04}$&$54.0_{-0.96}^{+1.02}$&$0.33_{-0.09}^{+0.0}$&$405.0_{-262.5}^{+112.5}$&$-1.8_{-0.2}^{+0.2}$&$-8.32_{-2.13}^{+0.41}$\\
			AzV175&$8.82_{-1.76}^{+10.85}$&$4.91_{-0.01}^{+0.01}$&$22.93_{-0.07}^{+0.07}$&$0.23_{-0.12}^{+0.06}$&$242.5_{-137.5}^{+87.5}$&$-0.5_{-0.05}^{+0.75}$&$-8.65_{-1.55}^{+0.76}$\\
			AzV18&$46.48_{-17.09}^{+18.04}$&$5.69_{-0.07}^{+0.04}$&$44.43_{-0.77}^{+1.54}$&$0.23_{-0.06}^{+0.12}$&$542.5_{-250.0}^{+200.0}$&$-1.45_{-0.4}^{+0.5}$&$-8.02_{-0.54}^{+0.26}$\\
			AzV374&$19.55_{-5.66}^{+6.86}$&$5.12_{-0.05}^{+0.04}$&$25.55_{-0.47}^{+0.64}$&$0.17_{-0.04}^{+0.06}$&$267.5_{-12.5}^{+400.0}$&$-0.1_{-0.55}^{+0.9}$&$-8.05_{-0.34}^{+0.97}$\\
			NGC330-ELS-04&$11.45_{-2.09}^{+4.49}$&$5.06_{-0.04}^{+0.04}$&$27.65_{-0.47}^{+0.5}$&$0.23_{-0.07}^{+0.04}$&$480.0_{-62.5}^{+350.0}$&$-1.8_{-0.1}^{+0.85}$&$-8.94_{-0.0}^{+1.21}$\\
			AzV234&$9.86_{-0.91}^{+3.97}$&$4.94_{-0.02}^{+0.06}$&$34.11_{-1.06}^{+0.37}$&$0.22_{-0.07}^{+0.01}$&$505.0_{-62.5}^{+275.0}$&$-1.65_{-0.2}^{+0.3}$&$-7.95_{-0.38}^{+0.2}$\\
			AzV22&$15.85_{-3.47}^{+5.12}$&$5.33_{-0.09}^{+0.05}$&$57.55_{-1.47}^{+2.81}$&$0.33_{-0.05}^{+0.01}$&N/A&N/A&N/A\\
			AzV445&$12.65_{-2.15}^{+2.51}$&$4.95_{-0.05}^{+0.05}$&$41.08_{-0.96}^{+1.03}$&$0.19_{-0.02}^{+0.03}$&N/A&N/A&N/A\\
			AzV314&$13.73_{-4.47}^{+3.11}$&$4.92_{-0.1}^{+0.04}$&$34.16_{-0.73}^{+1.91}$&$0.17_{-0.04}^{+0.05}$&N/A&N/A&N/A\\
			NGC330-ELS-02&$14.31_{-0.7}^{+3.63}$&$4.99_{-0.07}^{+0.06}$&$36.81_{-1.03}^{+1.27}$&$0.17_{-0.02}^{+0.02}$&$530.0_{-12.5}^{+312.5}$&$-1.15_{-0.25}^{+1.35}$&$-8.22_{-0.22}^{+1.56}$\\
			SK179&$12.56_{-1.6}^{+2.15}$&$4.68_{-0.04}^{+0.04}$&$35.13_{-0.75}^{+0.8}$&$0.1_{-0.01}^{+0.01}$&N/A&N/A&N/A\\
			AzV343&$10.39_{-1.69}^{+0.17}$&$4.7_{-0.05}^{+0.02}$&$43.06_{-0.42}^{+1.1}$&$0.13_{-0.01}^{+0.01}$&N/A&N/A&N/A\\
			AzV324&$9.88_{-2.21}^{+2.16}$&$4.68_{-0.04}^{+0.04}$&$42.88_{-0.84}^{+0.88}$&$0.12_{-0.02}^{+0.03}$&N/A&N/A&N/A\\
			\hline

		\end{tabular}
		\begin{tablenotes}
			\item \textbf{Notes}: For the cooler stars we could not converge models that include X-ray luminosities, and hence we do not include X-rays. $M_\text{spec}$ is computed using rotational velocity corrected $\log_{10}g_{\rm eff}$.
		\end{tablenotes}
		\label{tab:deriv_param}
	\end{table*}

	\section{Discussion}\label{sec:discussion}
    Some of the relevance of the results we show in the last section only become clear when put in the context of the current latest results. Here we discuss how our results compare to popular mass-loss prescriptions and the effect of metallicity on the wind. 
	\subsection{Comparison to current prescriptions}
\begin{figure}
		\centering
		\subfigure{\includegraphics[width=0.45\textwidth]{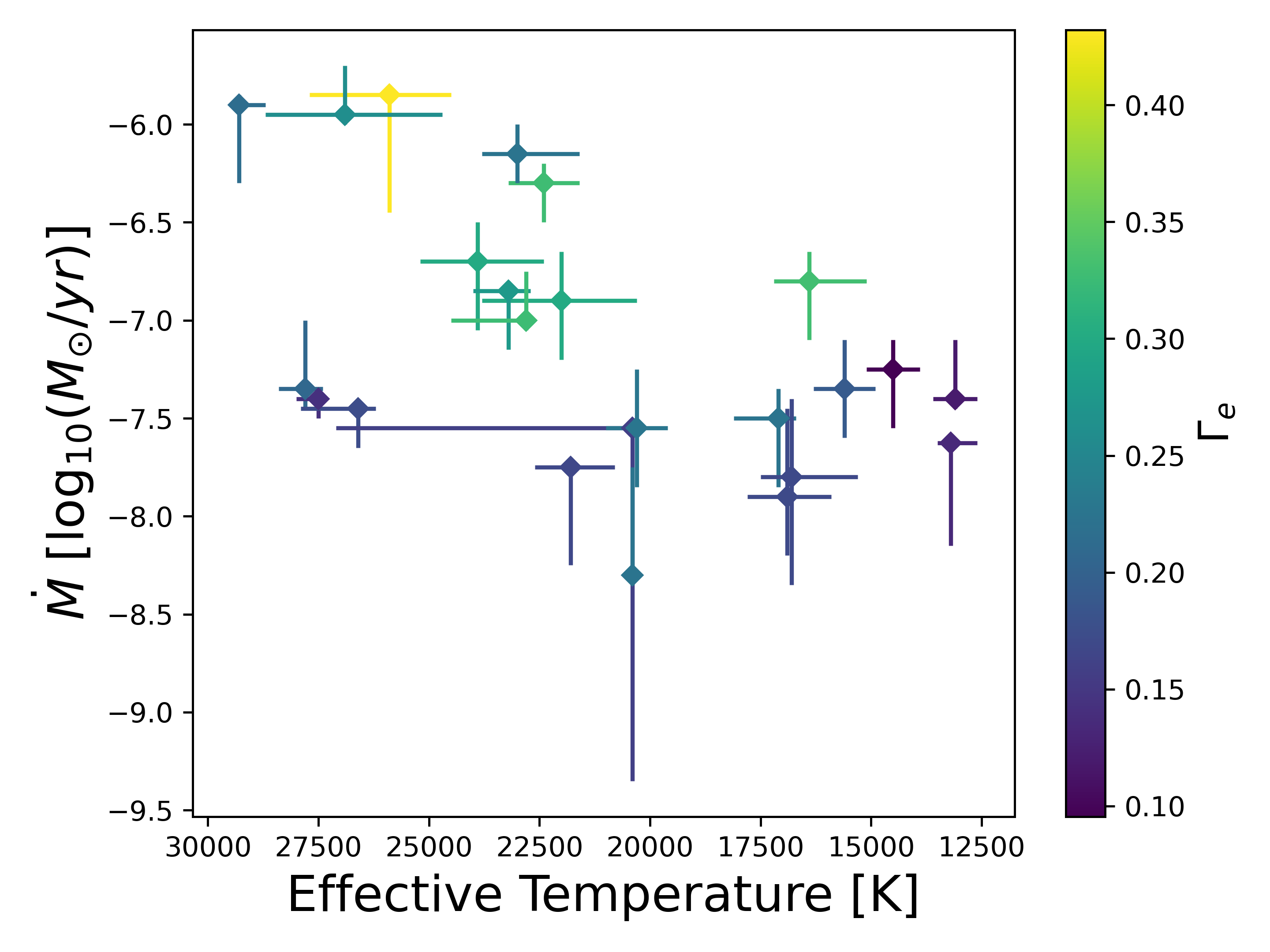}}
		\subfigure{\includegraphics[width=0.45\textwidth]{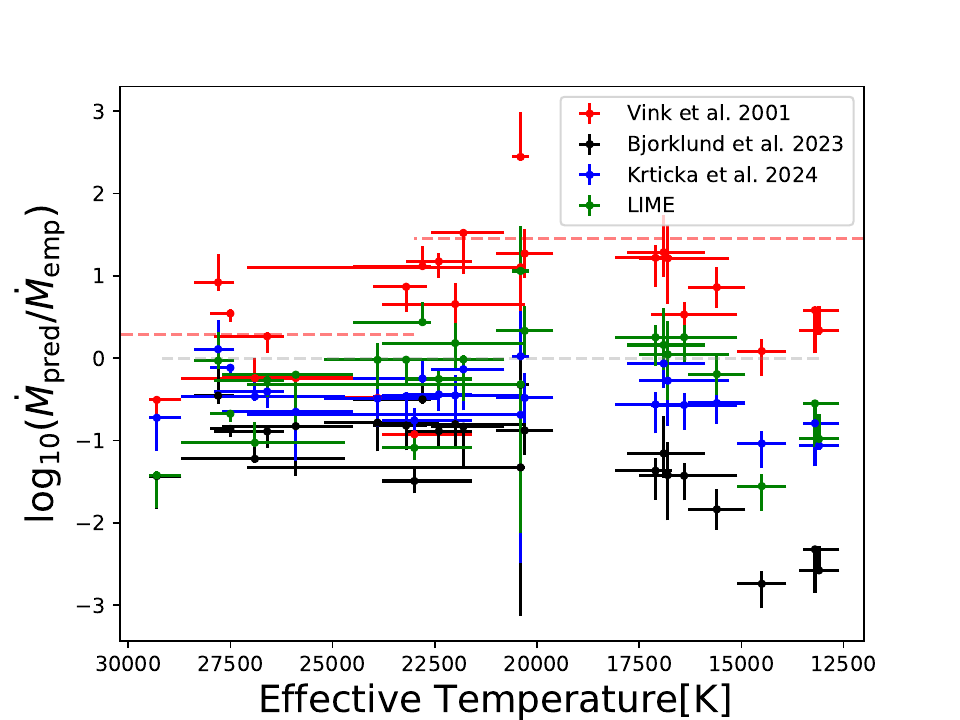}}
		\caption{Derived mass-loss rates for our sample as function of effective temperature. The top plot shows the derived mass-loss rates and the colour shows the derived classical Eddington parameter for each star. The bottom plot shows the logarithm of the ratio between the theoretical prescriptions by \cite{vink_mass-loss_2001, bjorklund_new_2023, krticka_new_2024, sundqvist_--fly_2025} and the empirically found values. 
        The red dashed line shows the average offset for the \citeauthor{vink_mass-loss_2001}, separated for stars on the hot side and the cool side of the jump.
        The grey line indicates where these values are equal.}
		\label{fig:Mass-loss rate emp}
	\end{figure}

    To further our understanding of mass loss in the SMC supergiant B-star regime, we compare our empirical $\dot{M}$ rates to several theoretical predictions. In the LMC and Galaxy, similar comparisons have already been made. 
    \citet{verhamme_x-shooting_2024} using optical and UV data did not identify a jump in mass loss at about spectral type B1 in a sample of LMC B supergiants. \citet{de_burgos_iacob_2024_mass_loss} also did not find such a jump in an optical-only study of a large sample of Galactic B stars. 
    In the bottom panel of figure~\ref{fig:Mass-loss rate emp}, we show the ratio of predicted to empirical mass-loss rate, i.e. $f_{\rm p/e} = \dot{M}_{\rm pred} / \dot{M}_{\rm emp}$, in order to test how well four different theories recover our empirical mass-loss rates.
    We discuss separately how the theories of \citet{vink_mass-loss_2001}, \citet{bjorklund_new_2023}, \citet{krticka_new_2024}, and \citet{sundqvist_--fly_2025} compare to our results.

    (1) \citet{vink_mass-loss_2001} is based on Monte Carlo simulations of photon--gas interactions, accounting for multiple-scattering effects, where the NLTE state of the gas is computed using the Sobolev approximation and the mass-loss rate is constrained from global energy conservation assuming a preset $\beta$-type velocity structure \citep[see also][]{vink_nature_1999,2000A&A...362..295V}.
    For $T_{\rm eff} \ga 25$\,K our empirical estimates are in fair agreement with these predictions, yielding $f_{\rm p/e} = 2.0 \pm 0.9$. 
	For stars cooler than 25\,kK, however, the prescription yields mass-loss rates that exceed the empirical values by an average factor of $30 \pm 17$. This ratio at temperatures below the bi-stability jump is similar to the LMC findings by \citet{verhamme_x-shooting_2024}. 
	This sizeable difference between the offset for stars on the cool and hot side of the jump location strongly suggests that the bi-stability jump is not present in this sample. 
    To allow a comparison to prescriptions that are limited in their prediction range, we also computed an average for all stars above 18\,kK: $f_{\rm p/e}=25 \pm 17$.

    (2) \citet{bjorklund_new_2023} computed dynamical models by deriving the radiative acceleration from NLTE radiative transfer in the co-moving frame and solving the steady-state equation of motion. Hence, they predict both wind mass-loss rate and velocity structure. A comparison to our empirical data does not reveal a clear discontinuous behaviour at about 25\,kK.  $f_{\rm p/e} = 0.12 \pm 0.02$, i.e. the predictions underestimate the mass loss by about a factor of eight.
    No temperature dependence in $f_{\rm p/e}$ was found also for the LMC sample of \citet{verhamme_x-shooting_2024} where the theoretical values are on average a factor of two lower then the empirical values. 
    The larger offset for SMC stars relative to LMC stars may point to a too strong dependence of mass loss on metallicity in the \citet{bjorklund_new_2023} models. 
    The \citeauthor{bjorklund_new_2023} models do not probe temperatures below 15\,kK, whereas three of our target stars are in this regime (SK179, AzV343, AzV324). 
    However, we cannot identify which part of this discrepancy may have a physical cause and which may be the result of extrapolation of the predictions. 
    For these three, the offset is several orders of magnitude.
    However, when only including stars with $T_{\rm eff}$ above 18\,kK, the average offset is very similar ($f_{\rm p/e} = 0.16 \pm 0.03$).

    (3) \citet{krticka_new_2024} compute dynamical models with similar approximations to those by \citet{bjorklund_new_2023}, for example though differences in treatment of atomic data, non-explicit energy levels, and the use of the Sobolev approximation in the inner region to compute bound-bound rates, neglecting velocity curvature effects, and the location of the wind critical point, might lead to slightly differing results. 
    For instance, while the fitting formula of \citeauthor{bjorklund_new_2023} does not have a bi-stability jump, the \citeauthor{krticka_new_2024} predictions feature a very localised bump setting in below 20\,kK and peaking at about 13\,kK, albeit with relatively small \citep[compared to ][]{vink_mass-loss_2001} amplitude. 
    We note that our sample poorly probes the location of this predicted bump at low temperatures. 
    For the full temperature range, $f_{\rm p/e} = 0.42 \pm 0.06$. 
    This is significantly better than for the previous two predictions. 
    For all stars hotter than 18\,kK, $f_{\rm p/e} = 0.48 \pm 0.08$. 
    The \citet{krticka_new_2024} models underpredict $\dot{M}$ in the full temperature range. Similarly to \citeauthor{bjorklund_new_2023}, this underestimation becomes worse for the coolest stars. 

    (4) \citet{sundqvist_--fly_2025} provide the web tool LIME\footnote{The line-driven-wind iterative mass-loss estimator tool LIME is available at \href{}{https://lime.ster.kuleuven.be/}} 
    to compute mass loss for arbitrary compositions from the modified \citep{pauldrach_radiation-driven_1986} Castor-Abbott-Klein theory \citep[CAK;][]{castor_radiation-driven_1975} as formulated by \citet{gayley_improved_1995}, adopting LTE flux-weighted line opacities \citep{2022A&A...667A.113P} to compute line-force parameters at the critical point. 
    For the full sample an average offset of  $f_{\rm p/e} = 1.3 \pm 0.3$ is found, when limiting the comparison to stars $T_{\rm eff}>18$\,kK, the comparison stays the same $f_{\rm p/e} = 1.3 \pm 0.3$. 
    For the coolest $T_{\rm eff}$ stars the mass loss is still underestimated. The latter may (in part) be the result of extrapolation, as the recommended $T_{\rm eff}$ parameter range is 18-60\,kK.
    On average in both regimes LIME performs significantly better than other predictions. 

\begin{figure}
		\centering
		\includegraphics[width=0.45\textwidth]{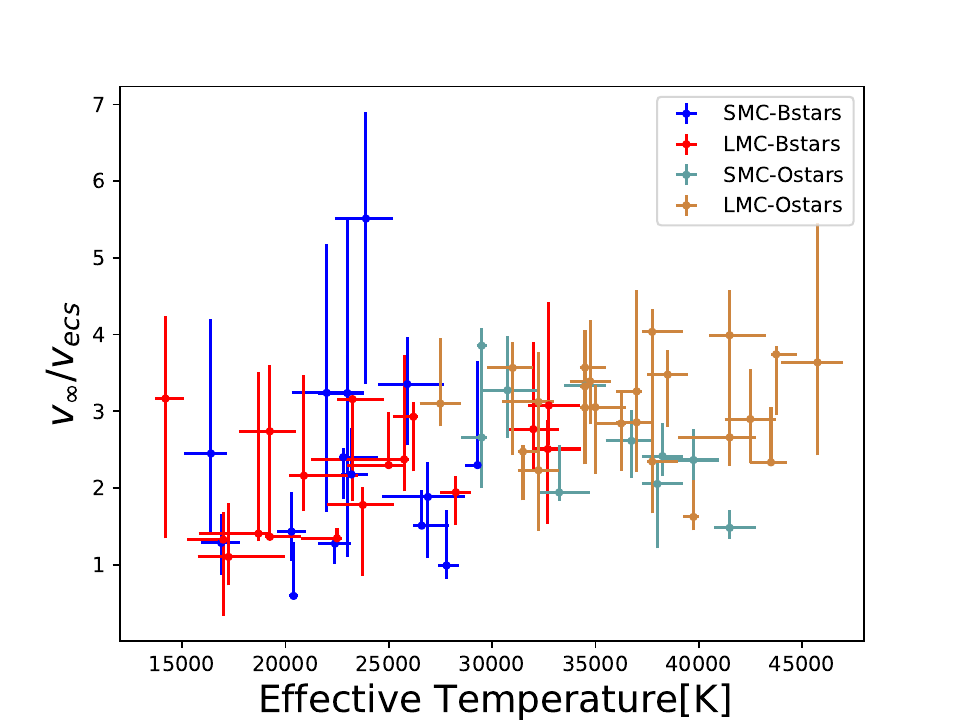}
		\caption{Terminal wind speed divided by escape speed plotted over temperature. The mass used to determine the escape speed is the effective spectral mass. The values shown here are a combination of the stars studied in \citet{backs_x-shooting_2024, hawcroft_empirical_2024, verhamme_x-shooting_2024, brands_x-shooting_2025}, and the sample studied here.}
		\label{fig:vinf_vesc}
	\end{figure}

    Empirical evidence for a bi-stable behaviour of stellar winds of hot stars was first presented by \citet{lamers_terminal_1995}, who identified a jump in the terminal wind velocity divided by the surface escape speed of galactic B1 supergiants. 
    A jump-like behaviour was also found for SMC stars \citep{bernini-peron_x-shooting_2024}. \citet{crowther_physical_2006}, however, reported a more gradual trend for galactic stars, though with large scatter.
    Here we combine the samples of all studies utilising the same collection data, and a similar methodology as utilised in this paper \citep{backs_x-shooting_2024, hawcroft_empirical_2024, verhamme_x-shooting_2024, brands_x-shooting_2025}. These are all for LMC and SMC samples, while \citet{lamers_terminal_1995,crowther_physical_2006} are galactic studies.
	Figure \ref{fig:vinf_vesc} shows $\varv_{\infty}/ \varv_{\rm esc}$ over $T_{\rm eff}$ for our full sample. 
    Here $\varv_{\rm esc}$ is defined as the escape speed from the stellar surface with the radiative acceleration provided by electrons subtracted ($\varv_{\rm esc}^2 = 2 G M_\star (1-\Gamma_{\rm e})/R_\star $).
    Unlike the study of 18 SMC B stars by \citet{bernini-peron_x-shooting_2024}, we do not find a jump in $\varv_{\infty}/\varv_{\rm esc}$.
    Instead, similar to the conclusion of \citeauthor{crowther_physical_2006}, we \blue{find a linear increase of $\varv_{\infty}/ \varv_{\rm esc}$ with $T_{\rm eff}$.}
    The linear trend shows no clear offset between LMC and SMC objects and appears to flatten somewhat at a value of about 3 for the hottest stars. 
    \blue{This behaviour can be seen in figure \ref{fig:fit_vinf_vesc}.}
    One should note that this parameter depends on the spectroscopic mass, for which both random and systematic uncertainties may be large \citep{herrero_mass_1992}. 
    This results in many observable trends within the uncertainties, making it a difficult empirical metric to use.

	\subsection{Effects of metallicity and temperature on mass loss}

    Combining the present work with the four studies mentioned above using a similar atmosphere analysis methodology, specifically the same detailed treatment of wind structure \blue{including a clumping description using $f_{\rm cl}$, $f_{\rm ic}$} and $f_{\rm vel}$ \citep{backs_x-shooting_2024, hawcroft_empirical_2024, verhamme_x-shooting_2024, brands_x-shooting_2025}, allows us to explore how mass loss depends on metallicity for different effective temperature and luminosity ranges. 
    These studies combined cover temperatures in the range 13$-$60\,kK and luminosities in the range $10^{4.7}-10^{6}\,L_{\odot}$, in the LMC and SMC metallicity environments at $Z_{\rm \small LMC} = 0.5 \,Z_{\odot}$ and $Z_{\rm \small SMC} = 0.2 \,Z_{\odot}$. 

    A practical way of doing this is by means of the mechanical momentum of the wind modified by the square root of the radius:
    \begin{eqnarray}
        D \equiv \dot{M} \varv_{\infty} \sqrt{R_*/R_{\odot}}.
    \end{eqnarray}
    The advantage of this modified wind momentum is that in tha CAK theory it can be expressed as a function of luminosity only for reasonable line-force parameters \citep{kudritzki_winds_2000}.

\begin{figure*}
		\centering
		\subfigure{\includegraphics[width=0.35\textwidth]{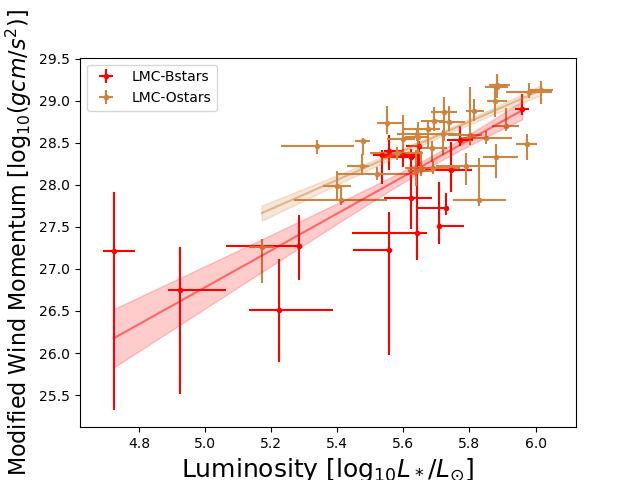}}
		\subfigure{\includegraphics[width=0.35\textwidth]{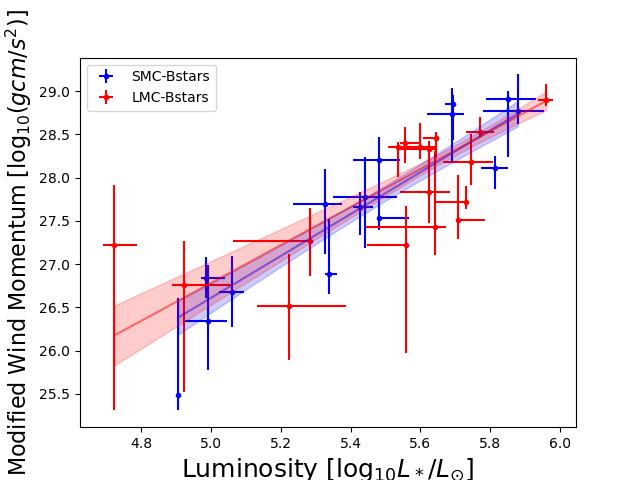}}
		\subfigure{\includegraphics[width=0.35\textwidth]{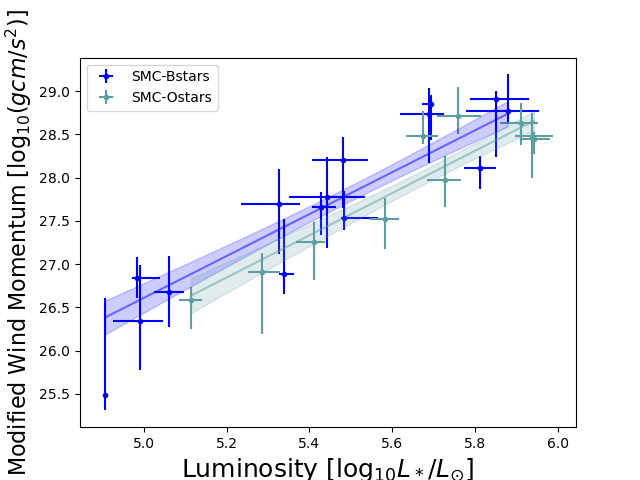}}
		\subfigure{\includegraphics[width=0.35\textwidth]{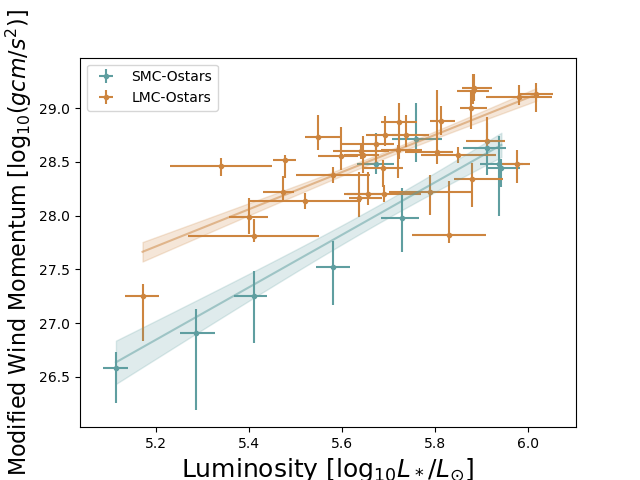}}
		\caption{Modified wind momentum vs luminosity from this study and previous SMC and LMC studies in this line of papers \citep{hawcroft_x-shooting_2024,backs_x-shooting_2024, verhamme_x-shooting_2024, brands_x-shooting_2025}. The fits were performed using an ODR method. The error margins shown reflect the 1$\sigma$ uncertainty.}
		\label{fig:momentum_plots}
	\end{figure*}

	Figure \ref{fig:momentum_plots} shows four comparison plots of the $D(L_*)$ of the subsets of the sample, 
    with all values in cgs units. 
	The orange crosses are all of the LMC O stars with a luminosity above $10^{5.2}L_{\odot}$ by \citet{brands_x-shooting_2025} and \citet{hawcroft_empirical_2024}, green crosses the SMC O stars from \citet{backs_x-shooting_2024}, the red crosses the LMC B-stars from \citet{verhamme_x-shooting_2024}, and the blue crosses are the SMC B stars presented in this paper. 
    For the LMC-O stars we removed all the stars from \citet{hawcroft_empirical_2024} that have a luminosity below $10^{5.2}L_{\odot}$ as they exhibit what is known as the weak wind problem.
    To quantify the differences between the subsamples we fit them with a linear wind momentum luminosity relation,
	\begin{align} \label{eq:Wind_mom_lum}
		\log_{10} D = \log_{10} D_0 + x \cdot \log_{10}(L_*/L_{\odot}).
	\end{align}
	As in \cite{backs_x-shooting_2024}, we use an orthogonal distance regression method \citep[ODR;][]{boggs_accurate_1987} to derive the best fit values and the 1-$\sigma$ error. 
    These are listed in Table \ref{tab:momentum_lin_fit} and shown in the respective panels of figure~\ref{fig:momentum_plots} using colours matching the star symbols.

    Two surprising findings immediately stand out.
    First, there is only a minor difference in wind momentum between O stars and B stars in the SMC (figure \ref{fig:momentum_plots} bottom left panel), and only a modest difference in the LMC (top left), suggesting a weak direct temperature dependence of the mass-loss rate.  
    Second, the wind momentum of LMC B stars matches the wind momentum of the SMC B stars for similar luminosities (top right), while there is a clear difference in $D$ between the LMC and SMC O stars (bottom right). 
	When making similar panels for the mass-loss rate versus luminosity, we see that the SMC B stars have a higher mass-loss rate when compared to the SMC O stars (figure \ref{fig:mass-loss_plots_full}), which is in conflict with all prescriptions as well.

	\begin{table}[]
		\centering
		\caption{Fit values for equation \ref{eq:Wind_mom_lum} for the four different datasets.}
        \small
		\begin{tabular}{c|cl}
			& $\log_{10} D_0$ & $x$ \\
			\hline
			LMC O-star  &$18.7\pm 0.9$& $1.7\pm 0.2$\\
			LMC B-star & $15.8 \pm 2.0$& $2.2 \pm 0.4$\\
			SMC O-star &$14.1 \pm 1.8$& $2.4\pm 0.3$ \\
			SMC B-star & $ 14.5\pm 1.6$ & $2.4 \pm 0.3$ \\
			\hline
		\end{tabular}
		
		\label{tab:momentum_lin_fit}
	\end{table}
	
    Our findings for the B-star regime are at odds with the mass-loss recipes discussed above. 
    The fit formula presented in \citet{vink_mass-loss_2001} yields $\dot{M} \propto Z^{0.64}$ for B stars and $\dot{M} \propto Z^{0.69}$ when taking into account the terminal wind speeds dependence on metallicity.
	The dependence of $\dot{M}$ on metallicity in the \citet{bjorklund_new_2023} prescription even steepens at lower temperatures: for B stars at 20\,kK it scales as $\dot{M} \propto Z^{1.4}$, while for O stars at 35\,kK the mass-loss rate scales as $\dot{M} \propto Z^{0.95}$.
	The \citet{krticka_new_2024} prescription has a more intricate relation with metallicity, which depends on both $T_{\rm eff}$ and luminosity. 
	At a typical value $\log_{10} (L/L_{\odot}) = 5.5$ and $T_{\rm eff} >$\,25\,kK it is found that $\dot{M} \propto Z^{0.41}$, and similarly to the \citet{bjorklund_new_2023} it becomes stronger towards lower temperatures. 
    Because of the differences between $\dot{M}(Z)$ trends in our findings and in theory, we have compared our results to those of previous studies that cover the same stars. 
    The direct comparisons are discussed in section \ref{sec:compare} and are overall in agreement.  

    Because ad libitum computations can be done with the web-tool LIME \citep{sundqvist_--fly_2025}, we can explore its $\dot{M}(Z)$ dependence in considerable depth.
	Figure \ref{fig:LIME_per_teff} shows the trends in different temperature brackets. Interestingly, the metallicity's effect on mass-loss rate has a clear temperature dependence.
	For instance, at $\log_{10}(L/L_{\odot}) = 5.8$ LMC stars with $T_{\rm eff} = 20$\,kK have a mass-loss rate only 1.2 times higher than their SMC counterparts. 
	In comparison, at the same luminosity at $T_{\rm eff} = 40$\,kK an LMC star is found to have a mass-loss rate 1.7 times higher than a similar SMC star. 
    We further note that LIME retrieves accurately the difference in mass-loss rate between LMC and SMC O stars also found in \cite{backs_x-shooting_2024}. 
    
	Although the weak metallicity dependence of LIME for B stars warrants deeper investigation, our considerations suggest it may be related to the conditions at the wind critical point. 
    As discussed in \citet{sundqvist_--fly_2025}, it is generally found from LIME that carbon, nitrogen and oxygen play a more important role in determining the mass-loss rate than previously thought, in particular for cooler objects that are some distance from the Eddington limit. As CNO lines are generally quite strong (and often saturated), the direct dependence on metallicity becomes less prominent than when the more numerous and weaker iron-group lines dominate.     
    \blue{Similarly, \citet{krticka_new_2025} find a weaker metallicity dependence of the mass-loss rate for stars with $T_{\rm eff}$ around 20-27\,kK.}
    \citet{krticka_new_2025} show that CNO are the dominant species for line driving in the $T_{\rm eff}$ range from 20\,kK to 27\,kK, where an Fe contribution seems particularly low. 
    This effect becomes more pronounced at lower metallicity.

	\subsection{Clumping parameters} 

    As introduced in Sect.~\ref{sec:model_fitting}, five parameters describe the clumping properties ($f_{\rm cl}$, $f_{\rm ic}$, $f_{\rm vel}$) and radial development of clumping ($\varv_{\rm cl,start}$, $\varv_{\rm cl,max}$).
	Figure \ref{fig:clumping} shows that the clumping parameters of the SMC sample are very scattered and generally have a high degree of uncertainty. 
	Similarly to the LMC sample \citep{verhamme_x-shooting_2024}, there are no clear trends between any stellar or wind parameters and clumping parameters. 
	Although all clumping parameters have a measurable impact on the mass-loss rate, comparatively the impact of $f_{\rm cl}$ is largest.
    As the errors in the GA fitting approach take into account degeneracies among the parameters, all trends --- including those in mass loss --- are real despite a possible degeneracy, notably between $\dot{M}$ and $f_{\rm cl}$.

    The large error margins and the lack of trends in clumping characteristics as a function of stellar and wind properties is possibly related to a clumping formalism that does not adequately characterise the physical structure of clumping. 
    Similar conclusions are drawn by \citet{verhamme_x-shooting_2024, brands_x-shooting_2025}.
    In contrast to the current two-component medium used in {\sc FASTWIND}, the results from multi-dimensional radiation hydrodynamics \citep{moens_first_2022, debnath_2d_2024} suggest a distribution of densities that more closely resembles a log-normal distribution around a mean. 
    A hint that such a local density distribution is indeed a better physical description may be visible in the relatively high values of $f_{\rm ic}$ (for which the weighted average is $\approx 0.4$). 
    A similar result for $f_{\rm ic}$ was also reported for LMC stars by \citet{verhamme_x-shooting_2024}.
    The importance of interclump density has also been marked by \citet{zsargo_importance_2008, puebla_x-ray_2016} for its role in producing super-ionised lines. 
    However, they required a low interclump density.  
	
	\section{Summary and conclusion}\label{sec:summary}

    The primary objective of the photosphere and wind analysis of the 24 B stars in the SMC presented in this work was to empirically determine the presence or absence of a bi-stability jump in mass loss at early B spectral types. 
    As our sample covers a wide range of luminosities, we searched for a jump behaviour with temperature in the ratio of predicted to empirical mass loss (see Fig.~\ref{fig:Mass-loss rate emp}). 
    No bi-stability feature in the spectral range B1-3 was found. 
    In the range 16$-$30\,kK, the predictions by the web tool LIME \citep{sundqvist_--fly_2025} lack a bi-stability jump or bump and recover the empirical $\dot{M}$ values most closely.

    For stars with spectral type later than B3 ($T_{\rm eff} \simeq 17$\,kK), for which only a few data points are available, theory tends to strongly underpredict $\dot{M}$. 
    For the \citet{vink_mass-loss_2001} rates, the differences between empirical and theoretical mass-loss rates are small at $T_{\rm eff} \simeq 17$\,K. 
    However, their strong overprediction at 20\,kK implies a decline that is too steep compared to observations.
    It remains to be explored whether this is the result of extrapolation beyond the parameter space of validity of the predictions or points to shortcomings in the theory or models used to analyse the observations.

    The derived wind clumping parameters are uncertain and show a large scatter. Found as well in other studies \citep[including the LMC B-star analysis by][]{verhamme_x-shooting_2024}, it was suggested in \citep{verhamme_x-shooting_2024,brands_x-shooting_2025} that this behaviour might point to shortcomings in the representation of clumping behaviour in {\sc fastwind}. 
    Specifically, the large value of the local interclump density relative to the local mean density, $f_{\rm ic} \sim 0.4$, potentially points to a breakdown of the assumption of a two-component density structure.
    Instead of high-density clumps and a very low-density interclump medium, a more continuous distribution may be more realistic. 
    Multi-dimensional hydrodynamical simulations, which naturally clump and need no ad hoc clumping parameters point to the same conclusion \citep[e.g.][]{driessen_2022}.

    One of the main strengths of the combined ULLYSES \citep{roman-duval_ultraviolet_2020,roman-duval_uv_2025} and XshootU \citep{vink_x-shooting_2023} programmes, which are securing ultraviolet and optical spectra of large samples of LMC and SMC stars, is the comprehensive coverage of hot stars in the upper Hertzsprung--Russell diagram. 
    Now that focussed studies have constrained the stellar and wind properties of over 80 stars in the temperature range 15$-$60\,kK using uniform methodologies that address both clumping and mass loss \citep{hawcroft_empirical_2024,backs_x-shooting_2024, verhamme_x-shooting_2024,brands_x-shooting_2025}, we were able to search for trends in wind properties with spectral type and metallicity. This yielded two surprising findings:
    (1) The modified wind-momenta ($D$) for SMC O-type and B-type stars are very similar. 
    In the LMC, the O stars have somewhat higher $D$ values.
    When considering mass-loss rate the SMC B stars even surpass the SMC O stars. 
    (2) The mass-loss rates of B-type stars only very weakly depend on metallicity. The physical meaning of this result requires further exploration. 
    Interestingly, the predictions by the recently published web tool LIME \citep{sundqvist_--fly_2025}, which computes mass-loss rates using line-force parameters specifically tailored to exact stellar and wind properties, predict a significantly weaker metallicity dependence for B stars relative to O stars.
    This offers a means to explore the physical reason(s) for this effect, which may be related to differences in the strength and elemental species of the dominant wind driving spectral lines near the wind critical point. Though for B stars the $Z$ dependence appears weak, this may change when going to sub-SMC metallicity.
    
    \section*{Data availability}
    The fits of all SMC B-supergiants discussed here and the distribution of the GA models are available at \url{https://zenodo.org/records/17725648}
    \begin{acknowledgements}
    The resources and services used in this work were provided by the VSC (Flemish Supercomputer Center), funded by the Research Foundation - Flanders (FWO) and the Flemish Government.
    We would also like to thank professor Leen Decin for her contribution to this work.  
    O.V. and J.S. acknowledge the support of 
    the Belgian Research Foundation Flanders (FWO) Odysseus programme under grant number G0H9218N and FWO grant G077822N and, KU Leuven C1 grant MAESTRO C16/17/007.
    O.V. would further like to thank the National Research Foundation of Korea (NRF) under grant number: 2021R1A2C1008928 and the KASI primary with program number: 2026183301.
    J.S., F.B., and P.S. further acknowledge the support of the European Research Council (ERC) Horizon Europe under grant agreement number 101044048. 
    ACGM thanks the support from project 10108195 MERIT (MSCA-COFUND Horizon Europe). 
    RK acknowledges financial support via the Heisenberg Research Grant funded by the Deutsche Forschungsgemeinschaft (DFG, German Research Foundation) under grant no.~KU 2849/9, project no.~445783058. We would also like to thank the anonymous referee for their detailed comments. 
    \end{acknowledgements}

	\addcontentsline{toc}{section}{Bibliography}
	\bibliography{Bibliography/aanda.bib}

@ARTICLE{2022A&A...667A.113P,
       author = {{Poniatowski}, L.~G. and {Kee}, N.~D. and {Sundqvist}, J.~O. and {Driessen}, F.~A. and {Moens}, N. and {Owocki}, S.~P. and {Gayley}, K.~G. and {Decin}, L. and {de Koter}, A. and {Sana}, H.},
        title = "{Method and new tabulations for flux-weighted line opacity and radiation line force in supersonic media}",
      journal = {\aap},
         year = 2022,
        month = nov,
       volume = {667},
          eid = {A113},
        pages = {A113},
          doi = {10.1051/0004-6361/202142888},
archivePrefix = {arXiv},
       eprint = {2204.09981},
 primaryClass = {astro-ph.SR},
       adsurl = {https://ui.adsabs.harvard.edu/abs/2022A&A...667A.113P}
}

@ARTICLE{2002A&A...393..543V,
       author = {{Vink}, Jorick S. and {de Koter}, A.},
        title = "{Predictions of variable mass loss for Luminous Blue Variables}",
      journal = {\aap},
         year = 2002,
        month = oct,
       volume = {393},
        pages = {543-553},
          doi = {10.1051/0004-6361:20021009},
archivePrefix = {arXiv},
       eprint = {astro-ph/0207170},
 primaryClass = {astro-ph},
       adsurl = {https://ui.adsabs.harvard.edu/abs/2002A&A...393..543V}
}

@ARTICLE{2011A&A...531L..10G,
       author = {{Groh}, J.~H. and {Vink}, J.~S.},
        title = "{The bi-stability jump as the origin for multiple P-Cygni absorption components in luminous blue variables}",
      journal = {\aap},
         year = 2011,
        month = jul,
       volume = {531},
          eid = {L10},
        pages = {L10},
          doi = {10.1051/0004-6361/201117087},
archivePrefix = {arXiv},
       eprint = {1106.3007},
 primaryClass = {astro-ph.SR},
       adsurl = {https://ui.adsabs.harvard.edu/abs/2011A&A...531L..10G}
}

@ARTICLE{2000A&A...362..295V,
       author = {{Vink}, J.~S. and {de Koter}, A. and {Lamers}, H.~J.~G.~L.~M.},
        title = "{New theoretical mass-loss rates of O and B stars}",
      journal = {\aap},
         year = 2000,
        month = oct,
       volume = {362},
        pages = {295-309},
          doi = {10.48550/arXiv.astro-ph/0008183},
archivePrefix = {arXiv},
       eprint = {astro-ph/0008183},
 primaryClass = {astro-ph},
       adsurl = {https://ui.adsabs.harvard.edu/abs/2000A&A...362..295V}
}

@ARTICLE{2011A&A...530A.115B,
       author = {{Brott}, I. and {de Mink}, S.~E. and {Cantiello}, M. and {Langer}, N. and {de Koter}, A. and {Evans}, C.~J. and {Hunter}, I. and {Trundle}, C. and {Vink}, J.~S.},
        title = "{Rotating massive main-sequence stars. I. Grids of evolutionary models and isochrones}",
      journal = {\aap},
         year = 2011,
        month = jun,
       volume = {530},
          eid = {A115},
        pages = {A115},
          doi = {10.1051/0004-6361/201016113},
archivePrefix = {arXiv},
       eprint = {1102.0530},
 primaryClass = {astro-ph.SR},
       adsurl = {https://ui.adsabs.harvard.edu/abs/2011A&A...530A.115B}
}

@ARTICLE{2022ARA&A..60..203V,
       author = {{Vink}, Jorick S.},
        title = "{Theory and Diagnostics of Hot Star Mass Loss}",
      journal = {\araa},
         year = 2022,
        month = aug,
       volume = {60},
        pages = {203-246},
          doi = {10.1146/annurev-astro-052920-094949},
archivePrefix = {arXiv},
       eprint = {2109.08164},
 primaryClass = {astro-ph.SR},
       adsurl = {https://ui.adsabs.harvard.edu/abs/2022ARA&A..60..203V}
}

@ARTICLE{2008A&ARv..16..209P,
       author = {{Puls}, Joachim and {Vink}, Jorick S. and {Najarro}, Francisco},
        title = "{Mass loss from hot massive stars}",
      journal = {\aapr},
         year = 2008,
        month = dec,
       volume = {16},
       number = {3-4},
        pages = {209-325},
          doi = {10.1007/s00159-008-0015-8},
archivePrefix = {arXiv},
       eprint = {0811.0487},
 primaryClass = {astro-ph},
       adsurl = {https://ui.adsabs.harvard.edu/abs/2008A&ARv..16..209P}
}

@ARTICLE{2007A&A...473..603M,
       author = {{Mokiem}, M.~R. and {de Koter}, A. and {Vink}, J.~S. and {Puls}, J. and {Evans}, C.~J. and {Smartt}, S.~J. and {Crowther}, P.~A. and {Herrero}, A. and {Langer}, N. and {Lennon}, D.~J. and {Najarro}, F. and {Villamariz}, M.~R.},
        title = "{The empirical metallicity dependence of the mass-loss rate of O- and early B-type stars}",
      journal = {\aap},
         year = 2007,
        month = oct,
       volume = {473},
       number = {2},
        pages = {603-614},
          doi = {10.1051/0004-6361:20077545},
archivePrefix = {arXiv},
       eprint = {0708.2042},
 primaryClass = {astro-ph},
       adsurl = {https://ui.adsabs.harvard.edu/abs/2007A&A...473..603M}
}

@article{shenar_binarity_2024,
	title = {Binarity at {LOw} {Metallicity} ({BLOeM}): {A} spectroscopic {VLT} monitoring survey of massive stars in the {SMC}},
	volume = {690},
	issn = {0004-6361},
	shorttitle = {Binarity at {LOw} {Metallicity} ({BLOeM})},
	url = {https://ui.adsabs.harvard.edu/abs/2024A&A...690A.289S},
	doi = {10.1051/0004-6361/202451586},
	urldate = {2025-10-31},
	journal = {Astronomy and Astrophysics},
	author = {Shenar, T. and Bodensteiner, J. and Sana, H. and Crowther, P. A. and Lennon, D. J. and Abdul-Masih, M. and Almeida, L. A. and Backs, F. and Berlanas, S. R. and Bernini-Peron, M. and Bestenlehner, J. M. and Bowman, D. M. and Bronner, V. A. and Britavskiy, N. and de Koter, A. and de Mink, S. E. and Deshmukh, K. and Evans, C. J. and Fabry, M. and Gieles, M. and Gilkis, A. and González-Torà, G. and Gräfener, G. and Götberg, Y. and Hawcroft, C. and Hénault-Brunet, V. and Herrero, A. and Holgado, G. and Janssens, S. and Johnston, C. and Josiek, J. and Justham, S. and Kalari, V. M. and Katabi, Z. Z. and Keszthelyi, Z. and Klencki, J. and Kubát, J. and Kubátová, B. and Langer, N. and Lefever, R. R. and Ludwig, B. and Mackey, J. and Mahy, L. and Maíz Apellániz, J. and Mandel, I. and Maravelias, G. and Marchant, P. and Menon, A. and Najarro, F. and Oskinova, L. M. and O'Grady, A. J. G. and Ovadia, R. and Patrick, L. R. and Pauli, D. and Pawlak, M. and Ramachandran, V. and Renzo, M. and Rocha, D. F. and Sander, A. A. C. and Sayada, T. and Schneider, F. R. N. and Schootemeijer, A. and Schösser, E. C. and Schürmann, C. and Sen, K. and Shahaf, S. and Simón-Díaz, S. and Stoop, M. and Toonen, S. and Tramper, F. and van Loon, J. Th. and Valli, R. and van Son, L. A. C. and Vigna-Gómez, A. and Villaseñor, J. I. and Vink, J. S. and Wang, C. and Willcox, R.},
	month = oct,
	year = {2024},
	note = {Publisher: EDP
ADS Bibcode: 2024A\&A...690A.289S},
	pages = {A289},
}

@article{krticka_new_2025,
	title = {New line-driven wind mass-loss rates for {OB} stars with metallicities down to 0.01 {Z}⊙},
	volume = {702},
	issn = {0004-6361},
	url = {https://ui.adsabs.harvard.edu/abs/2025A&A...702A...9K},
	doi = {10.1051/0004-6361/202556234},
	urldate = {2026-01-30},
	journal = {\aap},
	publisher = {EDP},
	author = {Krtička, J. and Kubát, J. and Krtičková, I.},
	month = oct,
	year = {2025},
	note = {ADS Bibcode: 2025A\&A...702A...9K},
	pages = {A9},
}

@article{britavskiy_binarity_2025,
	title = {Binarity at {LOw} {Metallicity} ({BLOeM}): {Multiplicity} of early {B}-type supergiants in the {Small} {Magellanic} {Cloud}},
	volume = {698},
	issn = {0004-6361},
	shorttitle = {Binarity at {LOw} {Metallicity} ({BLOeM})},
	url = {https://ui.adsabs.harvard.edu/abs/2025A&A...698A..40B},
	doi = {10.1051/0004-6361/202452963},
	urldate = {2025-10-31},
	journal = {\aap},
	author = {Britavskiy, N. and Mahy, L. and Lennon, D. J. and Patrick, L. R. and Sana, H. and Villaseñor, J. I. and Shenar, T. and Bodensteiner, J. and Bernini-Peron, M. and Berlanas, S. R. and Bowman, D. M. and Crowther, P. A. and de Mink, S. E. and Evans, C. J. and Götberg, Y. and Holgado, G. and Johnston, C. and Keszthelyi, Z. and Klencki, J. and Langer, N. and Mandel, I. and Menon, A. and Moe, M. and Oskinova, L. M. and Pauli, D. and Pawlak, M. and Ramachandran, V. and Renzo, M. and Sander, A. A. C. and Schneider, F. R. N. and Schootemeijer, A. and Sen, K. and Simón-Díaz, S. and van Loon, J. Th. and Vink, J. S.},
	month = jun,
	year = {2025},
	note = {Publisher: EDP
ADS Bibcode: 2025A\&A...698A..40B},
	pages = {A40},
}

@article{santolaya-rey_atmospheric_1997,
	title = {Atmospheric {NLTE}-models for the spectroscopic analysis of luminous blue stars with winds.},
	volume = {323},
	issn = {0004-6361},
	url = {https://ui.adsabs.harvard.edu/abs/1997A%26A...323..488S/abstract},
	language = {en},
	urldate = {2021-10-27},
	journal = {\aap, v.323, p.488-512},
	author = {Santolaya-Rey, A. E. and Puls, J. and Herrero, A.},
	month = jul,
	year = {1997},
	pages = {488},
}

@article{puls_atmospheric_2005,
	title = {Atmospheric {NLTE}-models for the spectroscopic analysis of blue stars with winds. {II}. {Line}-blanketed models},
	volume = {435},
	issn = {0004-6361},
	url = {https://ui.adsabs.harvard.edu/abs/2005A%26A...435..669P/abstract},
	doi = {10.1051/0004-6361:20042365},
	language = {en},
	number = {2},
	urldate = {2021-09-16},
	journal = {\aap, Volume 435, Issue 2, May IV 2005, pp.669-698},
	author = {Puls, J. and Urbaneja, M. A. and Venero, R. and Repolust, T. and Springmann, U. and Jokuthy, A. and Mokiem, M. R.},
	month = may,
	year = {2005},
	pages = {669},
}

@article{owocki_rybicki_1984,
	title = {Instabilities in line-driven stellar winds. {I}. {Dependence} on perturbation wavelength.},
	volume = {284},
	issn = {0004-637X},
	url = {https://ui.adsabs.harvard.edu/abs/1984ApJ...284..337O},
	doi = {10.1086/162412},
	urldate = {2021-09-09},
	journal = {\apj},
	author = {Owocki, S. P. and Rybicki, G. B.},
	month = sep,
	year = {1984},
	note = {ADS Bibcode: 1984ApJ...284..337O},
	pages = {337--350},
}

@article{castor_radiation-driven_1975,
	title = {Radiation-driven winds in {Of} stars},
	volume = {195},
	issn = {0004-637X, 1538-4357},
	url = {http://adsabs.harvard.edu/doi/10.1086/153315},
	doi = {10.1086/153315},
	language = {en},
	urldate = {2021-09-09},
	journal = {\apj},
	author = {Castor, J. I. and Abbott, D. C. and Klein, R. I.},
	month = jan,
	year = {1975},
	pages = {157},
}

@article{roman-duval_ultraviolet_2020,
	title = {Ultraviolet {Legacy} {Library} of {Young} {Stars} as {Essential} {Standards} ({ULLYSES}): {Data} {Release} {I}},
	volume = {4},
	issn = {2515-5172},
	shorttitle = {Ultraviolet {Legacy} {Library} of {Young} {Stars} as {Essential} {Standards} ({ULLYSES})},
	url = {https://doi.org/10.3847/2515-5172/abca2f},
	doi = {10.3847/2515-5172/abca2f},
	language = {en},
	number = {11},
	urldate = {2021-08-19},
	journal = {Research Notes of the AAS},
	author = {Roman-Duval, Julia and Proffitt, Charles R. and Taylor, Joanna M. and Monroe, TalaWanda R. and Fischer, Travis C. and Fischer, William J. and Fullerton, A. W. and Aloisi, Alessandra and Britt, Christopher T. and Busko, Ivo and Carlberg, Joleen K. and Rosa, Gisella De and Jedrzejewski, R. I. and Lockwood, Sean and Frazer, Elaine M. and Hernandez, Svea and James, Bethan L. and Oliveira, Cristina and Plesha, Rachel J. and Riedel, Adric R. and Riley, Allyssa and Sahnow, David J. and Sankrit, Ravi and Shaw, Richard A. and Smith, Linda J. and Sohn, Sangmo Tony and Som, Debopam and Ubeda, Leonardo and Welty, Daniel E.},
	month = nov,
	year = {2020},
	note = {Publisher: American Astronomical Society},
	pages = {205},
}

@article{brands_r136_2022,
	title = {The {R136} star cluster dissected with {Hubble} {Space} {Telescope}/{STIS} - {III}. {The} most massive stars and their clumped winds},
	volume = {663},
	copyright = {© ESO 2022},
	issn = {0004-6361, 1432-0746},
	url = {https://www.aanda.org/articles/aa/abs/2022/07/aa42742-21/aa42742-21.html},
	doi = {10.1051/0004-6361/202142742},
	language = {en},
	urldate = {2024-10-18},
	journal = {\aap},
	author = {Brands, Sarah A. and Koter, Alex de and Bestenlehner, Joachim M. and Crowther, Paul A. and Sundqvist, Jon O. and Puls, Joachim and Caballero-Nieves, Saida M. and Abdul-Masih, Michael and Driessen, Florian A. and García, Miriam and Geen, Sam and Gräfener, Götz and Hawcroft, Calum and Kaper, Lex and Keszthelyi, Zsolt and Langer, Norbert and Sana, Hugues and Schneider, Fabian R. N. and Shenar, Tomer and Vink, Jorick S.},
	month = jul,
	year = {2022},
	note = {Publisher: EDP Sciences},
	pages = {A36},
}

@article{mokiem_spectral_2005,
	title = {Spectral analysis of early-type stars using a genetic algorithm based fitting method},
	volume = {441},
	copyright = {© ESO, 2005},
	issn = {0004-6361, 1432-0746},
	url = {https://www.aanda.org/articles/aa/abs/2005/38/aa3522-05/aa3522-05.html},
	doi = {10.1051/0004-6361:20053522},
	language = {en},
	number = {2},
	urldate = {2022-04-14},
	journal = {\aap},
	author = {Mokiem, M. R. and Koter, A. de and Puls, J. and Herrero, A. and Najarro, F. and Villamariz, M. R.},
	month = oct,
	year = {2005},
	note = {Number: 2
Publisher: EDP Sciences},
	pages = {711--733},
}

@article{bjorklund_new_2023,
	title = {New predictions for radiation-driven, steady-state mass-loss and wind-momentum from hot, massive stars. {III}. {Updated} mass-loss rates for stellar evolution},
	volume = {676},
	issn = {0004-6361},
	url = {https://ui.adsabs.harvard.edu/abs/2023A&A...676A.109B},
	doi = {10.1051/0004-6361/202141948},
	urldate = {2024-04-03},
	journal = {\aap},
	author = {Björklund, R. and Sundqvist, J. O. and Singh, S. M. and Puls, J. and Najarro, F.},
	month = aug,
	year = {2023},
	note = {ADS Bibcode: 2023A\&A...676A.109B},
	pages = {A109},
}

@article{bjorklund_new_2021,
	title = {New predictions for radiation-driven, steady-state mass-loss and wind-momentum from hot, massive stars. {II}. {A} grid of {O}-type stars in the {Galaxy} and the {Magellanic} {Clouds}},
	volume = {648},
	issn = {0004-6361},
	url = {https://ui.adsabs.harvard.edu/abs/2021A&A...648A..36B},
	doi = {10.1051/0004-6361/202038384},
	urldate = {2024-10-18},
	journal = {\aap},
	author = {Björklund, R. and Sundqvist, J. O. and Puls, J. and Najarro, F.},
	month = apr,
	year = {2021},
	note = {ADS Bibcode: 2021A\&A...648A..36B},
	pages = {A36},
}

@article{puls_atmospheric_2020,
	title = {Atmospheric {NLTE} models for the spectroscopic analysis of blue stars with winds. {V}. {Complete} comoving frame transfer, and updated modeling of {X}-ray emission},
	volume = {642},
	issn = {0004-6361, 1432-0746},
	url = {http://arxiv.org/abs/2011.02310},
	doi = {10.1051/0004-6361/202038464},
	language = {en},
	urldate = {2022-06-26},
	journal = {\aap},
	author = {Puls, J. and Najarro, F. and Sundqvist, J. O. and Sen, K.},
	month = oct,
	year = {2020},
    note = {ADS Bibcode: 2020A\&A...642A.172P},
	pages = {A172},
}

@article{vink_mass-loss_2001,
	title = {Mass-loss predictions for {O} and {B} stars as a function of metallicity},
	volume = {369},
	issn = {0004-6361, 1432-0746},
	url = {http://arxiv.org/abs/astro-ph/0101509},
	doi = {10.1051/0004-6361:20010127},
	language = {en},
	number = {2},
	urldate = {2022-06-27},
	journal = {\aap},
	author = {Vink, Jorick S. and de Koter, Alex and Lamers, Henny J. G. L. M.},
	month = apr,
	year = {2001},
	note = {ADS Bibcode: 2001A\&A...369..574V},
	pages = {574--588},
}

@article{rubio-diez_upper_2022,
	title = {Upper {Mass}-{Loss} {Limits} and {Clumping} in the {Intermediate} and {Outer} {Wind} {Regions} of {OB} stars},
	volume = {658},
	issn = {0004-6361, 1432-0746},
	url = {http://arxiv.org/abs/2108.11734},
	doi = {10.1051/0004-6361/202040116},
	language = {en},
	urldate = {2022-06-29},
	journal = {\aap},
	author = {Rubio-Díez, M. M. and Sundqvist, J. O. and Najarro, F. and Traficante, A. and Puls, J. and Calzoletti, L. and Figer, D.},
	month = feb,
	year = {2022},
	note = {arXiv:2108.11734 [astro-ph]},
	pages = {A61},
}

@article{markova_bright_2008,
	title = {Bright {OB} stars in the {Galaxy}. {IV}. {Stellar} and wind parameters of early to late {B} supergiants},
	volume = {478},
	issn = {0004-6361},
	url = {https://ui.adsabs.harvard.edu/abs/2008A%26A...478..823M/abstract},
	doi = {10.1051/0004-6361:20077919},
	language = {en},
	number = {3},
	urldate = {2022-06-29},
	journal = {\aap, Volume 478, Issue 3, February II 2008, pp.823-842},
	author = {Markova, N. and Puls, J.},
	month = feb,
	year = {2008},
	pages = {823},
}

@article{hawcroft_empirical_2021,
	title = {Empirical mass-loss rates and clumping properties of {Galactic} early-type {O} supergiants},
	volume = {655},
	issn = {0004-6361, 1432-0746},
	url = {http://arxiv.org/abs/2108.08340},
	doi = {10.1051/0004-6361/202140603},
	language = {en},
	urldate = {2022-07-10},
	journal = {\aap},
	author = {Hawcroft, C. and Sana, H. and Mahy, L. and Sundqvist, J. O. and Abdul-Masih, M. and Bouret, J. C. and Brands, S. A. and de Koter, A. and Driessen, F. A. and Puls, J.},
	month = nov,
	year = {2021},
	note = {arXiv:2108.08340 [astro-ph]},
	pages = {A67},
}

@article{trundle_understanding_2004,
	title = {Understanding {B}-type supergiants in the low metallicity environment of the {SMC}},
	volume = {417},
	issn = {0004-6361, 1432-0746},
	url = {http://www.aanda.org/10.1051/0004-6361:20034325},
	doi = {10.1051/0004-6361:20034325},
	language = {en},
	number = {1},
	urldate = {2022-12-19},
	journal = {\aap},
	author = {Trundle, C. and Lennon, D. J. and Puls, J. and Dufton, P. L.},
	month = apr,
	year = {2004},
	pages = {217--234},
}

@article{trundle_understanding_2005,
	title = {Understanding {B}-type supergiants in the low metallicity environment of the {SMC} {II}},
	volume = {434},
	issn = {0004-6361, 1432-0746},
	url = {http://www.aanda.org/10.1051/0004-6361:20042061},
	doi = {10.1051/0004-6361:20042061},
	language = {en},
	number = {2},
	urldate = {2022-12-19},
	journal = {\aap},
	author = {Trundle, C. and Lennon, D. J.},
	month = may,
	year = {2005},
	pages = {677--689},
}

@article{evans_ultraviolet_2004,
	title = {The {Ultraviolet} and {Optical} {Spectra} of {Luminous} {B}-{Type} {Stars} in the {Small} {Magellanic} {Cloud}},
	volume = {116},
	issn = {0004-6280},
	url = {https://ui.adsabs.harvard.edu/abs/2004PASP..116..909E},
	doi = {10.1086/425563},
	urldate = {2023-01-10},
	journal = {Publications of the Astronomical Society of the Pacific},
	author = {Evans, C. J. and Lennon, D. J. and Walborn, N. R. and Trundle, C. and Rix, S. A.},
	month = oct,
	year = {2004},
	note = {ADS Bibcode: 2004PASP..116..909E},
	pages = {909--919},
}

@article{lamers_terminal_1995,
	title = {Terminal {Velocities} and the {Bistability} of {Stellar} {Winds}},
	volume = {455},
	issn = {0004-637X},
	url = {https://ui.adsabs.harvard.edu/abs/1995ApJ...455..269L},
	doi = {10.1086/176575},
	urldate = {2023-01-16},
	journal = {\apj},
	author = {Lamers, Henny J. G. L. M. and Snow, Theodore P. and Lindholm, Douglas M.},
	month = dec,
	year = {1995},
	note = {ADS Bibcode: 1995ApJ...455..269L},
	pages = {269},
}

@article{carneiro_atmospheric_2016,
	title = {Atmospheric {NLTE} models for the spectroscopic analysis of blue stars with winds. {III}. {X}-ray emission from wind-embedded shocks},
	volume = {590},
	issn = {0004-6361},
	url = {https://ui.adsabs.harvard.edu/abs/2016A&A...590A..88C},
	doi = {10.1051/0004-6361/201527718},
	urldate = {2023-04-27},
	journal = {\aap},
	author = {Carneiro, L. P. and Puls, J. and Sundqvist, J. O. and Hoffmann, T. L.},
	month = may,
	year = {2016},
	note = {ADS Bibcode: 2016A\&A...590A..88C},
	pages = {A88},
}

@article{krticka_new_2021,
	title = {New mass-loss rates of {B} supergiants from global wind models},
	volume = {647},
	issn = {0004-6361},
	url = {https://ui.adsabs.harvard.edu/abs/2021A&A...647A..28K},
	doi = {10.1051/0004-6361/202039900},
	urldate = {2023-06-11},
	journal = {\aap},
	author = {Krtička, J. and Kubát, J. and Krtičková, I.},
	month = mar,
	year = {2021},
	note = {ADS Bibcode: 2021A\&A...647A..28K},
	pages = {A28},
}

@article{hawcroft_x-shooting_2024,
	title = {X-{Shooting} {ULLYSES}: {Massive} stars at low metallicity. {III}. {Terminal} wind speeds of {ULLYSES} massive stars},
	volume = {688},
	issn = {0004-6361},
	shorttitle = {X-{Shooting} {ULLYSES}},
	url = {https://ui.adsabs.harvard.edu/abs/2024A&A...688A.105H},
	doi = {10.1051/0004-6361/202245588},
	urldate = {2024-09-20},
	journal = {\aap},
	author = {Hawcroft, C. and Sana, H. and Mahy, L. and Sundqvist, J. O. and de Koter, A. and Crowther, P. A. and Bestenlehner, J. M. and Brands, S. A. and David-Uraz, A. and Decin, L. and Erba, C. and Garcia, M. and Hamann, W. -R. and Herrero, A. and Ignace, R. and Kee, N. D. and Kubátová, B. and Lefever, R. and Moffat, A. and Najarro, F. and Oskinova, L. and Pauli, D. and Prinja, R. and Puls, J. and Sander, A. A. C. and Shenar, T. and St-Louis, N. and ud-Doula, A. and Vink, J. S.},
	month = aug,
	year = {2024},
	note = {ADS Bibcode: 2024A\&A...688A.105H},
	pages = {A105},
}

@article{hawcroft_empirical_2024,
	title = {Empirical mass-loss rates and clumping properties of {O}-type stars in the {Large} {Magellanic} {Cloud}},
	volume = {690},
	issn = {0004-6361},
	url = {https://ui.adsabs.harvard.edu/abs/2024A&A...690A.126H},
	doi = {10.1051/0004-6361/202348478},
	urldate = {2024-10-18},
	journal = {\aap},
	author = {Hawcroft, C. and Mahy, L. and Sana, H. and Sundqvist, J. O. and Abdul-Masih, M. and Brands, S. A. and Decin, L. and de Koter, A. and Puls, J.},
	month = oct,
	year = {2024},
	note = {Publisher: EDP
ADS Bibcode: 2024A\&A...690A.126H},
	pages = {A126},
}

@article{vink_nature_1999,
	title = {On the nature of the bi-stability jump in the winds of early-type supergiants},
	volume = {350},
	issn = {0004-6361},
	url = {https://ui.adsabs.harvard.edu/abs/1999A&A...350..181V},
	doi = {10.48550/arXiv.astro-ph/9908196},
	urldate = {2023-06-15},
	journal = {\aap},
	author = {Vink, J. S. and de Koter, A. and Lamers, H. J. G. L. M.},
	month = oct,
	year = {1999},
	note = {ADS Bibcode: 1999A\&A...350..181V},
	pages = {181--196},
}

@article{moens_first_2022,
	title = {First {3D} radiation-hydrodynamic simulations of {Wolf}-{Rayet} winds},
	volume = {665},
	issn = {0004-6361},
	url = {https://ui.adsabs.harvard.edu/abs/2022A&A...665A..42M},
	doi = {10.1051/0004-6361/202243451},
	urldate = {2023-06-18},
	journal = {\aap},
	author = {Moens, N. and Poniatowski, L. G. and Hennicker, L. and Sundqvist, J. O. and El Mellah, I. and Kee, N. D.},
	month = sep,
	year = {2022},
	note = {ADS Bibcode: 2022A\&A...665A..42M},
	pages = {A42},
}

@article{steiger_notes_2016,
	title = {Notes on the {Steiger}–{Lind} (1980) {Handout}},
	volume = {23},
	issn = {1070-5511},
	url = {https://doi.org/10.1080/10705511.2016.1217487},
	doi = {10.1080/10705511.2016.1217487},
	number = {6},
	urldate = {2023-08-18},
	journal = {Structural Equation Modeling: A Multidisciplinary Journal},
	author = {Steiger, James H.},
	month = nov,
	year = {2016},
	note = {Publisher: Routledge
\_eprint: https://doi.org/10.1080/10705511.2016.1217487},
	pages = {777--781},
}

@article{vink_x-shooting_2023,
	title = {X-{Shooting} {ULLYSES}: {Massive} stars at low metallicity - {I}. {Project} description},
	volume = {675},
	copyright = {© The Authors 2023},
	issn = {0004-6361, 1432-0746},
	shorttitle = {X-{Shooting} {ULLYSES}},
	url = {https://www.aanda.org/articles/aa/abs/2023/07/aa45650-22/aa45650-22.html},
	doi = {10.1051/0004-6361/202245650},
	language = {en},
	urldate = {2023-08-14},
	journal = {\aap},
	author = {Vink, Jorick S. and Mehner, A. and Crowther, P. A. and Fullerton, A. and Garcia, M. and Martins, F. and Morrell, N. and Oskinova, L. M. and St-Louis, N. and ud-Doula, A. and Sander, A. a. C. and Sana, H. and Bouret, J.-C. and Kubátová, B. and Marchant, P. and Martins, L. P. and Wofford, A. and Loon, J. Th van and Telford, O. Grace and Götberg, Y. and Bowman, D. M. and Erba, C. and Kalari, V. M. and Abdul-Masih, M. and Alkousa, T. and Backs, F. and Barbosa, C. L. and Berlanas, S. R. and Bernini-Peron, M. and Bestenlehner, J. M. and Blomme, R. and Bodensteiner, J. and Brands, S. A. and Evans, C. J. and David-Uraz, A. and Driessen, F. A. and Dsilva, K. and Geen, S. and Gómez-González, V. M. A. and Grassitelli, L. and Hamann, W.-R. and Hawcroft, C. and Herrero, A. and Higgins, E. R. and Hillier, D. John and Ignace, R. and Istrate, A. G. and Kaper, L. and Kee, N. D. and Kehrig, C. and Keszthelyi, Z. and Klencki, J. and Koter, A. de and Kuiper, R. and Laplace, E. and Larkin, C. J. K. and Lefever, R. R. and Leitherer, C. and Lennon, D. J. and Mahy, L. and Apellániz, J. Maíz and Maravelias, G. and Marcolino, W. and McLeod, A. F. and Mink, S. E. de and Najarro, F. and Oey, M. S. and Parsons, T. N. and Pauli, D. and Pedersen, M. G. and Prinja, R. K. and Ramachandran, V. and Ramírez-Tannus, M. C. and Sabhahit, G. N. and Schootemeijer, A. and Serantes, S. Reyero and Shenar, T. and Stringfellow, G. S. and Sudnik, N. and Tramper, F. and Wang, L.},
	month = jul,
	year = {2023},
	note = {Publisher: EDP Sciences},
	pages = {A154},
}

@article{sundqvist_atmospheric_2018,
	title = {Atmospheric {NLTE} models for the spectroscopic analysis of blue stars with winds. {IV}. {Porosity} in physical and velocity space},
	volume = {619},
	issn = {0004-6361},
	url = {https://ui.adsabs.harvard.edu/abs/2018A&A...619A..59S},
	doi = {10.1051/0004-6361/201832993},
	urldate = {2024-10-18},
	journal = {\aap},
	author = {Sundqvist, J. O. and Puls, J.},
	month = nov,
	year = {2018},
	note = {ADS Bibcode: 2018A\&A...619A..59S},
	pages = {A59},
}

@ARTICLE{rivero-gonzalez2012,
       author = {{Rivero Gonz{\'a}lez}, J.~G. and {Puls}, J. and {Najarro}, F. and {Brott}, I.},
        title = "{Nitrogen line spectroscopy of O-stars. II. Surface nitrogen abundances for O-stars in the Large Magellanic Cloud}",
      journal = {\aap},
         year = 2012,
        month = jan,
       volume = {537},
          eid = {A79},
        pages = {A79},
          doi = {10.1051/0004-6361/201117790},
archivePrefix = {arXiv},
       eprint = {1110.5148},
 primaryClass = {astro-ph.SR},
       adsurl = {https://ui.adsabs.harvard.edu/abs/2012A&A...537A..79R}
}

@ARTICLE{asplund_2009,
       author = {{Asplund}, Martin and {Grevesse}, Nicolas and {Sauval}, A. Jacques and {Scott}, Pat},
        title = "{The Chemical Composition of the Sun}",
      journal = {\araa},
         year = 2009,
        month = sep,
       volume = {47},
       number = {1},
        pages = {481-522},
          doi = {10.1146/annurev.astro.46.060407.145222},
archivePrefix = {arXiv},
       eprint = {0909.0948},
 primaryClass = {astro-ph.SR},
       adsurl = {https://ui.adsabs.harvard.edu/abs/2009ARA&A..47..481A}
}

@BOOK{sobolev_1960,
       author = {{Sobolev}, V.~V.},
        title = "{Moving Envelopes of Stars}",
         year = 1960,
          doi = {10.4159/harvard.9780674864658},
       adsurl = {https://ui.adsabs.harvard.edu/abs/1960mes..book.....S},
    publisher = {HARVARD UNIVERSITY PRESS} 
}

@ARTICLE{driessen_2022,
       author = {{Driessen}, F.~A. and {Sundqvist}, J.~O. and {Dagore}, A.},
        title = "{Theoretical wind clumping predictions from 2D LDI models of O-star winds at different metallicities}",
      journal = {\aap},
         year = 2022,
        month = jul,
       volume = {663},
          eid = {A40},
        pages = {A40},
          doi = {10.1051/0004-6361/202142844},
archivePrefix = {arXiv},
       eprint = {2204.05670},
 primaryClass = {astro-ph.SR},
       adsurl = {https://ui.adsabs.harvard.edu/abs/2022A&A...663A..40D}
}

@ARTICLE{puls_2006,
       author = {{Puls}, J. and {Markova}, N. and {Scuderi}, S. and {Stanghellini}, C. and {Taranova}, O.~G. and {Burnley}, A.~W. and {Howarth}, I.~D.},
        title = "{Bright OB stars in the Galaxy. III. Constraints on the radial stratification of the clumping factor in hot star winds from a combined H$_{{\ensuremath{\alpha}}}$, IR and radio analysis}",
      journal = {\aap},
         year = 2006,
        month = aug,
       volume = {454},
       number = {2},
        pages = {625-651},
          doi = {10.1051/0004-6361:20065073},
archivePrefix = {arXiv},
       eprint = {astro-ph/0604372},
 primaryClass = {astro-ph},
       adsurl = {https://ui.adsabs.harvard.edu/abs/2006A&A...454..625P}
}

@article{sundqvist_rotation_2013,
	title = {The rotation rates of massive stars. {How} slow are the slow ones?},
	volume = {559},
	issn = {0004-6361},
	url = {https://ui.adsabs.harvard.edu/abs/2013A&A...559L..10S},
	doi = {10.1051/0004-6361/201322761},
	urldate = {2024-09-13},
	journal = {\aap},
	author = {Sundqvist, J. O. and Simón-Díaz, S. and Puls, J. and Markova, N.},
	month = nov,
	year = {2013},
	note = {ADS Bibcode: 2013A\&A...559L..10S},
	pages = {L10},
}

@article{krticka_new_2024,
	title = {New mass-loss rates of {Magellanic} {Cloud} {B} supergiants from global wind models},
	volume = {681},
	issn = {0004-6361},
	url = {https://ui.adsabs.harvard.edu/abs/2024A&A...681A..29K},
	doi = {10.1051/0004-6361/202347916},
	urldate = {2024-02-05},
	journal = {\aap},
	author = {Krtička, J. and Kubát, J. and Krtičková, I.},
	month = jan,
	year = {2024},
	note = {ADS Bibcode: 2024A\&A...681A..29K},
	pages = {A29},
}

@article{carneiro_carbon_2018,
	title = {Carbon line formation and spectroscopy in {O}-type stars},
	volume = {615},
	issn = {0004-6361},
	url = {https://ui.adsabs.harvard.edu/abs/2018A&A...615A...4C},
	doi = {10.1051/0004-6361/201731839},
	urldate = {2024-02-07},
	journal = {\aap},
	author = {Carneiro, L. P. and Puls, J. and Hoffmann, T. L.},
	month = jul,
	year = {2018},
	note = {ADS Bibcode: 2018A\&A...615A...4C},
	pages = {A4},
}

@article{debnath_2d_2024,
	title = {{2D} unified atmosphere and wind simulations of {O}-type stars},
	volume = {684},
	issn = {0004-6361},
	url = {https://ui.adsabs.harvard.edu/abs/2024A&A...684A.177D},
	doi = {10.1051/0004-6361/202348206},
	urldate = {2024-05-09},
	journal = {\aap},
	author = {Debnath, D. and Sundqvist, J. O. and Moens, N. and Van der Sijpt, C. and Verhamme, O. and Poniatowski, L. G.},
	month = apr,
	year = {2024},
	note = {ADS Bibcode: 2024A\&A...684A.177D},
	pages = {A177},
}

@article{sana_x-shooting_2024,
	title = {X-{Shooting} {ULLYSES}: {Massive} stars at low metallicity. {II}. {DR1}: {Advanced} optical data products for the {Magellanic} {Clouds}},
	volume = {688},
	issn = {0004-6361},
	shorttitle = {X-{Shooting} {ULLYSES}},
	url = {https://ui.adsabs.harvard.edu/abs/2024A&A...688A.104S},
	doi = {10.1051/0004-6361/202347479},
	urldate = {2024-09-19},
	journal = {\aap},
	author = {Sana, H. and Tramper, F. and Abdul-Masih, M. and Blomme, R. and Dsilva, K. and Maravelias, G. and Martins, L. and Mehner, A. and Wofford, A. and Banyard, G. and Barbosa, C. L. and Bestenlehner, J. and Hawcroft, C. and John Hillier, D. and Todt, H. and Larkin, C. J. K. and Mahy, L. and Najarro, F. and Ramachandran, V. and Ramírez-Tannus, M. C. and Rubio-Díez, M. M. and Sander, A. A. C. and Shenar, T. and Vink, J. S. and Backs, F. and Brands, S. A. and Crowther, P. and Decin, L. and de Koter, A. and Hamann, W. -R. and Kehrig, C. and Kuiper, R. and Oskinova, L. and Pauli, D. and Sundqvist, J. and Verhamme, O. and {Xshoot-U Collaboration}},
	month = aug,
	year = {2024},
	note = {ADS Bibcode: 2024A\&A...688A.104S},
	pages = {A104},
}

@article{boggs_accurate_1987,
	title = {Accurate determination of molecular structure and vibrational force constants by computation},
	volume = {14},
	issn = {0342-1791},
	url = {https://ui.adsabs.harvard.edu/abs/1987PCM....14..407B},
	doi = {10.1007/BF00628817},
	urldate = {2025-04-23},
	journal = {Physics and Chemistry of Minerals},
	author = {Boggs, James E.},
	month = sep,
	year = {1987},
	note = {ADS Bibcode: 1987PCM....14..407B},
	pages = {407--412},
}

@article{gayley_improved_1995,
	title = {An {Improved} {Line}-{Strength} {Parameterization} in {Hot}-{Star} {Winds}},
	volume = {454},
	issn = {0004-637X},
	url = {https://ui.adsabs.harvard.edu/abs/1995ApJ...454..410G},
	doi = {10.1086/176492},
	urldate = {2024-05-22},
	journal = {\apj},
	author = {Gayley, Kenneth G.},
	month = nov,
	year = {1995},
	note = {Publisher: IOP
ADS Bibcode: 1995ApJ...454..410G},
	pages = {410},
}

@article{sundqvist_--fly_2025,
	title = {An on-the-fly line-driven wind iterative mass-loss estimator ({LIME}) for hot, massive stars of arbitrary chemical compositions},
	volume = {703},
	issn = {0004-6361},
	url = {https://ui.adsabs.harvard.edu/abs/2025A&A...703A.284S},
	doi = {10.1051/0004-6361/202554940},
	urldate = {2026-01-30},
	journal = {\aap},
	publisher = {EDP},
	author = {Sundqvist, J. O. and Debnath, D. and Backs, F. and Verhamme, O. and Moens, N. and Delbroek, L. and Dickson, D. and Schillemans, P. and Van der Sijpt, C. and Dirickx, M.},
	month = nov,
	year = {2025},
	note = {ADS Bibcode: 2025A\&A...703A.284S},
	pages = {A284},
}

@article{bernini-peron_x-shooting_2024,
	title = {X-{Shooting} {ULLYSES}: {Massive} stars at low metallicity: {VII}. {Stellar} and wind properties of {B} supergiants in the {Small} {Magellanic} {Cloud}},
	volume = {692},
	issn = {0004-6361},
	shorttitle = {X-{Shooting} {ULLYSES}},
	url = {https://ui.adsabs.harvard.edu/abs/2024A&A...692A..89B},
	doi = {10.1051/0004-6361/202450475},
	urldate = {2026-01-30},
	journal = {\aap},
	publisher = {EDP},
	author = {Bernini-Peron, M. and Sander, A. A. C. and Ramachandran, V. and Oskinova, L. M. and Vink, J. S. and Verhamme, O. and Najarro, F. and Josiek, J. and Brands, S. A. and Crowther, P. A. and Gómez-González, V. M. A. and Gormaz-Matamala, A. C. and Hawcroft, C. and Kuiper, R. and Mahy, L. and Marcolino, W. L. F. and Martins, L. P. and Mehner, A. and Parsons, T. N. and Pauli, D. and Shenar, T. and Schootemeijer, A. and Todt, H. and van Loon, J. Th. and {XShootU Collaboration}},
	month = dec,
	year = {2024},
	note = {ADS Bibcode: 2024A\&A...692A..89B},
	pages = {A89},
}

@article{bestenlehner_x-shooting_2025,
	title = {X-{Shooting} {ULLYSES}: {Massive} stars at low metallicity: {XI}. {Pipeline}-determined physical properties of {Magellanic} {Cloud} {OB} stars},
	volume = {695},
	issn = {0004-6361},
	shorttitle = {X-{Shooting} {ULLYSES}},
	url = {https://ui.adsabs.harvard.edu/abs/2025A&A...695A.198B},
	doi = {10.1051/0004-6361/202452491},
	urldate = {2025-04-23},
	journal = {\aap},
	author = {Bestenlehner, J. M. and Crowther, P. A. and Hawcroft, C. and Sana, H. and Tramper, F. and Vink, J. S. and Brands, S. A. and Sander, A. A. C. and {XShootU Collaboration}},
	month = mar,
	year = {2025},
	note = {Publisher: EDP
ADS Bibcode: 2025A\&A...695A.198B},
	pages = {A198},
}

@article{gormaz-matamala_evolution_2024,
	title = {Evolution of rotating massive stars adopting a newer, self-consistent wind prescription at {Small} {Magellanic} {Cloud} metallicity},
	volume = {687},
	issn = {0004-6361},
	url = {https://ui.adsabs.harvard.edu/abs/2024A&A...687A.290G},
	doi = {10.1051/0004-6361/202449782},
	urldate = {2025-05-22},
	journal = {\aap},
	author = {Gormaz-Matamala, A. C. and Cuadra, J. and Ekström, S. and Meynet, G. and Curé, M. and Belczynski, K.},
	month = jul,
	year = {2024},
	note = {Publisher: EDP
ADS Bibcode: 2024A\&A...687A.290G},
	pages = {A290},
}

@article{josiek_impact_2024,
	title = {Impact of main sequence mass loss on the appearance, structure, and evolution of {Wolf}-{Rayet} stars},
	volume = {688},
	issn = {0004-6361},
	url = {https://ui.adsabs.harvard.edu/abs/2024A&A...688A..71J},
	doi = {10.1051/0004-6361/202449281},
	urldate = {2024-10-18},
	journal = {\aap},
	author = {Josiek, J. and Ekström, S. and Sander, A. A. C.},
	month = aug,
	year = {2024},
	note = {ADS Bibcode: 2024A\&A...688A..71J},
	pages = {A71},
}

@article{alkousa_x-shooting_2025,
	title = {X-{Shooting} {ULLYSES}: {Massive} stars at low metallicity: {XIII}. {Testing} the bi-stability jump in the {Large} {Magellanic} {Cloud}},
	volume = {699},
	issn = {0004-6361},
	shorttitle = {X-{Shooting} {ULLYSES}},
	url = {https://ui.adsabs.harvard.edu/abs/2025A&A...699A.314A},
	doi = {10.1051/0004-6361/202553799},
	urldate = {2025-11-05},
	journal = {\aap},
	author = {Alkousa, T. and Crowther, P. A. and Bestenlehner, J. M. and Sana, H. and Tramper, F. and Vink, J. S. and Pauli, D. and van Loon, J. Th. and Najarro, F. and Kuiper, R. and Sander, A. A. C. and Bernini-Peron, M.},
	month = jul,
	year = {2025},
	note = {Publisher: EDP
ADS Bibcode: 2025A\&A...699A.314A},
	pages = {A314},
}

@article{pauli_new_2025,
	title = {New empirical mass-loss recipe for {UV} radiation line-driven winds of hot stars across various metallicities},
	volume = {697},
	issn = {0004-6361},
	url = {https://ui.adsabs.harvard.edu/abs/2025A&A...697A.114P},
	doi = {10.1051/0004-6361/202553910},
	urldate = {2025-11-13},
	journal = {\aap},
	author = {Pauli, D. and Oskinova, L. M. and Hamann, W.-R. and Sander, A. A. C. and Vink, J. S. and Bernini-Peron, M. and Josiek, J. and Lefever, R. R. and Sana, H. and Ramachandran, V.},
	month = may,
	year = {2025},
	note = {Publisher: EDP
ADS Bibcode: 2025A\&A...697A.114P},
	pages = {A114},
}

@article{zsargo_importance_2008,
	title = {On the {Importance} of the {Interclump} {Medium} for {Superionization}: {O} {VI} {Formation} in the {Wind} of ζ {Puppis}},
	volume = {685},
	issn = {0004-637X},
	shorttitle = {On the {Importance} of the {Interclump} {Medium} for {Superionization}},
	url = {https://ui.adsabs.harvard.edu/abs/2008ApJ...685L.149Z},
	doi = {10.1086/592568},
	urldate = {2025-11-05},
	journal = {\apj},
	author = {Zsargó, J. and Hillier, D. J. and Bouret, J.-C. and Lanz, T. and Leutenegger, M. A. and Cohen, D. H.},
	month = oct,
	year = {2008},
	note = {Publisher: IOP
ADS Bibcode: 2008ApJ...685L.149Z},
	pages = {L149},
}

@article{puebla_x-ray_2016,
	title = {X-ray, {UV} and optical analysis of supergiants: ɛ {Ori}},
	volume = {456},
	issn = {0035-8711},
	shorttitle = {X-ray, {UV} and optical analysis of supergiants},
	url = {https://ui.adsabs.harvard.edu/abs/2016MNRAS.456.2907P},
	doi = {10.1093/mnras/stv2783},
	urldate = {2025-11-05},
	journal = {\mnras},
	author = {Puebla, Raul E. and Hillier, D. John and Zsargó, Janos and Cohen, David H. and Leutenegger, Maurice A.},
	month = mar,
	year = {2016},
	note = {Publisher: OUP
ADS Bibcode: 2016MNRAS.456.2907P},
	pages = {2907--2936},
}

@article{pauldrach_radiation-driven_1986,
	title = {Radiation-driven winds of hot luminous stars. {Improvements} of the theory and first results.},
	volume = {164},
	issn = {0004-6361},
	url = {https://ui.adsabs.harvard.edu/abs/1986A&A...164...86P},
	urldate = {2025-11-21},
	journal = {\aap},
	author = {Pauldrach, A. and Puls, J. and Kudritzki, R. P.},
	month = aug,
	year = {1986},
	note = {Publisher: EDP
ADS Bibcode: 1986A\&A...164...86P},
	pages = {86--100},
}

@article{de_burgos_iacob_2024_survey,
	title = {The {IACOB} project. {X}. {Large}-scale quantitative spectroscopic analysis of {Galactic} luminous blue stars},
	volume = {687},
	issn = {0004-6361},
	url = {https://ui.adsabs.harvard.edu/abs/2024A&A...687A.228D},
	doi = {10.1051/0004-6361/202348808},
	urldate = {2025-11-14},
	journal = {\aap},
	author = {de Burgos, A. and Simón-Díaz, S. and Urbaneja, M. A. and Puls, J.},
	month = jul,
	year = {2024},
	note = {Publisher: EDP
ADS Bibcode: 2024A\&A...687A.228D},
	pages = {A228},
}

@article{crowther_physical_2006,
	title = {Physical parameters and wind properties of galactic early {B} supergiants},
	volume = {446},
	issn = {0004-6361},
	url = {https://ui.adsabs.harvard.edu/abs/2006A&A...446..279C},
	doi = {10.1051/0004-6361:20053685},
	urldate = {2024-01-21},
	journal = {\aap},
	author = {Crowther, P. A. and Lennon, D. J. and Walborn, N. R.},
	month = jan,
	year = {2006},
	note = {ADS Bibcode: 2006A\&A...446..279C},
	pages = {279--293},
}

@article{vink_metallicity-dependent_2021,
	title = {Metallicity-dependent wind parameter predictions for {OB} stars},
	volume = {504},
	issn = {0035-8711},
	url = {https://ui.adsabs.harvard.edu/abs/2021MNRAS.504.2051V},
	doi = {10.1093/mnras/stab902},
	urldate = {2024-05-21},
	journal = {\mnras},
	author = {Vink, Jorick S. and Sander, Andreas A. C.},
	month = jun,
	year = {2021},
	note = {Publisher: OUP
ADS Bibcode: 2021MNRAS.504.2051V},
	pages = {2051--2061},
}

@article{sander_x-shooting_2024,
	title = {X-{Shooting} {ULLYSES}: {Massive} stars at low metallicity: {IV}. {Spectral} analysis methods and exemplary results for {O} stars},
	volume = {689},
	issn = {0004-6361},
	shorttitle = {X-{Shooting} {ULLYSES}},
	url = {https://ui.adsabs.harvard.edu/abs/2024A&A...689A..30S},
	doi = {10.1051/0004-6361/202449829},
	urldate = {2024-09-20},
	journal = {\aap},
	author = {Sander, A. A. C. and Bouret, J. -C. and Bernini-Peron, M. and Puls, J. and Backs, F. and Berlanas, S. R. and Bestenlehner, J. M. and Brands, S. A. and Herrero, A. and Martins, F. and Maryeva, O. and Pauli, D. and Ramachandran, V. and Crowther, P. A. and Gómez-González, V. M. A. and Gormaz-Matamala, A. C. and Hamann, W. -R. and Hillier, D. J. and Kuiper, R. and Larkin, C. J. K. and Lefever, R. R. and Mehner, A. and Najarro, F. and Oskinova, L. M. and Schösser, E. C. and Shenar, T. and Todt, H. and ud-Doula, A. and Vink, J. S.},
	month = sep,
	year = {2024},
	note = {ADS Bibcode: 2024A\&A...689A..30S},
	pages = {A30},
}

@article{de_burgos_iacob_2024_mass_loss,
	title = {The {IACOB} project. {XI}. {No} increase in mass-loss rates over the bistability region},
	volume = {687},
	issn = {0004-6361},
	url = {https://ui.adsabs.harvard.edu/abs/2024A&A...687L..16D},
	doi = {10.1051/0004-6361/202450301},
	urldate = {2024-10-18},
	journal = {\aap},
	author = {de Burgos, A. and Keszthelyi, Z. and Simón-Díaz, S. and Urbaneja, M. A.},
	month = jul,
	year = {2024},
	note = {ADS Bibcode: 2024A\&A...687L..16D},
	pages = {L16},
}

@article{backs_x-shooting_2024,
	title = {X-{Shooting} {ULLYSES}: {Massive} stars at low metallicity: {VI}. {Atmosphere} and mass-loss properties of {O}-type giants in the {Small} {Magellanic} {Cloud}},
	volume = {692},
	issn = {0004-6361},
	shorttitle = {X-{Shooting} {ULLYSES}},
	url = {https://ui.adsabs.harvard.edu/abs/2024A&A...692A..88B},
	doi = {10.1051/0004-6361/202451893},
	urldate = {2025-04-16},
	journal = {\aap},
	author = {Backs, F. and Brands, S. A. and de Koter, A. and Kaper, L. and Vink, J. S. and Puls, J. and Sundqvist, J. and Tramper, F. and Sana, H. and Bernini-Peron, M. and Bestenlehner, J. M. and Crowther, P. A. and Hawcroft, C. and Ignace, R. and Kuiper, R. and van Loon, J. Th. and Mahy, L. and Marcolino, W. and Najarro, F. and Oskinova, L. M. and Pauli, D. and Ramachandran, V. and Sander, A. A. C. and Verhamme, O.},
	month = dec,
	year = {2024},
	pages = {A88},
}

@article{brands_x-shooting_2025,
	title = {X-{Shooting} {ULLYSES}: {Massive} stars at low metallicity: {XII}. {Clumped} winds of {O}-type (super)giants in the {Large} {Magellanic} {Cloud}},
	volume = {697},
	issn = {0004-6361},
	shorttitle = {X-{Shooting} {ULLYSES}},
	url = {https://ui.adsabs.harvard.edu/abs/2025A&A...697A..54B},
	doi = {10.1051/0004-6361/202452784},
	urldate = {2025-07-16},
	journal = {\aap},
	author = {Brands, Sarah A. and Backs, Frank and de Koter, Alex and Puls, Joachim and Crowther, Paul A. and Sana, Hugues and Tramper, Frank and Kaper, Lex and Sundqvist, Jon O. and Bestenlehner, Joachim M. and Driessen, Florian A. and Erba, Christiana and Hawcroft, Calum and Herrero, Artemio and John Hillier, D. and Ignace, Richard and Lefever, Roel R. and Dylan Kee, N. and Kubátová, Brankica and Mahy, Laurent and Moffat, Anthony F. J. and Najarro, Francisco and Prinja, Raman K. and Ramachandran, Varsha and Sander, Andreas A. C. and Vink, Jorick S. and {XShootU Collaboration}},
	month = may,
	year = {2025},
	note = {Publisher: EDP
ADS Bibcode: 2025A\&A...697A..54B},
	pages = {A54},
}

@article{verhamme_x-shooting_2024,
	title = {X-{Shooting} {ULLYSES}: {Massive} {Stars} at low metallicity: {IX}. {Empirical} constraints on mass-loss rates and clumping parameters for {OB} supergiants in the {Large} {Magellanic} {Cloud}},
	volume = {692},
	issn = {0004-6361},
	shorttitle = {X-{Shooting} {ULLYSES}},
	url = {https://ui.adsabs.harvard.edu/abs/2024A%26A...692A..91V/abstract},
	doi = {10.1051/0004-6361/202451169},
	language = {en},
	urldate = {2024-12-21},
	journal = {Astronomy \&amp; Astrophysics, Volume 692, id.A91, 24 pp.},
	author = {Verhamme, O. and Sundqvist, J. and de Koter, A. and Sana, H. and Backs, F. and Brands, S. A. and Najarro, F. and Puls, J. and Vink, J. S. and Crowther, P. A. and Kubátová, B. and Sander, A. a. C. and Bernini-Peron, M. and Kuiper, R. and Prinja, R. K. and Schillemans, P. and Shenar, T. and van Loon, J. Th and Collaboration, XShootu},
	month = dec,
	year = {2024},
	pages = {A91},
}

@incollection{benesty_pearson_2009,
	address = {Berlin, Heidelberg},
	title = {Pearson {Correlation} {Coefficient}},
	isbn = {978-3-642-00296-0},
	url = {https://doi.org/10.1007/978-3-642-00296-0_5},
	doi = {10.1007/978-3-642-00296-0_5},
	language = {en},
	urldate = {2026-02-02},
	booktitle = {Noise {Reduction} in {Speech} {Processing}},
	publisher = {Springer},
	author = {Benesty, Jacob and Chen, Jingdong and Huang, Yiteng and Cohen, Israel},
	editor = {Cohen, Israel and Huang, Yiteng and Chen, Jingdong and Benesty, Jacob},
	year = {2009},
	pages = {1--4},
}

@article{kudritzki_winds_2000,
	title = {Winds from {Hot} {Stars}},
	volume = {38},
	issn = {0066-4146},
	url = {https://ui.adsabs.harvard.edu/abs/2000ARA&A..38..613K},
	doi = {10.1146/annurev.astro.38.1.613},
	urldate = {2025-07-30},
	journal = {\araa},
	author = {Kudritzki, Rolf-Peter and Puls, Joachim},
	month = jan,
	year = {2000},
	note = {ADS Bibcode: 2000ARA\&A..38..613K},
	pages = {613--666},
}

@article{cure_influence_2004,
	title = {The {Influence} of {Rotation} in {Radiation}-driven {Wind} from {Hot} {Stars}: {New} {Solutions} and {Disk} {Formation} in {Be} {Stars}},
	volume = {614},
	issn = {0004-637X},
	shorttitle = {The {Influence} of {Rotation} in {Radiation}-driven {Wind} from {Hot} {Stars}},
	url = {https://ui.adsabs.harvard.edu/abs/2004ApJ...614..929C},
	doi = {10.1086/423776},
	urldate = {2025-06-25},
	journal = {\apj},
	author = {Curé, Michel},
	month = oct,
	year = {2004},
	note = {Publisher: IOP
ADS Bibcode: 2004ApJ...614..929C},
	pages = {929--941},
}

@article{cure_slow_2011,
	title = {{SLOW} {RADIATION}-{DRIVEN} {WIND} {SOLUTIONS} {OF} {A}-{TYPE} {SUPERGIANTS}},
	volume = {737},
	issn = {0004-637X},
	url = {https://dx.doi.org/10.1088/0004-637X/737/1/18},
	doi = {10.1088/0004-637X/737/1/18},
	language = {en},
	number = {1},
	urldate = {2025-06-25},
	journal = {\apj},
	author = {Curé, M. and Cidale, L. and Granada, A.},
	month = jul,
	year = {2011},
	note = {Publisher: The American Astronomical Society},
	pages = {18},
}

@article{Schramm_1977_abundance,
   author = "Schramm, D N and Wagoner, R V",
   title = "Element Production in the Early Universe", 
   journal= "Annual Review of Nuclear and Particle Science",
   year = "1977",
   volume = "27",
   number = "Volume 27, 1977",
   pages = "37-74",
   doi = "https://doi.org/10.1146/annurev.ns.27.120177.000345",
   url = "https://www.annualreviews.org/content/journals/10.1146/annurev.ns.27.120177.000345",
   publisher = "Annual Reviews",
   issn = "1545-4134",
   type = "Journal Article",
  }

@article{urbaneja_metallicity_2023,
	title = {The {Metallicity} and {Distance} of {Leo} {A} from {Blue} {Supergiants}},
	volume = {959},
	issn = {0004-637X},
	url = {https://ui.adsabs.harvard.edu/abs/2023ApJ...959...52U},
	doi = {10.3847/1538-4357/acfc3d},
	urldate = {2025-07-22},
	journal = {\apj},
	author = {Urbaneja, Miguel A. and Bresolin, Fabio and Kudritzki, Rolf-Peter},
	month = dec,
	year = {2023},
	note = {Publisher: IOP
ADS Bibcode: 2023ApJ...959...52U},
	pages = {52},
}

@article{urbaneja_lmc_2017,
	title = {{LMC} {Blue} {Supergiant} {Stars} and the {Calibration} of the {Flux}-weighted {Gravity}-{Luminosity} {Relationship}},
	volume = {154},
	issn = {0004-6256},
	url = {https://ui.adsabs.harvard.edu/abs/2017AJ....154..102U},
	doi = {10.3847/1538-3881/aa79a8},
	urldate = {2025-07-22},
	journal = {The Astronomical Journal},
	author = {Urbaneja, M. A. and Kudritzki, R. -P. and Gieren, W. and Pietrzyński, G. and Bresolin, F. and Przybilla, N.},
	month = sep,
	year = {2017},
	note = {Publisher: IOP
ADS Bibcode: 2017AJ....154..102U},
	pages = {102},
}

@article{przybilla_cosmic_2008,
	title = {A {Cosmic} {Abundance} {Standard}: {Chemical} {Homogeneity} of the {Solar} {Neighborhood} and the {ISM} {Dust}-{Phase} {Composition}},
	volume = {688},
	issn = {0004-637X},
	shorttitle = {A {Cosmic} {Abundance} {Standard}},
	url = {https://ui.adsabs.harvard.edu/abs/2008ApJ...688L.103P},
	doi = {10.1086/595618},
	urldate = {2025-07-22},
	journal = {\apj},
	author = {Przybilla, Norbert and Nieva, Maria-Fernanda and Butler, Keith},
	month = dec,
	year = {2008},
	note = {Publisher: IOP
ADS Bibcode: 2008ApJ...688L.103P},
	pages = {L103},
}

@article{przybilla_quantitative_2006,
	title = {Quantitative spectroscopy of {BA}-type supergiants},
	volume = {445},
	issn = {0004-6361},
	url = {https://ui.adsabs.harvard.edu/abs/2006A&A...445.1099P},
	doi = {10.1051/0004-6361:20053832},
	urldate = {2025-07-22},
	journal = {\aap},
	author = {Przybilla, N. and Butler, K. and Becker, S. R. and Kudritzki, R. P.},
	month = jan,
	year = {2006},
	note = {Publisher: EDP
ADS Bibcode: 2006A\&A...445.1099P},
	pages = {1099--1126},
}

@article{roman-duval_uv_2025,
	title = {The {UV} {Legacy} {Library} of {Young} {Stars} as {Essential} {Standards} ({ULLYSES}) {Large} {Director}'s {Discretionary} {Program} with {Hubble}. {I}. {Goals}, {Design}, and {Initial} {Results}},
	volume = {985},
	issn = {0004-637X},
	url = {https://ui.adsabs.harvard.edu/abs/2025ApJ...985..109R},
	doi = {10.3847/1538-4357/adc45b},
	urldate = {2025-07-22},
	journal = {\apj},
	author = {Roman-Duval, Julia and Fischer, William J. and Fullerton, Alexander W. and Taylor, Jo and Plesha, Rachel and Proffitt, Charles and Monroe, TalaWanda and Fischer, Travis C. and Aloisi, Alessandra and Bouret, Jean-Claude and Britt, Christopher and Calvet, Nuria and Carlberg, Joleen K. and Crowther, Paul A. and De Rosa, Gisella and Dixon, William V. and Espaillat, Catherine C. and Evans, Christopher J. and Fox, Andrew J. and France, Kevin and Garcia, Miriam and Fleming, Scott W. and Frazer, Elaine M. and Gómez de Castro, Ana I. and Herczeg, Gregory J. and Hernandez, Svea and Hirschauer, Alec S. and James, Bethan L. and Johns-Krull, Christopher M. and Leitherer, Claus and Lockwood, Sean and Najita, Joan and Oey, M. S. and Oliveira, Cristina and Pauly, Tyler and Reid, I. Neill and Riedel, Adric and Rodriguez, David R. and Sahnow, David and Sankrit, Ravi and Sembach, Kenneth R. and Shaw, Richard and Smith, Linda J. and Sohn, S. Tony and Som, Debopam and Úbeda, Leonardo and Welty, Daniel E.},
	month = may,
	year = {2025},
	note = {Publisher: IOP
ADS Bibcode: 2025ApJ...985..109R},
	pages = {109},
}

@article{herrero_mass_1992,
	title = {The mass and helium discrepancy in massive young stars},
	volume = {401},
	url = {https://ui.adsabs.harvard.edu/abs/1992LNP...401...21H},
	urldate = {2025-04-02},
	journal = {The Atmospheres of Early-Type Stars},
	author = {Herrero, A. and Kudritzki, R. P. and Vilchez, J. M. and Kunze, D. and Butler, K. and Haser, S.},
	month = jan,
	year = {1992},
	doi = {10.1007/3-540-55256-1_269},
	note = {ADS Bibcode: 1992LNP...401...21H},
	pages = {21},
}

@article{tramper_properties_2014,
	title = {The properties of ten {O}-type stars in the low-metallicity galaxies {IC} 1613, {WLM}, and {NGC} 3109},
	volume = {572},
	issn = {0004-6361},
	url = {https://ui.adsabs.harvard.edu/abs/2014A&A...572A..36T},
	doi = {10.1051/0004-6361/201424312},
	urldate = {2025-09-04},
	journal = {\aap},
	author = {Tramper, F. and Sana, H. and de Koter, A. and Kaper, L. and Ramírez-Agudelo, O. H.},
	month = dec,
	year = {2014},
	note = {ADS Bibcode: 2014A\&A...572A..36T},
	pages = {A36},
}

@article{abdul-masih_clues_2019,
	title = {Clues on the {Origin} and {Evolution} of {Massive} {Contact} {Binaries}: {Atmosphere} {Analysis} of {VFTS} 352},
	volume = {880},
	issn = {0004-637X},
	shorttitle = {Clues on the {Origin} and {Evolution} of {Massive} {Contact} {Binaries}},
	url = {https://ui.adsabs.harvard.edu/abs/2019ApJ...880..115A},
	doi = {10.3847/1538-4357/ab24d4},
	urldate = {2025-09-04},
	journal = {\apj},
	author = {Abdul-Masih, Michael and Sana, Hugues and Sundqvist, Jon and Mahy, Laurent and Menon, Athira and Almeida, Leonardo A. and De Koter, Alex and de Mink, Selma E. and Justham, Stephen and Langer, Norbert and Puls, Joachim and Shenar, Tomer and Tramper, Frank},
	month = aug,
	year = {2019},
	note = {ADS Bibcode: 2019ApJ...880..115A},
	pages = {115},
}

@article{schootemeijer_constraining_2019,
	title = {Constraining mixing in massive stars in the {Small} {Magellanic} {Cloud}},
	volume = {625},
	issn = {0004-6361},
	url = {https://ui.adsabs.harvard.edu/abs/2019A&A...625A.132S},
	doi = {10.1051/0004-6361/201935046},
	urldate = {2025-10-17},
	journal = {\aap},
	author = {Schootemeijer, A. and Langer, N. and Grin, N. J. and Wang, C.},
	month = may,
	year = {2019},
	note = {Publisher: EDP
ADS Bibcode: 2019A\&A...625A.132S},
	pages = {A132},
}

@article{bellinger_potential_2024,
	title = {The {Potential} of {Asteroseismology} to {Resolve} the {Blue} {Supergiant} {Problem}},
	volume = {967},
	issn = {0004-637X},
	url = {https://ui.adsabs.harvard.edu/abs/2024ApJ...967L..39B},
	doi = {10.3847/2041-8213/ad4990},
	urldate = {2026-04-01},
	journal = {\apj},
	publisher = {IOP},
	author = {Bellinger, Earl Patrick and de Mink, Selma E. and van Rossem, Walter E. and Justham, Stephen},
	month = jun,
	year = {2024},
	note = {ADS Bibcode: 2024ApJ...967L..39B},
	pages = {L39},
}
	\clearpage
	\begin{appendix}
	\section{Fitting example}
    Figure \ref{fig:line_compare} shows an example of some of the lines fit over the temperature domain. These stars range from B0-B8.
    \begin{figure*}[t]
		\centering
		\includegraphics[width=0.75\textwidth]{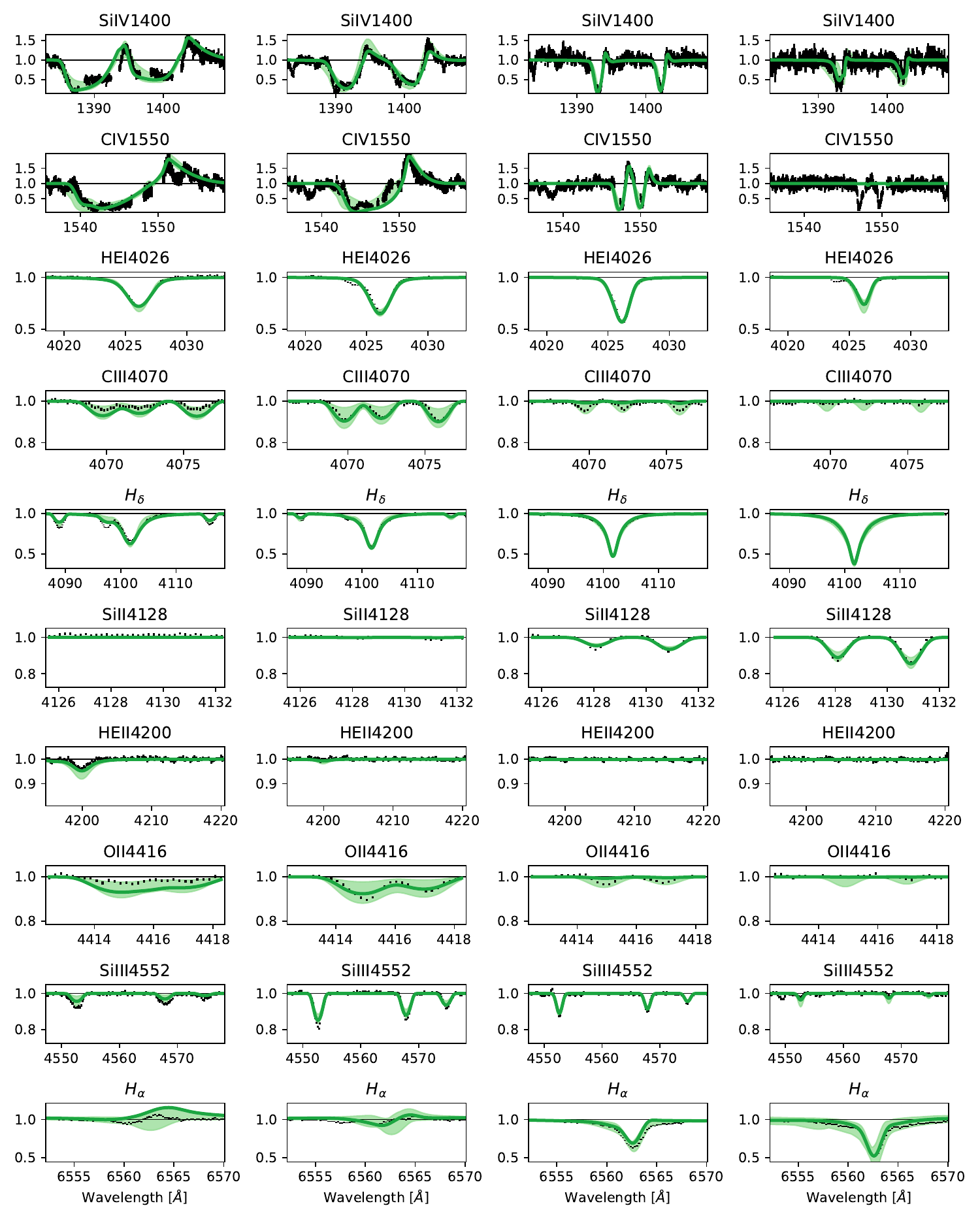}
		\caption{Presentation of normalised line fits to (from left to right) AzV215 (B0\,Ia), AzV96 (B1\,Iab), AzV234 (B2.5\,Ib), and AzV324 (B8\,Ib). 
        The corresponding effective temperatures are 29.3, 23.2, 17.1, and 13.1\,kK, respectively. 
        Data is shown in black; the dark green line denotes the best fit and the light green region the 1-$\sigma$ uncertainty region.}
		\label{fig:line_compare}
	\end{figure*}

    \section{Exploration of potential correlations}
    In this section we have compiled additional figures which explore the clumping behaviour of the sample closer. 

    \blue{Figure \ref{fig:correlation_plot} shows the correlation between many of the resulting wind parameters. 
    This plot shows the Pearson correlation co\"efficient \citep{benesty_pearson_2009} between the different derived parameters not including the stars for which we could not determine $\varv_{\infty}$.
    We account for the error margins on the parameters by computing the correlation matrix 1000 times each time using randomly selected values according to a Guassian distribution based on the best fit and the error margins. 
    When averaging this result we can determine an error margin on this correlation by computing the standard deviation.
    Figure \ref{fig:correlation_plot} uses red to show positive linear correlations and blue to show negative correlation.
    If the numbers are printed in grey the standard deviation of the correlation is larger than the determined mean correlation.}

    \begin{figure}
        \centering
        \includegraphics[width=1.0\linewidth]{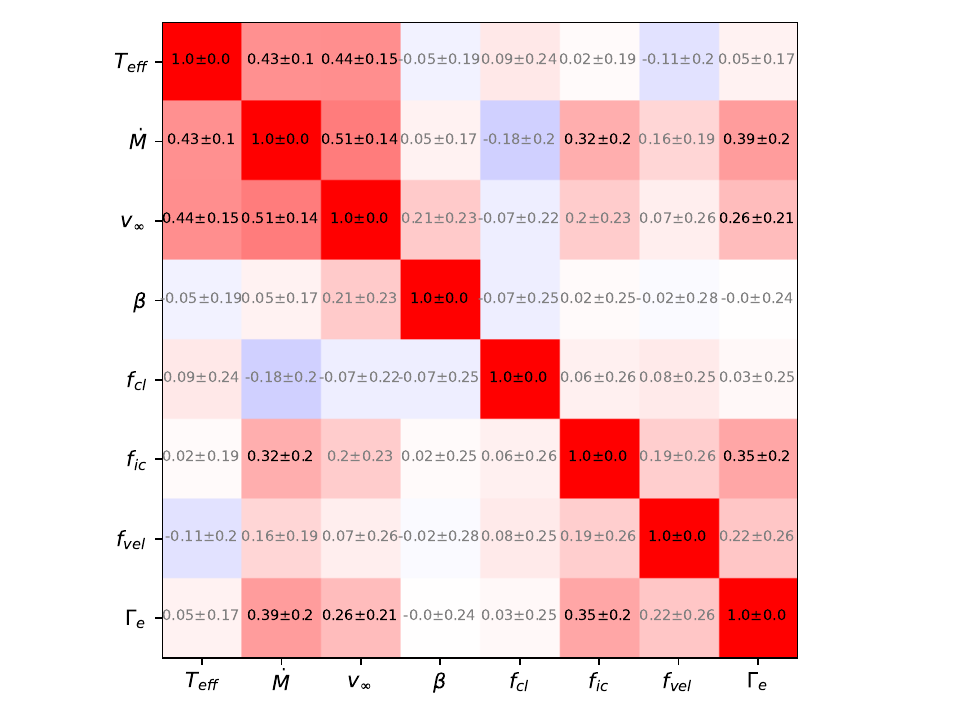}
        \caption{Correlation plot for wind parameters. The colour shows how positive (red) or negative (blue) the Pearson correlation co\"efficient is. If the standard deviation is larger than the derived correlation the values are written in grey. }
        \label{fig:correlation_plot}
    \end{figure}
    \blue{
    Some of the expected correlation are clearly retrieved.
    There is a positive clear mass-loss rate, $T_{\rm eff}$ and $\varv_{\infty}$, $T_{\rm eff}$ correlation as expected. 
    This results, logically, in a correlation between $\varv_{\infty}$ and $\dot{M}$ as well.
    The positive correlation between $f_{\rm ic}$, $\dot{M}$, and $\Gamma_{\rm e}$ is not immediately obvious.
    The two positive correlations are related as the mass-loss rate correlates with the Eddington factor for obvious reasons. 
    A high mass-loss rate resulting in a high $f_{\rm ic}$ might be related to the amplification rate of line de-shadowing instability being inversely correlated to the optical depth of the line \citep{owocki_rybicki_1984}. 
    Therefore, increasing the mass-loss rate reduces the effect of the line de-shadowing instability.  
    This correlation seems to mostly be driven by the highest $\Gamma_{\rm e}$ star, when removed the mean correlation drops to 0.25 only slightly above the standard deviation.
    }

    \begin{figure}[htp!]
		\centering
		\includegraphics[width=0.4\textwidth]{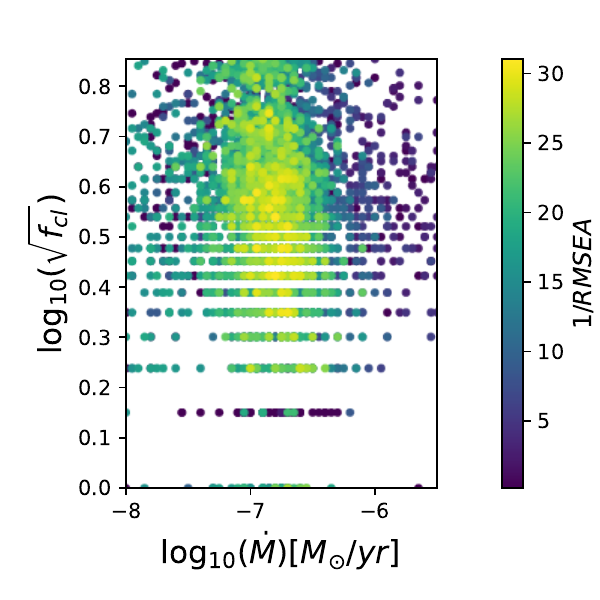}
		\caption{Mass loss versus clumping factor correlation plot for AzV22 (B3\,Ia). The symbol colour yields the quality of the fit expressed as the inverse of the root mean square error of approximation. \blue{We plot the $\sqrt{f_{\rm cl}}$ to emulate the expected clumping relation $f_{\rm cl}(\rho_{\rm uncl})^2 = \langle \rho^2 \rangle$ \citep{puls_2006}, although this correlation only holds for optically thin clumping and a constant clumping factor, both of which are not true here. 
        For the same reason we plot this relation in a log-log scale.} }
		\label{fig:corr_mass_fcl}
	\end{figure}
    
        \begin{figure}[htp!]
			\centering
			\subfigure{\includegraphics[width=0.45\textwidth]{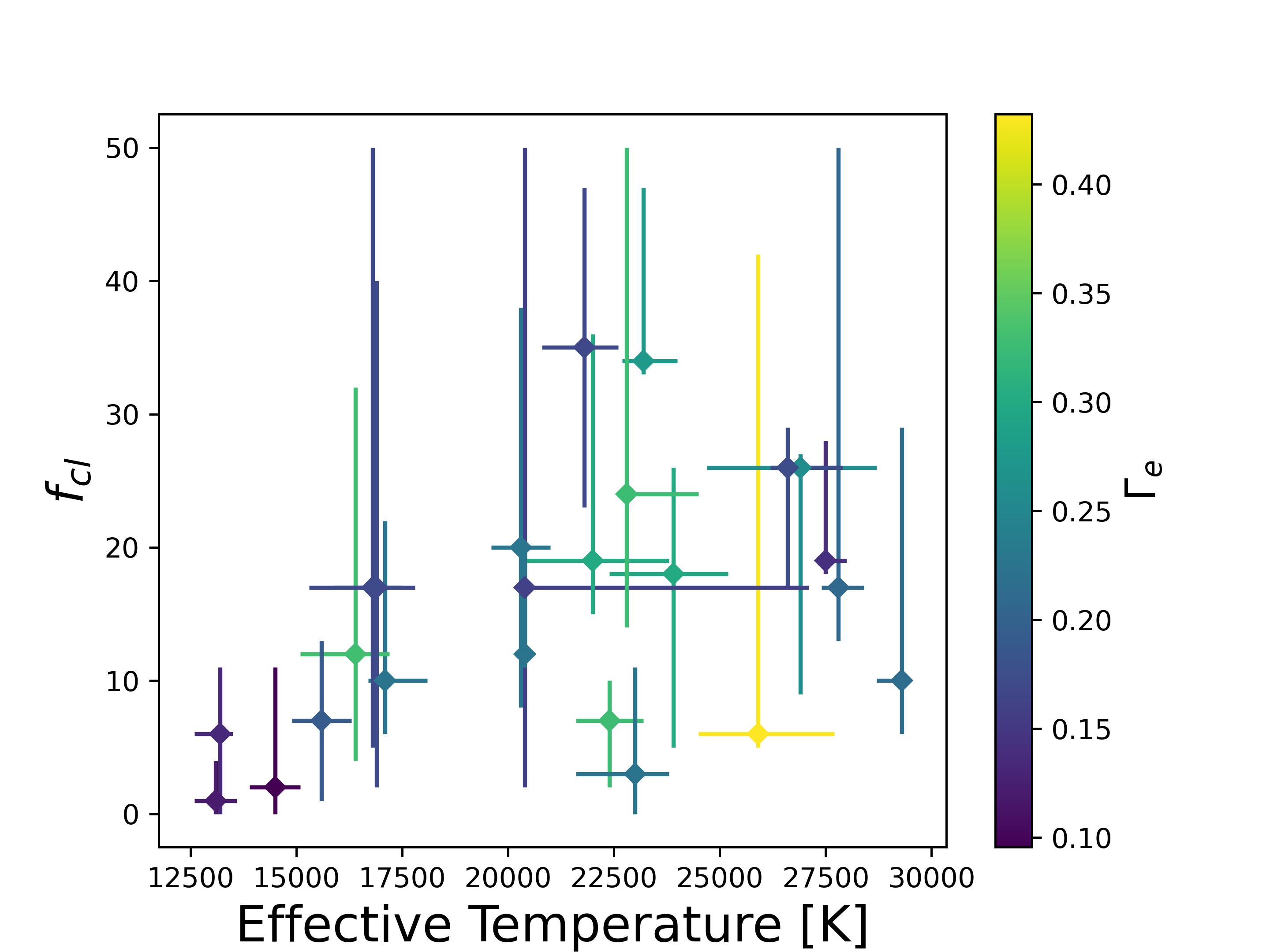}}
			\subfigure{\includegraphics[width=0.45\textwidth]{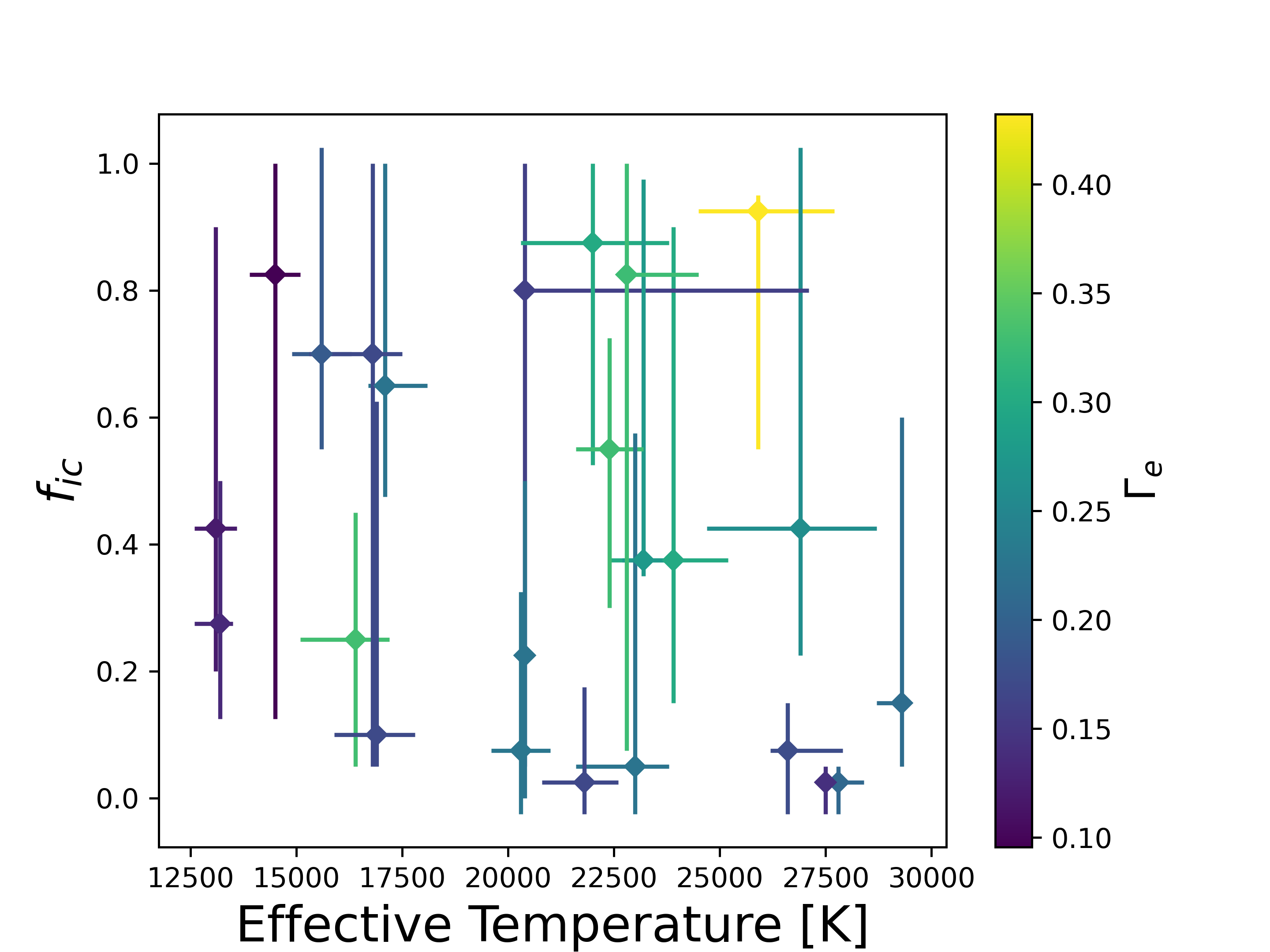}}
			\subfigure{\includegraphics[width=0.45\textwidth]{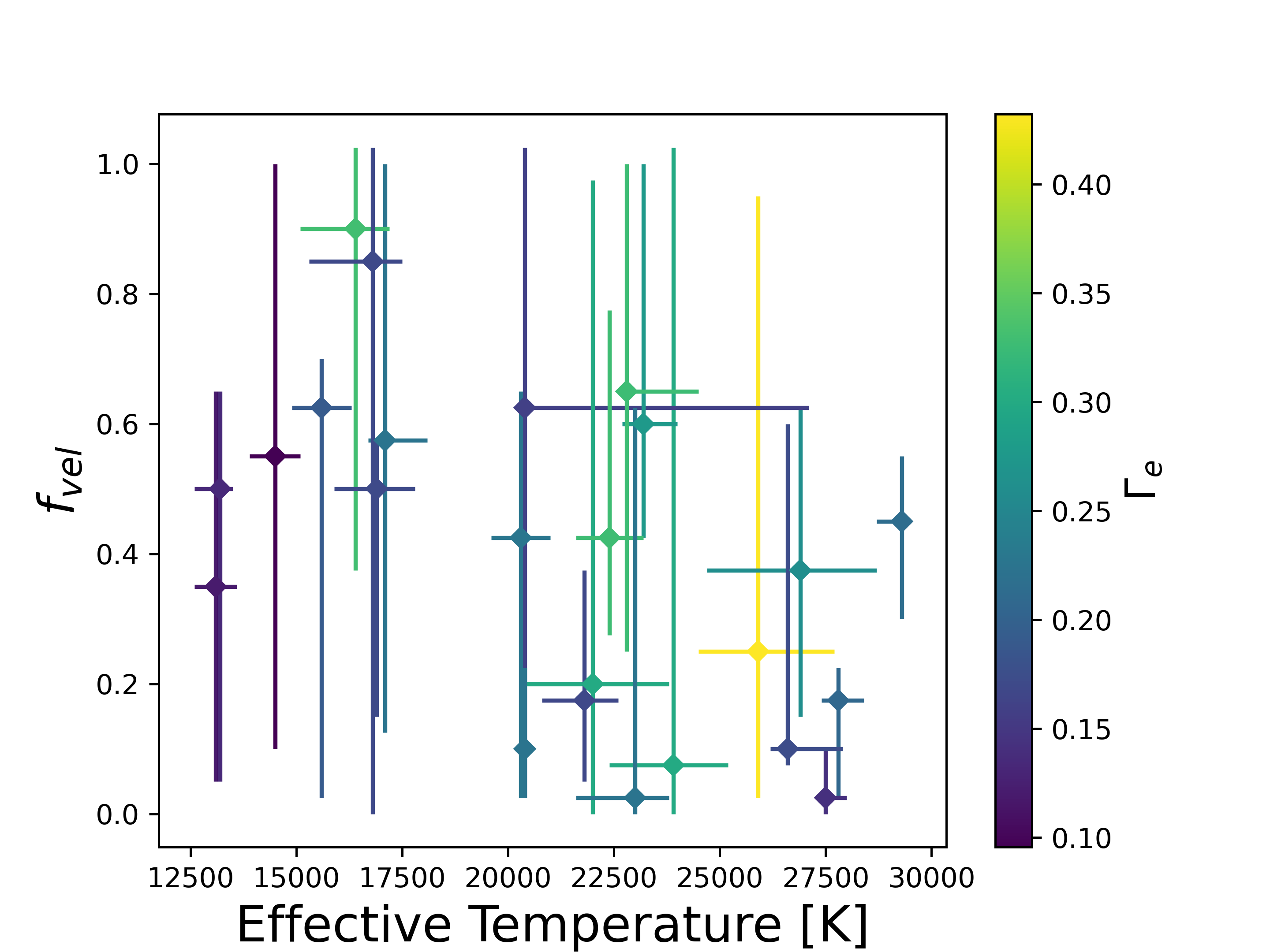}}
			\caption{Clumping parameters with respect to the effective temperature. The panels from top to bottom show the clumping factor, interclump density (in units of mean density), and velocity filling factor. The colour indicates the derived $\Gamma_{\rm e}$ for the stars.  }
			\label{fig:clumping}
		\end{figure}

    \section{Comparative tests to previous studies}\label{sec:compare}
	In this paper, we \blue{appear to have shown that the mass-loss rates of B stars do not depend on metallicity.}
    This result is not in line with some of the basic assumptions of line-driven winds. 
    As a result, we extensively compared our results to other papers which study similar stars.
    Here we show how our results compare to several other studies which studied some of the same stars.
    
    The SMC B star results of \citet{bernini-peron_x-shooting_2024} uses identical data.
    In total there are 13 stars which are studied in both this paper and the sample of \citet{bernini-peron_x-shooting_2024}. 
    Even though that paper uses a very different method we can still compare the mass-loss rates.
    \citet{bernini-peron_x-shooting_2024} does not use a GA method or another systematic fitting method instead using a 'by-eye' fit manually tuning models to find a good fit.
    As a result, deriving error margins of the 'by-eye' fit is difficult.
    \citet{bernini-peron_x-shooting_2024} use their expertise to estimate an error margin which they apply to the full sample.
    Instead of {\sc fastwind} \citet{bernini-peron_x-shooting_2024} use CMFGEN to produce the spectra and assume optically thin clumps.
    \blue{CMFGEN, like {\sc fastwind}, takes into account a whole range of elements in different ionisation stages to determine the temperature stratification, but unlike the version of {\sc fastwind} we used, CMFGEN is able to produce spectral lines of heavier elements such as aluminium and iron to use in determining wind behaviour. 
    By using the 'by-eye' method they are still forced to focus on a limited number of lines, most of which are also included here, with the notable exception of a UV Al\,III line and magnesium lines as we mentioned above. 
    The UV Al\,III line could allow for $\varv_{\infty}$ determination to cooler temperatures and the Mg lines give a more accurate $T_{\rm eff}$ at $T_{\rm eff} <15$\,kK as we already mentioned.}
    Despite all these differences, figure \ref{fig:matheus_compare} shows that there is in general a very good agreement between the derived mass-loss rates between this paper and \citet{bernini-peron_x-shooting_2024}. 
    For all stars except SK 179 there is significant overlap in the error margin between this paper's mass-loss rates and those derived by \citet{bernini-peron_x-shooting_2024}.
    \blue{The difference between our derived $\varv_{\infty}$ and \citet{bernini-peron_x-shooting_2024} is higher than the error margins in 4 cases. 
    Certainly for the cooler stars this can be explained by our lacking of the Al\,III lines, which is represented by our large error margins as well.}
    We also compared other parameters which agree well, however we find that $\log_{10}g_{\rm eff}$ and $T_{\rm eff}$ is consistently higher in this paper when compared to \citet{bernini-peron_x-shooting_2024} (see figure \ref{fig:matheus_compare_stellar}). However for all but 3 of the stars they do still have an overlap in error margins.
    A similar agreement between the methods was also shown on a sample of three O stars \citep{sander_x-shooting_2024}.
    \begin{figure}
        \centering
        \subfigure{\includegraphics[width=1.0\linewidth]{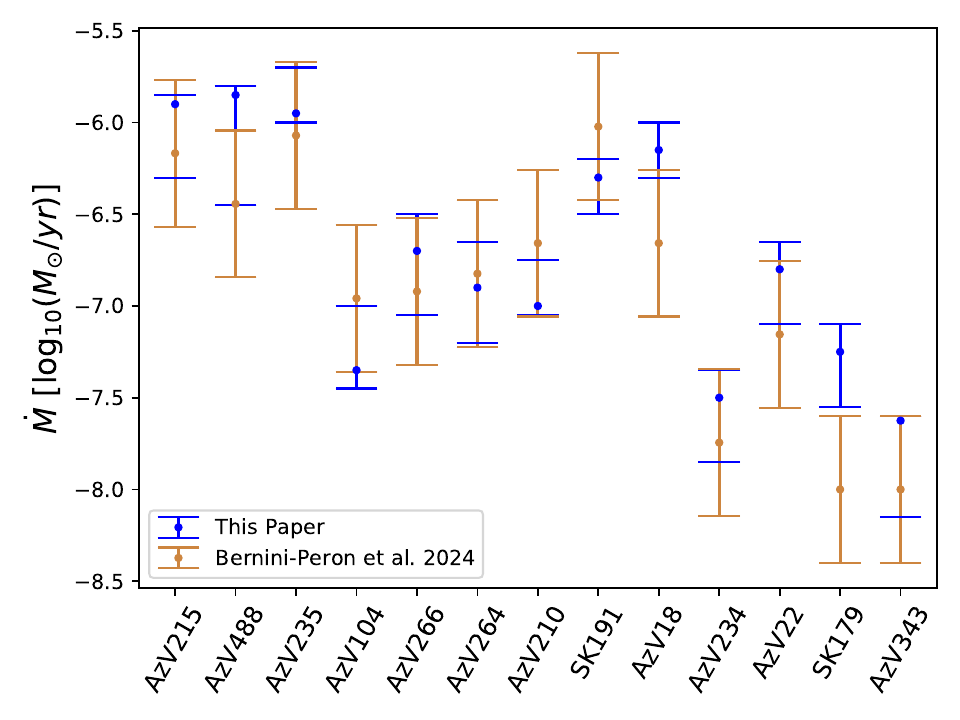}}
        \subfigure{\includegraphics[width=1.0\linewidth]{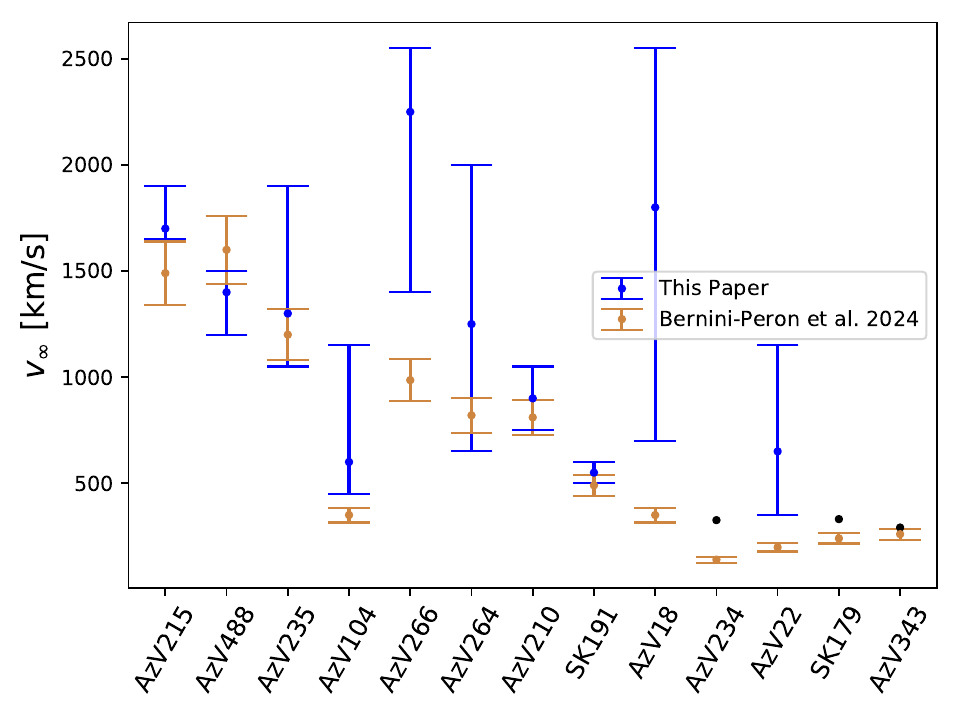}}
        \caption{Comparison between derived mass-loss rates and terminal velocities. Here we show a comparison of the derived mass-loss rates and terminal wind speeds between all stars this paper shares with \citet{bernini-peron_x-shooting_2024}. In blue we show the results of this paper, while orange shows the results of \citet{bernini-peron_x-shooting_2024}. The black points in the $\varv_{\infty}$ show the assumed terminal wind speeds, where we could not derive a satisfactory $\varv_{\infty}$.}
        \label{fig:matheus_compare}
    \end{figure}

        \begin{figure}
        \centering
        \subfigure{\includegraphics[width=1.0\linewidth]{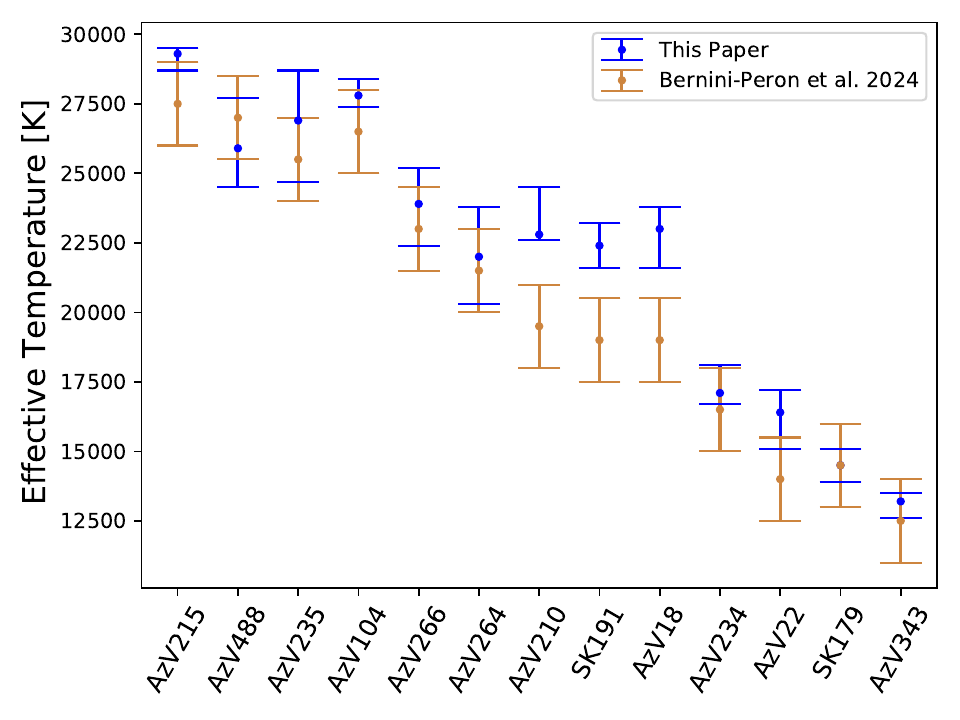}}
        \subfigure{\includegraphics[width=1.0\linewidth]{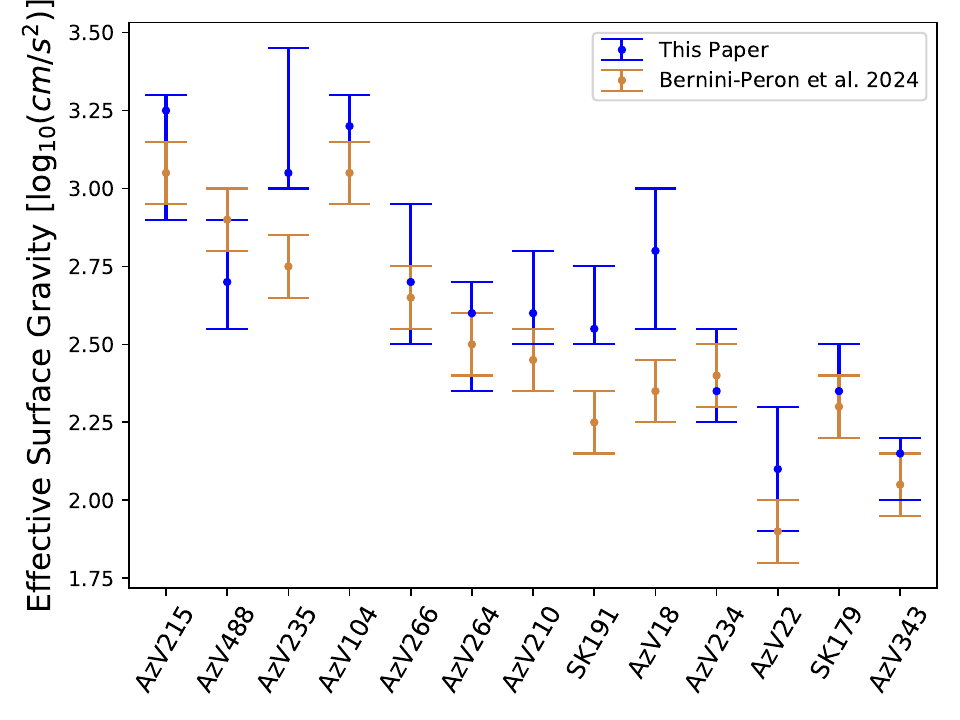}}
        
        \caption{Comparison between derived $T_{\rm eff}$ and $\log_{10} g_{\rm eff}$. Here we show a comparison of the derived $T_{\rm eff}$ (top figure) and the surface gravity (bottom figure) between all stars this paper shares with \citet{bernini-peron_x-shooting_2024}. In blue we show the results of this paper, while orange shows the results of \citet{bernini-peron_x-shooting_2024}}
        \label{fig:matheus_compare_stellar}
    \end{figure}

	\begin{figure*}
		\centering
		\subfigure{\includegraphics[width=0.35\textwidth]{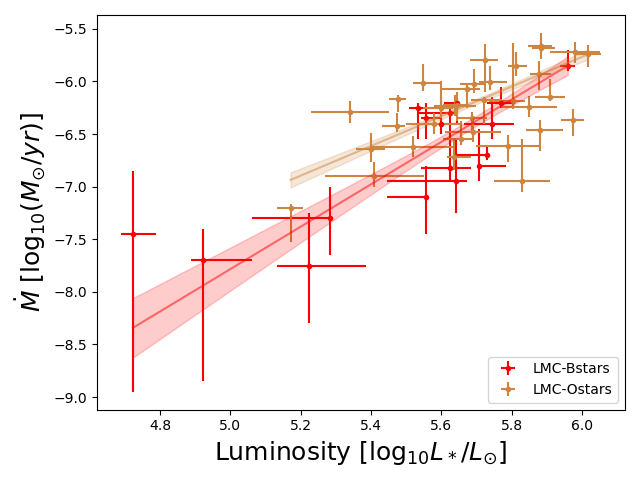}}
		\subfigure{\includegraphics[width=0.35\textwidth]{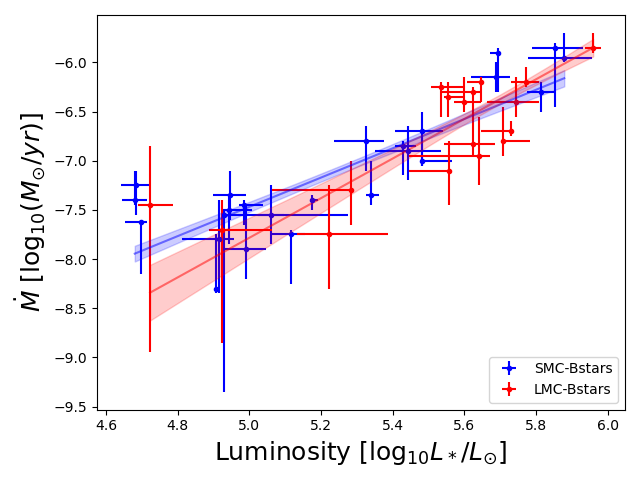}}
		\subfigure{\includegraphics[width=0.35\textwidth]{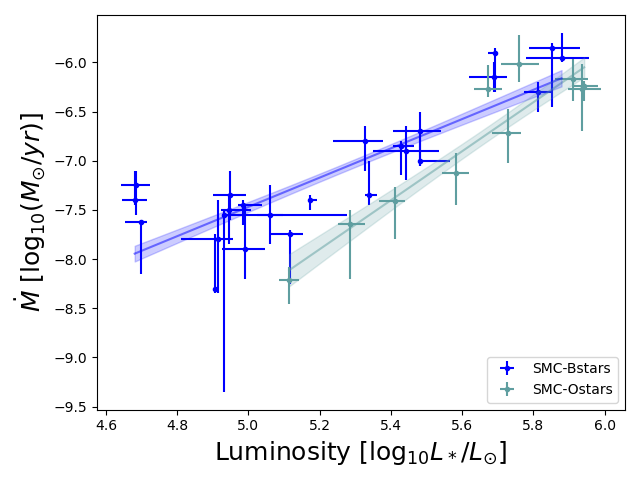}}
		\subfigure{\includegraphics[width=0.35\textwidth]{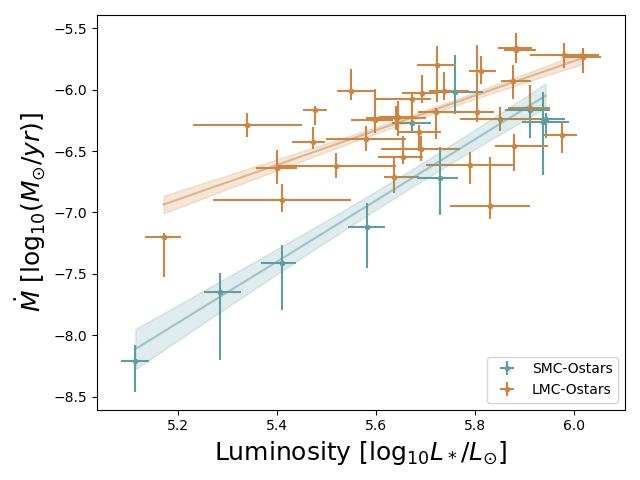}}
		\caption{Mass-loss rate comparisons. These plots all show the mass-loss rate vs the luminosity from this study and the previous SMC and LMC studies in this series of papers \citep{hawcroft_x-shooting_2024,backs_x-shooting_2024, verhamme_x-shooting_2024, brands_x-shooting_2025}. The fits were performed using an ODR method. The error margins shown are the 1$\sigma$ uncertainty. }
		\label{fig:mass-loss_plots_full}
	\end{figure*}

Even when comparing to previous studies using different data we find relatively strong agreement. 
\citet{trundle_understanding_2004} is an optical only study on 8 SMC B stars of which 6 have also been studied in this paper (AzV215, AzV104, SK191, AzV210, AzV18, AzV22).
The $\log_{10}g_{\rm eff}$ and $T_{\rm eff}$ agree within error margin for all stars, but AzV215 for which we find $T_{\rm eff} = 29300_{-600}^{+200}$\,K, $\log_{10}(g) = 3.25_{-0.35}^{+0.05}$, while \citet{trundle_understanding_2004} finds $T_{\rm eff} = 27000 \pm 1000$\,K, and $\log_{10}(g_{\rm eff}) = 2.9 \pm 0.1$.
Here we see the effective surface gravity still overlaps, but we find a higher $T_{\rm eff}$ with a very low uncertainty.
The mass-loss rates agree within error margins for all stars, but AzV18. 
We find $\log_{10} (\dot{M}) = -6.15 \pm 0.15 [\log_{10}M_{\odot}/\text{yr}]$ and \citet{trundle_understanding_2004} derives  $\log_{10} (\dot{M}) = -6.64 \pm 0.05 [\log_{10}M_{\odot}/\text{yr}]$.
Even this difference is not worrying, considering the difference in studied lines and would not effect our conclusions. 
\citet{trundle_understanding_2005} is another optical only study for which some terminal wind speeds were known from \citet{evans_ultraviolet_2004}.
This study has 2 stars in common with this study (AzV264, AzV96).
Within error margins the $T_{\rm eff}$, surface gravity, and the mass-loss rates agree. 
These comparisons show that although the resulting conclusions of our derived mass-loss rates are surprising the derived parameters themselves are similar to previous studies of these stars.

Finally we also decided to compare to the empirical prescription by \citet{pauli_new_2025}, which uses a $\Gamma_e$ dependence instead of a $T_{\rm eff}$, mass, and luminosity dependence. 
We note an average $f_{\rm p/e} =1.3 \pm 0.4$ for the SMC B stars, and an average $f_{\rm p/e} =1.2\pm 0.9$ for the SMC O stars. 
The LMC systematically does worse with an average ratio of LMC O stars of $f_{\rm p/e} =3.6 \pm 1.0$ and $f_{\rm p/e} =3.4 \pm 2.4$ for LMC B stars. 
The results for the SMC seem to be comparable with the results LIME \citep{sundqvist_--fly_2025} provides.
The LMC, in comparison, is overestimated and with a significant scatter.
The success of this prescriptions is not entirely surprising as it is based on previous empirical studies which this paper, as shown above, is in agreement with.

\begin{figure}
		\centering
		\includegraphics[width=0.7\linewidth]{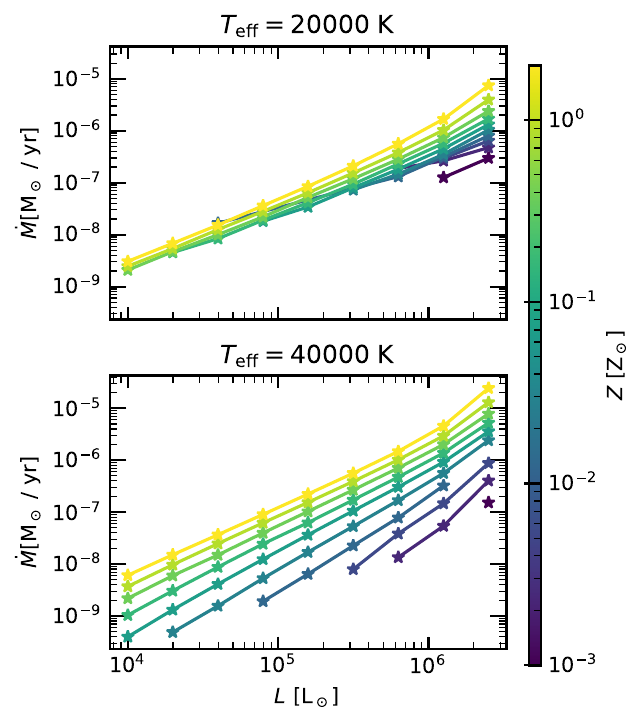}
		\caption{LIME mass-loss predictions as function of luminosity for different metallicities and temperatures. Each panel shows a $T_{\rm eff}$ value. The coloured lines from blue to yellow are for $\log_{10}(Z/Z_{\odot}) = -3,-2.6, -2.3, -1.9, -1.5, -1.2, -0.8, -0.4, -0.07, 0.3$.}
		\label{fig:LIME_per_teff}
\end{figure}

\section{$\varv_{\infty}$ behaviour}\label{sec:vinf_teff_discussion}
\blue{From previous theoretical works we expect an existing, but weak $\varv_{\infty}$ metallicity dependence. 
\citet{vink_metallicity-dependent_2021}, using a Monte Carlo approach, find $\varv_{\infty} \propto Z^{0.19}$. 
\citet{bjorklund_new_2021}, on the other hand find a relation indistinguishable from zero ($\varv_{\infty} \propto Z^{-0.10 \pm 0.18}$). 
However, they do note some potential issues with their terminal wind speeds.}

We compare the previously determined $\varv_{\infty}$, $T_{\rm eff}$ relation through a Sobolev exact integration technique \citep{hawcroft_x-shooting_2024} with the combined results of this work with \citet{backs_x-shooting_2024, hawcroft_empirical_2024, verhamme_x-shooting_2024,brands_x-shooting_2025}. 
Table \ref{tab:teff_vinf_fit} shows the results of our ODR fit and the fit of \citet{hawcroft_x-shooting_2024}, both using a simple linear fitting law,
\begin{align}
\varv_{\infty} = a \cdot T_{\rm eff} -b. 
\end{align}
It is clear that both the slope and the offset for the LMC is almost identical although we retrieve a lower 1-$\sigma$ error margin.
The SMC slope is different between the two studies, but for the \citet{hawcroft_x-shooting_2024} results still within error margins. 
The fits are plotted in the upper panel of figure \ref{fig:fit_full_SMC+LMC}, which show that the smaller slope is caused mostly by some hot SMC O stars which have a lower $\varv_{\infty}$ in comparison to the LMC O stars with the same $T_{\rm eff}$. Up to 35\,kK the $\varv_{\infty}$ dependence on $T_{\rm eff}$ of the SMC appears mostly equal to that of the LMC. 
We also plot the \citet{hawcroft_x-shooting_2024} results over the same results in the lower panel of figure \ref{fig:fit_full_SMC+LMC}, showing that their retrieved results still function within error margins.
When fitting the full sample with a new Z dependence, 
\begin{align}
    \varv_{\infty} = aT_{\rm eff} \cdot (Z/Z_\odot)^{c}-b,
\end{align} 
we find that $a =(8.5 \pm 0.2)\cdot 10^{-2} \frac{km}{sK} $, b = $1080 \pm 55 \frac{km}{s}$ , and  c = $(-0.9\pm 1.1) \cdot 10^{-2} $. 
We note here that the Z-dependence is within error margins equal to zero, showing the lack of Z dependence of the terminal wind speed. 
\blue{This result was also hinted at in \citet{bernini-peron_x-shooting_2024}, where the found $\varv_{\infty}$, $T_{\rm eff}$ relation was suggested to be not impacted by metallicity.
Here they compared their found relation for the SMC B stars and found it remarkably similar to the LMC relation of \citet{hawcroft_x-shooting_2024}.}

\begin{table}[]
    \caption{Fits of $\varv_{\infty}$ with $T_{\rm eff}$. }
    \centering
    \tiny
    \begin{tabular}{c|cc}
              This work        & $a$ [km $\cdot$s$^{-1}$K$^{-1}$]& $b$ [km$\cdot$s$^{-1}$]\\   
              \hline
         SMC stars & $(7.9\pm 0.3) \cdot 10^{-2}$ & $993 \pm 99$\\
         LMC stars &  $(8.8\pm 0.2)\cdot 10^{-2}$ & $1083 \pm 66$ \\
         \hline
         Hawcroft et al. 2024 & &\\
         \hline
         SMC stars & $(8.9\pm 1.1) \cdot 10^{-2}$ & $1560 \pm 420$ \\ 
         LMC stars & $(8.8\pm 0.4) \cdot 10^{-2}$ & $1200 \pm 150$ \\ 

    \end{tabular}
    \begin{tablenotes}
		\item \textbf{Note:} We compared the $\varv_{\infty}$, $T_{\rm eff}$ fit of this sample with the results of \citet{hawcroft_x-shooting_2024}. The fit formula is given by formula \ref{eq:hawcroft}.
	\end{tablenotes}
    \label{tab:teff_vinf_fit}
\end{table}

\begin{figure}
    \centering
    \subfigure{\includegraphics[width=0.9\linewidth]{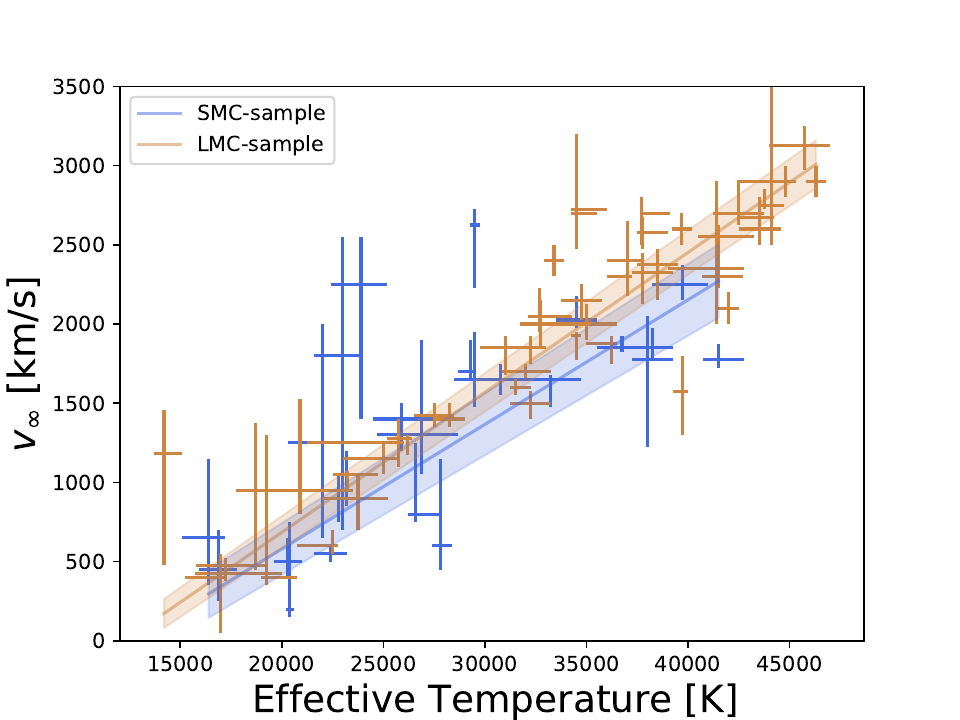}}
    \subfigure{\includegraphics[width=0.9\linewidth]{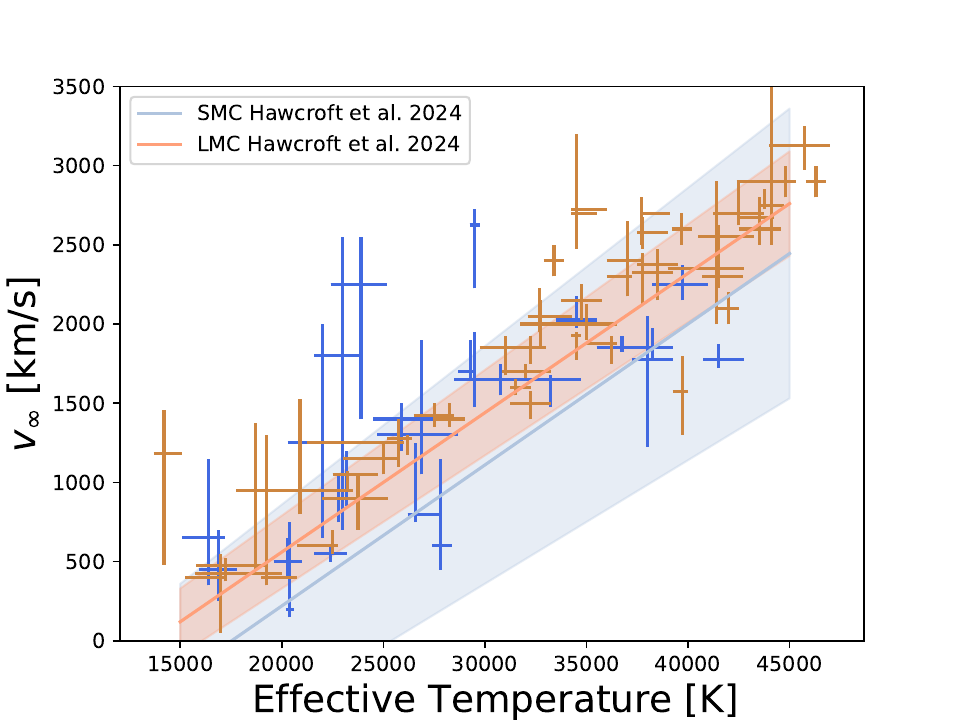}}
    \caption{Relation between $T_{\rm eff}$ and $\varv_{\infty}$. 
    The top panel shows the combined results from \citet{backs_x-shooting_2024, verhamme_x-shooting_2024, hawcroft_empirical_2024, brands_x-shooting_2025} and a fit separating the SMC and LMC sample.
    The bottom panel shows the same sample but with the results of \citet{hawcroft_x-shooting_2024} overplotted.}
    \label{fig:fit_full_SMC+LMC}
\end{figure}

Although, as noted in the main text, the indicator of $\varv_{\infty}/\varv_{\rm esc}$ is a tenuous one, due to major uncertainties. 
We also plot here two fits of the relation in figure \ref{fig:fit_vinf_vesc}. 
\blue{We fit a linear function to the full sample regardless of metallicity up to $T_{\rm eff} = 25$\,kK as a very rough estimate to where we expect the increase to stop. 
We note that data points with low error margins have a higher weight when determining the linear fit.
Therefore the most noticeable points at low $T_{\rm eff}$, with a high $\varv_{\infty}/\varv_{\rm esc}$ and high error margins are only weakly impacting the fit.
The fit of stars with $T_{\rm eff}>25$\,kK are also plotted. 
We also overplot the original \citet{lamers_terminal_1995} result.}

\begin{figure}
    \centering
    \includegraphics[width=0.9\linewidth]{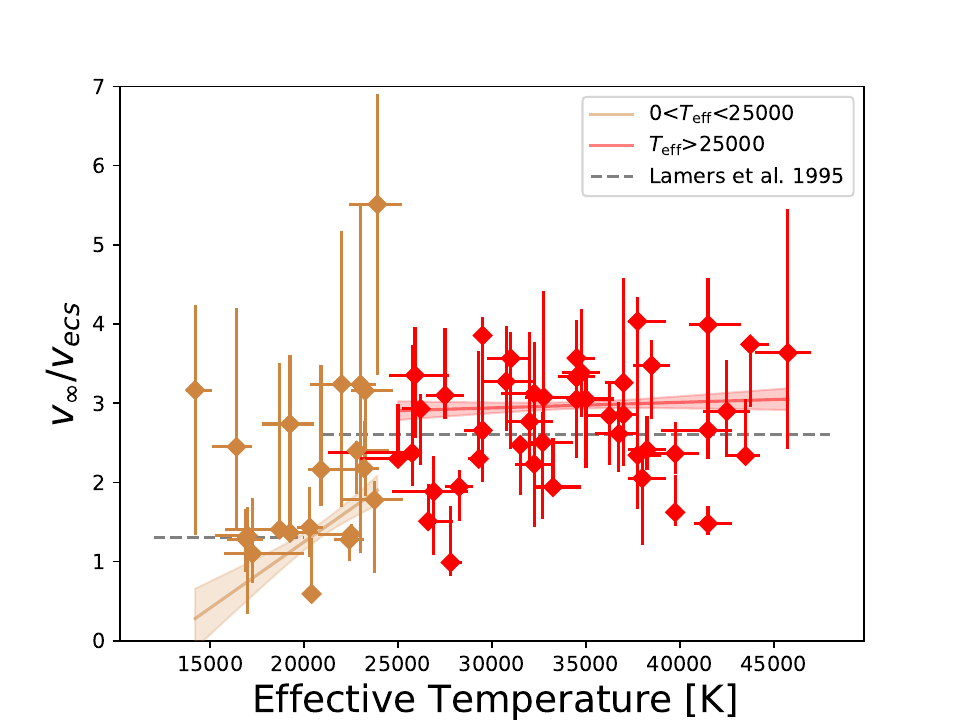}
    \caption{$\varv_{\infty}/\varv_{\rm esc}$ over $T_{\rm eff}$. We show the full sample including all results from \citet{backs_x-shooting_2024, hawcroft_empirical_2024, verhamme_x-shooting_2024, brands_x-shooting_2025} which are fitted by two linear functions.
    The orange line fits only the stars up to $T_{\rm eff}=25$\,kK, while the red line fit stars with $T_{\rm eff}>25$\,kK. The dashed lines are the initial results from \citep{lamers_terminal_1995}.}
    \label{fig:fit_vinf_vesc}
\end{figure}

    \clearpage

	\section{Line list}

	\begin{table}[htp!]
		\caption{\normalsize{Detailed list of all spectral lines which are fitted. }}\label{tab:linelist}
		\begin{tabular}{ccc}
			\hline
			Ion& Wavelength $[\AA]$&Line window\\
			\hline
			C\,{\sc iv} & 1168.9, 1169.0 & C\,{\sc iv}\,1196b\\
			C\,{\sc iii} & 1174.9, 1175.3, 1175.6  & C\,{\sc iv}\,1169b\\
			& 1175.7, 1176.0, 1176.4 & \\
			Si\,{\sc iv} & 1393.8, 1402.8 & Si\,{\sc iv}\,1400\\
			C\,{\sc iv} & 1548.2, 1550.8& C\,{\sc iv}\,1550\\
			\hline
			O\,{\sc iii} & 3961.6 & H$\epsilon$\\
			He\,{\sc i} &  3964.7 &H$\epsilon$\\
			H\,{\sc i} & 3970.1 & H$\epsilon$\\
			He\,{\sc ii} & 4025.4 & He\,{\sc i} 4026\\
			He\,{\sc i} & 4026.2 & He\,{\sc i} 4026\\
			C\,{\sc iii} & 4068.9, 4070.3 & C\,{\sc iii} 4070\\
			O\,{\sc ii} & 4069.6, 4069.9, & C\,{\sc iii} 4070 \\
			& 4072.16, 4075.86 & \\
			S\,{\sc iv} & 4088.9, 4116.1 & H$\delta$ \\
			N\,{\sc iii} & 4097.4, 4103.4 & H$\delta$ \\
			H\,{\sc i} & 4101.7 & H$\delta$ \\
			Si\,{\sc ii} & 4128.1, 4130.9 & Si\,{\sc ii} 4128\\
			N\,{\sc iii} & 4195.8, 4200.1, 4215.77 & He\,{\sc ii} 4200\\ 
			He\,{\sc ii} & 4199.6 & He\,{\sc ii} 4200\\
			C\,{\sc ii} & 4267.0, 4267.3 & C\,{\sc ii} 4267\\
			He\,{\sc ii} & 4338.7 & H$\gamma$\\
			H\,{\sc i} & 4340.5 & H$\gamma$\\
			O\,{\sc ii} & 4317.1,4319.6, 4366.9 & H$\gamma$\\
			N\,{\sc iii} & 4345.7, 4332.91 &  H$\gamma$\\
			N\,{\sc iii} & 4379.0, 4379.2 & He\,{\sc i} 4387 \\
			He\,{\sc i} & 4387.9 & HeI4387 \\
			O\,{\sc ii} & 4414.9, 4417.0 & O\,{\sc ii} 4416\\
			He\,{\sc i} & 4471.5 & He\,{\sc i} 4471\\
			N\,{\sc ii} & 4613.9, 4621.4, 4630.5 & N\,{\sc ii} 4601\\
			& 4601.5, 4607.2, 4643.1 & \\
			N\,{\sc iii} & 4534.6 & He\,{\sc ii} 4541\\
			He\,{\sc ii} & 4541.4 & He\,{\sc ii} 4541\\
		\end{tabular}
	\end{table}
	\begin{table}
		\begin{tabular}{ccc}
			Si\,{\sc iii} &4552.6, 4567.8, 4574.8 & Si\,{\sc iii} 4552\\
			C\,{\sc iii} & 4647.4, 4650.2, 4651.5 & C\,{\sc iii} N\,{\sc iii} COLD/HOT\\
			N\,{\sc iii} & 4634.1, 4640.6, 4641.9 & C\,{\sc iii} N\,{\sc iii} COLD/HOT \\
			O\,{\sc ii} & 4638.9, 4641.8, & C\,{\sc iii} N\,{\sc iii} COLD\\
			&  4661.6, 4676.2& \\
			He\,{\sc ii} & 4685.6 & He\,{\sc ii} 4686 \\
			N\,{\sc iii} & 4858.7, 4859.0, 4861.3,  & H$\beta$\\
			& 4867.1, 4867.2, 4873.6 &\\
			He\,{\sc ii} & 4859.1 & H$\beta$ \\
			H\,{\sc i} & 4861.4 & H$\beta$\\
			He\,{\sc i} & 4921.9 & He\,{\sc i} 4922\\
			He\,{\sc ii} & 5411.3 & He\,{\sc ii} 5411\\
			O\,{\sc iii} & 5592.3 & O\,{\sc iii} 5592\\
			C\,{\sc iii} & 5695.9 & C\,{\sc iii} 5695 \\
			He\,{\sc i} & 5875.6 & He\,{\sc i} 5875 \\ 
			He\,{\sc ii} & 6527.1 & He\,{\sc ii} 6527\\
			He\,{\sc ii} & 6559.8 & H\,${\alpha}$\\
			H\,{\sc i} & 6562.8 & H\,${\alpha}$\\
			C\,{\sc ii} & 6578.1, 6582.9 & C\,{\sc ii} 6578\\
			He\,{\sc i} & 6678.2 & He\,{\sc ii} 6683\\
			He\,{\sc ii} & 6682.8 & He\,{\sc ii} 6683\\
			He\,{\sc i} & 7065.2 & He\,{\sc i} 7065\\
			\hline
			\normalsize
		\end{tabular}
    \begin{tablenotes}
		\item \textbf{Note:} The first column shows the atom and its ionisation stage responsible for the transition. The second column shows the corresponding wavelength with possible multiplets. The third column shows where to find this line in the fit summary which is available on Zenodo.
	\end{tablenotes}
	\end{table}
\end{appendix}
\end{document}